\documentclass[twocolumn]{aastex631}

\newcommand\swift{{\it Swift}}
\newcommand\hst{{\it HST}}
\newcommand\chandra{{\it Chandra}}

\newcommand\astrosat{{\it AstroSat}}

\newcommand\s{{\rm~s}}

\newcommand\kev{{\rm~keV}}
\newcommand\keV{{\rm~keV}}
\newcommand\ev{{\rm~eV}}
\newcommand\angs{{\rm~\AA}}

\usepackage{enumitem}
\usepackage{booktabs}
\usepackage{multirow}
\usepackage{array}
\newcolumntype{H}{>{\setbox0=\hbox\bgroup}c<{\egroup}@{}}

\shorttitle{Far UV spectroscopy of AGN}
\shortauthors{Kumar et al.}
\graphicspath{{./}{figures/}}

\usepackage{booktabs, makecell}
\usepackage{booktabs}
\usepackage{array}
\usepackage{verbatim}
\usepackage{amsmath}
\begin{document}

\title{FAR-ULTRAVIOLET SPECTROSCOPY OF ACTIVE GALACTIC NUCLEI WITH \textit{ ASTROSAT}/UVIT}

\correspondingauthor{Shrabani Kumar}
\email{shrabani@iucaa.in}

\author{Shrabani Kumar }
\affiliation{Inter-University Centre for Astronomy and Astrophysics, Pune, 411007, India}

\author{G. C. Dewangan}
\affiliation{Inter-University Centre for Astronomy and Astrophysics, Pune, 411007, India}
\author{K. P. Singh}
\affiliation{Education and Research Mohali, Knowledge City, Sector 81, Manauli P.O., SAS Nagar, 140306, Punjab, India}
\affiliation{Department of Astronomy and Astrophysics, Tata Institute of Fundamental Research, 1 Homi Bhabha Road, Mumbai 400005, India}
\author{P. Gandhi}
\affiliation{School of Physics \& Astronomy, University of Southampton, Highfield SO17 1BJ, UK}
\author{I. E. Papadakis}
\affiliation{Department of Physics and Institute of Theoretical and Computational Physics, University of Crete, 71003 Heraklion, Greece}
\affiliation{Institute of Astrophysics—FORTH, N. Plastira 100, 70013 Vassilika Vouton, Greece}
\author{P. Tripathi}
\affiliation{Inter-University Centre for Astronomy And Astrophysics, Pune, 411007, India}
\author{L. Mallick}
\affiliation{Cahill Center for Astronomy and Astrophysics, California Institute of Technology, Pasadena, CA 91125, USA}

%% AASTeX 6.31 has the new \collaboration and \nocollaboration commands to
%% provide the collaboration status of a group of authors. These commands 

\begin{abstract}
We study accretion disk emission from eight Seyfert $1 - 1.5$ active galactic nuclei (AGN) using far ultra-violet ($1300-1800$\angs) slit-less grating spectra acquired with \astrosat{}/UVIT. We correct for the Galactic and intrinsic extinction, contamination from the host galaxies, narrow and broad-line regions, \ion{Fe}{2} emission and Balmer continuum, and derive the intrinsic continua.
We use \hst{} COS/FOS spectra to account for the emission/absorption lines in the low-resolution UVIT spectra.
 We find generally redder power-law ($f_\nu \propto \nu^{\alpha}$) slopes ($\alpha \sim -1.1 - 0.3$) in the far UV band than predicted by the standard accretion disk model in the optical/UV band. We fit accretion disk models such as the multi-temperature disk blackbody (\texttt{DISKBB}) and  relativistic disk (\texttt{ZKERRBB}, \texttt{OPTXAGNF}) models to the observed intrinsic continuum emission.  
 We measure the inner disk temperatures using the \texttt{DISKBB} model for seven AGN. These temperatures in the range $\sim 3.6-5.8\ev$ are lower than the peak temperatures predicted for standard disks around maximally spinning super-massive black holes accreting at Eddington rates. The inner disks in two AGN, NGC~7469 and  Mrk~352, appear to be truncated at $\sim 35-125r_g$ and $50-135r_g$, respectively. While our results show that the intrinsic FUV emission from the AGN are consistent with the standard disks, it is possible that UV continua may be affected by the presence of soft X-ray excess emission, X-ray reprocessing, and thermal Comptonisation in the hot corona.   
 Joint spectral modeling of simultaneously acquired UV/X-ray data may be necessary to further investigate the nature of accretion disks in AGN.

\end{abstract}
\keywords{accretion, accretion disks -- galaxies: Seyfert -- techniques: spectroscopic -- ultraviolet: galaxies}

\section{Introduction} \label{sec:intro}
The large luminosity ($\sim 10^{40} - 10^{48}{\rm~erg~ s^{-1}}$; \citealt{ho1999spectral,woo2002active,duras2020universal}) observed from  active galactic nuclei (AGN) is thought to arise due to the accretion of matter, in the form of disks, onto supermassive black holes (SMBHs) at the centers of active galaxies. Luminous AGN are thought to host radiatively efficient standard accretion disks that are geometrically thin and optically thick \citep[see e.g.,][]{koratkar1999ultraviolet, netzer_2013}.
%
%The spectral energy distributions (SEDs) of radio-quiet AGN comprise of mainly three components -- the primary X-ray continuum, the big blue bump (BBB), and the infra-red bump. The primary X-ray emission is believed to arise from a hot ($kT_e \sim$ 100keV), optically thin ($ \tau \lesssim 1$) corona as a result of repeated inverse Compton scattering of optical/UV photons from the accretion disk by hot electrons in the corona \citep{haardt1991two,10.1093/mnras/stv1218}. 
 The standard disk  model, also known as the $\alpha$ disk model, describes the accretion flow onto black holes and predicts the emission spectrum from geometrically-thin and optically-thick accretion disks \citep{1973A&A....24..337S, 1973blho.conf..343N}. According to the standard disk model, each annulus of the disk emits as a  blackbody with temperature varying as a function of radius ($T\propto r^{-3/4}$) and the spectrum of the entire disk varying as $f_\nu \propto \nu^\alpha, ~ \rm where ~\alpha = {\frac{1}{3}}$ in the optical/UV band of AGN emission. 
 
 Observationally, the big blue bump (BBB) emission in the optical/UV band is believed to be arising from the accretion disks in AGN,  thus providing a direct probe of the accretion flow. A number of studies have investigated the optical/UV spectral shape and found varying spectral slopes in different AGN. 
 
 \begin{deluxetable*}{lccccccc}
\tablenum{1}
\tablecaption{List of AGN and their general properties\label{tab:genpar}}
\tablewidth{0pt}
\tablehead{
\colhead{Sources} & \colhead{ Redshift } & \colhead{ B} &\colhead{N$_H^\dagger$} & \colhead{$M_{BH}^{\dagger\dagger}$ (ref)}   & \colhead{$f_{14-195\kev}^{\dagger\dagger\dagger}$} & \colhead{$\rm E_{B-V}$}
\\
\colhead{} & \colhead{} & \colhead{} &\colhead{($10^{20}{\rm~cm^{-2}}$)} & \colhead{($10^7{\rm~M_{\odot}}$)} & \colhead{}  & \colhead{}}
\startdata
Mrk~841 &  $0.036$ &  $14.50$ & $2.0$  &$4.9$ (1) &$33.3$& $0.032$ \\
MR~2251--178   &  $0.064$ & $14.99$ &$2.6$  &$31.6$ (2)&$102.6$& $0.038$ \\
PG 0804+761 & $0.1$ & $15.03$ & $3.3$&  $54$ (3)& $16.7$ & $0.048$\\
NGC~7469 &  $0.016$ & $13.0$ &$4.5$  &$1.0$ (4)& $83.0$  & $0.065$  \\
I~Zw~1 & $0.06$ &$14.4 $& $4.6$ & $3.1$ (5)& -- & $0.067$\\
SWIFT~J1921.1--5842 &  $0.037$ & $14.36$ &$4.9$  &$1.25$  (6)  &$54.4$  & $0.071$ \\
Mrk~352  &  $0.015$ & $15.25$  & $5.6$ &$0.6$ (2)&$41.6$  & $0.082$ \\
SWIFT~J1835.0+3240 &  $0.058$ & $16.50$ & $6.2$ &$100$ (7) & $81.0 $ & $0.09$\\
\enddata
$^\dagger$ The Galactic \ion{H}{1} column density (N$_H$)  obtained from the  N$_H$ calculator available at the HEASARC website \url{https://heasarc.gsfc.nasa.gov/cgi-bin/Tools/w3nh/w3nh.pl}.  \\
$^{\dagger\dagger}$ Black hole masses are taken from (1)~\citet{Vestergaard_2006}, (2)~\citet{Winter_2012}, (3)~\citet{2015PASP..127...67B}, (4)~\citet{2004ApJ...613..682P}, (5)~\citet{wilkins2021light}, (6)~\citet{Wang_2007} and
(7)~\citet{10.1111/j.1365-2966.2004.07822.x}. 
		\\
$^{\dagger\dagger\dagger}$ \swift{} BAT X-ray flux in units of $10^{-12}{\rm~ergs~cm^{-2}~s^{-1}}$ and in the $14-195\kev$ band.\\
 $\rm E_{B-V}$ is calculated using Eq.~\ref{eq3:nhgal}.
%	h. Flux calculated in $0.3$ to $10$ keV band.\\ 
\end{deluxetable*}

 \citet{10.1093/mnras/233.4.801} found the UV slope ($\alpha$) of 81 quasars varying from $0$ to $-1.5$ with a median value of $-0.7$ using  the \textit{International Ultraviolet Explorer (IUE)} observations in the  $1215 - 1900$\angs~ band. They found the ultraviolet continuum to become harder with increasing redshift. \citet{1991ApJ...373..465F} accounted for  \ion{Fe}{2} emission and Balmer continuum, and found the median $\alpha \sim -0.32$ in the  $1450 - 5050$\angs~ band  for their sample of 718 quasars observed with the  \textit{Multiple Mirror Telescope (MMT)} and \textit{Du Pont Telescope}  at Las Campanas Observatory. Using the \textit{Hubble Space Telescope}, \citet{1997ApJ...475..469Z} found the UV slope  $\alpha \sim -0.99$ for their composite spectrum of 101 quasars in the $1050 - 2200$\angs~  band. In the sample of 220 AGN in \citet{2003astro.ph.10165K}, the UV continuum slope has a large scatter varying from $0$ to $-2$. 
 %A Flatter optical/UV continuum slope with an average value of $-1$ has been observed for bright AGN and $ -0.3$ for quasars \citep{1991ApJ...373..465F, neugebauer1987continuum}. 
 All these results show that quasars have different spectral slopes which are redder than that predicted from the standard disk model.  However, a number of effects can alter the observed continuum slopes.

AGN are embedded in their host galaxies, the emission from the accretion disks is not resolved, and  not well separated from the host galaxies.  %from  medium surrounding the disk and along the sight lines hinder the direct view of the intrinsic disk emission.
The intrinsic emission from the accretion disks is subjected to contamination by the host galaxy emission, complex \ion{Fe}{2} emission, Balmer continuum, and emission lines from the broad and narrow line regions.  Additionally, the disk emission   suffers reddening due to the  host galaxy as well as our own Galaxy.  These effects altogether can modify the disk spectra substantially. Theoretical
models of accretion disks can only be tested with the intrinsic continuum spectra after accounting for the various effects.

The inner regions of accretion disks can be probed by using the broad, relativistic  Fe $\rm K_\alpha$ line observed  near $\sim 6.4\kev$ in the X-ray spectra of AGN. The line is thought to arise due to the reflection of coronal X-rays irradiated onto the disk \citep[see e.g.,][]{laor1991line,tanaka1995gravitationally,fabian2000broad}. The broad iron line can be used to probe the ionization state, special/general relativistic effects and the line emissivity profile of the disk, and to measure the inner edge of the disk and thereby the black hole spin.  
%The broadening and the skewed shape of the iron line ($\rm line~ width \sim 50000  ~km~s^{-1}$) can be used to investigate special and general relativistic effects that arise due to fast rotation and strong gravity near the central SMBH \citep{laor1991line,fabian2000broad}. 
However, the broad iron line cannot be used to probe the temperature profile of the disk that can be used to test predictions of the standard disk models. Thus, the study of the intrinsic continuum emission in the optical/UV band remains the best way to probe the nature of accretion disks in AGN.

In this paper, we study intrinsic accretion disk emission from eight Seyfert type 1 AGN  based on observations performed with the \astrosat{} mission. Here we study the intrinsic far UV continuum emission using slit-less grating spectra, and in a future paper, we aim to study UV/X-ray broadband spectra. The general properties of the sources are listed in Table~\ref{tab:genpar}. These AGN are nearby with redshifts $z=0.015-0.1$ and cover a range of black hole masses $\rm M_{BH} \sim 6.3\times10^6 - 10^9 M_{\odot}$. 
%We have used UVIT grating observations in the far UV band ($1200-1800$\angs). 
The limited bandwidth ($1200-1800$\angs) and low  resolution of the slit-less grating spectra make it difficult to separate the emission/absorption lines and the continuum emission. We, therefore, make use of the available \textit{HST} data for seven out of eight AGN from the \textit{HST} archive to identify the emission/absorption lines. We organize our paper as follows. We describe our \astrosat{} observations and  data reduction in Section~\ref{sec:obs};   \hst{} spectra in Section~\ref{sec:hst_spec_data}; spectral complexities in Section \ref{sec:obs_complx}; spectral analysis and results in Section \ref{sec:spec_analys}. We discuss our results in Section \ref{sec:discuss}, followed by conclusions in \ref{sec:concl}. Unless otherwise mentioned, we assumed the cosmological parameters, $H_0 = 69.6~\rm km~s^{-1}~Mpc^{-1}, \Omega_{M} = 0.286 ~and~ \Omega_{\Lambda} = 0.714$, to calculate the co-moving radial distances of the sources.

\begin{deluxetable*}{lccccc}
\tablenum{2}
\tablecaption{List of \astrosat{}/UVIT and \hst{}/COS or FOS observations. The last column is the background-corrected net count rate of sources in the $-2$ order of FUV gratings  or $-1$ order of NUV grating. \label{tab:log}}
\tablewidth{0pt}
\tablehead{
\colhead{Sources} & \colhead{ Obs. ID } & \colhead{Instrument} &\colhead{Date of Obs.} & \colhead{Exposure time} & \colhead{Count rate}  \\
\colhead{} & \colhead{} & \colhead{} &\colhead{} & \colhead{(ks)} & \colhead{(counts~s$^{-1}$)}
} 

\startdata
Mrk~841 &A09\_008T05\_9000003724 &UVIT/FUV-G1 & 2020-06-18 \iffalse21:55:58.19 \fi & 5.8& $6.67\pm 0.04$  \\
& & UVIT/FUV-G2  & 2020-06-19 \iffalse14:08:25.93 \fi &6.0 & $7.39\pm 0.04$  \\
&LC8Y15010&HST/COS 130M/1291 &2014-07-06&1.7 & \\
&LC8Y15020& HST/COS 160M/1600 &2014-07-06&2.4 & \\
&&  && & \\
MR~2251-178    &A04\_218T03\_9000002214  &UVIT/FUV-G1 &2018-07-09 \iffalse17:26:09.18 \fi  & 5.2 & $6.73\pm 0.04$\\
 &Y3AI2006T& HST/FOS 130H &1996-08-02&2.3 & \\
&Y3AI2007T& HST/FOS 190H &1996-08-02&2.3& \\
&&  && & \\
PG0804 &G07\_062T01\_9000001560 & UVIT/FUV G2 & 2017-09-25 &4.1 &$8.9\pm 0.05$\\ 
& G07\_062T01\_9000001560& UVIT/NUV Grating  & 2017-09-25 & 4.0 & $59\pm 0.1$\\
& LB4F08030 & HST/COS 130M/1291 & 2010-06-12 & 1.6 & \\
&LB4F08080 & HST/COS 160M/1623 &2010-06-12 & 1.6\\
&&  && & \\
NGC~7469 & G08\_071T02\_9000001620 & UVIT/FUV-G1  &2017-10-18 \iffalse16:25:49.84 \fi & 3.4& $5.73\pm 0.04$  \\
\multicolumn{2}{c}{} &UVIT/FUV-G2  &  2017-10-18 \iffalse11:56:11.81 \fi &4.0 & $7.88\pm 0.05$    \\
& & UVIT/FUV-Silica & 2017-10-18& 45& \\
&Y3B60106T& HST/FOS 130H &1996-06-18&2.2 & \\
&Y3B60107T& HST/FOS 190H &1996-06-18&1.6 & \\
&&  && & \\
I~Zw~1 & A10\_101T02\_9000004084 & UVIT/FUV G2 & 2020-12-26& 4.8 & $1.03\pm 0.02$ \\
&LCKR02030 & HST/COS 130M/1309 & 2015-01-21 & 1.9 & \\
&LCKR01020 & HST/COS 130M/1327 &2015-01-20 &0.9 &\\
&LCKR01050 & HST/COS 160M/1589 &2015-01-20 & 2.5 &\\
&&  && & \\
SWIFT1921 &A04\_218T08\_9000002236 & UVIT/FUV-G1 & 2018-07-17 \iffalse20:35:41.05 \fi & 5.7 & $8.50\pm 0.04$\\
\multicolumn{2}{c}{} & UVIT/FUV-G2  & 2018-07-18 \iffalse09:26:31.6 \fi &5.4 & $9.72\pm 0.04$  \\
&LC1202010&HST/COS 130M/1309 &2012-11-20&2.2 & \\
&LC1202020& HST/COS 160M/1589 &2012-11-20&2.0 & \\
&&  && & \\
Mrk~352        &  A09\_008T06\_9000003748 & UVIT/FUV-G1 &2020-06-30  \iffalse12:58:38.16 \fi& 5.3 & $1.48\pm 0.02$  \\
\multicolumn{2}{c}{} & UVIT/FUV-G2 & 2020-07-01 \iffalse10:04:45.29 \fi &6.2 & $1.59\pm 0.02$   \\
&LCXV07060&COS 130M/1327 &2016-07-20&1.7 & \\
&LCXV07070& COS 160M/1589 &2016-07-20&2.2 & \\
&&  && & \\
SWIFT1835 &  A04\_218T04\_9000002086 & UVIT/FUV-G1  & 2018-05-10 \iffalse00:51:16.4 \fi &3.2 & $1.04\pm 0.02$ \\
\enddata 
\end{deluxetable*}

\section{\textit{AstroSat}/UVIT Observations and Data Reduction} \label{sec:obs}
%{\color{blue}\astrosat{}, launched by the Indian Space Research Organisation (ISRO) on 28 September 2015 into a 650 km near-equatorial orbit with $6^\circ$ inclination, carries several co-aligned payloads for multiwavelength studies \citep{singh2014astrosat}.
%is the first dedicated Indian multi-wavelength astronomy space observatory  \citep{singh2014astrosat}. It was launched by the Indian Space Research Organisation (ISRO) on 28 September 2015 into a $650{\rm~km}$ circular,  near-equatorial orbit with $6^\circ$ inclination. \astrosat{} is a proposal-driven observatory and is operated by the ISRO.
\astrosat{} is a multi-wavelength space observatory \citep{singh2014astrosat}. It carries
four co-aligned and simultaneously-operating instruments  --  the Ultraviolet Imaging Telescope (UVIT) \citep{tandon2017orbit,tandon2020additional}, the Soft X-Ray Telescope (SXT) \citep{singh2016orbit,singh2017soft},  the Large Area X-ray Proportional Counters (LAXPC) \citep{yadav2016large, antia2017calibration}  and the Cadmium-Zinc-Telluride Imager (CZTI) \citep{czti}. In this paper, we utilized the data acquired with the UVIT only as our aim is to study the intrinsic continuum emission from the AGN disks.
%\subsection{UVIT}

 The UVIT consists of twin telescopes.  One of them is sensitive in the far ultraviolet ($1200-1800$\angs) band and is referred to as the FUV channel. The light from the second telescope is split into near ultraviolet ($2000-3000 $\angs)   and visible ($3200-5500 $\angs) bands,  forming the NUV and VIS channels, respectively.  The VIS channel is mainly used for tracking satellite pointing, while the FUV and NUV channels are used for scientific observations. Both FUV and NUV channels  are equipped with a number of broadband filters. The UVIT has excellent spatial resolution with a point spread function (PSF) in the range of $1-1.5{\rm~arcsec}$. The UV images are obtained in photon counting mode at a rate of 28 frames s$^{-1}$ in the full window mode.
 %In FUV the broadband filters are  BaF$_2$, CaF$_2$, Silica and Sapphire. In NUV broadband filters are Silica, NUVB4, NUVB13, NUVB15, NUVN2. 

 In addition, the FUV channel is equipped with two slit-less gratings, FUV-Grating1 and FUV-Grating2 (hereafter FUV-G1 and FUV-G2),  and the NUV channel carries a single slit-less grating, referred as the NUV-Grating (hereafter NUV-G). The two FUV gratings are oriented orthogonal to each other to avoid possible contamination along the dispersion direction due to the presence of neighboring sources in the dispersed image. More details on the performance and calibration of the UVIT gratings can be found in \cite{Dewangan_2021}.
 
 The peak effective area  of FUV-G1 and FUV-G2 in the $-2$ order are $\sim 4.5{\rm~cm^2}$ at $1390$\angs~ and $\sim 4.3{\rm~cm^2}$ at $1500 $\angs, respectively. The NUV grating has a peak effective area of  $\sim 18.7{\rm~cm^2}$ ($2325$\angs) in the $-1$ order. The spectral resolution for the FUV gratings in the $-2$ order is FWHM $\sim 14.3$\angs~ and that for  the NUV grating  in the $-1$ order is $\sim 33$\angs. The NUV channel stopped functioning in 2018 March \citep{ghosh2021orbit}.
 %The spatial resolution for both FUV and NUV ranges from  $1.2^{\prime\prime}$ to $1.5^{\prime\prime}$.
 %The Visible band is used to track stars to make correction for drifting of satellite during observation.\\

\begin{figure*}
\centering
\gridline{\fig{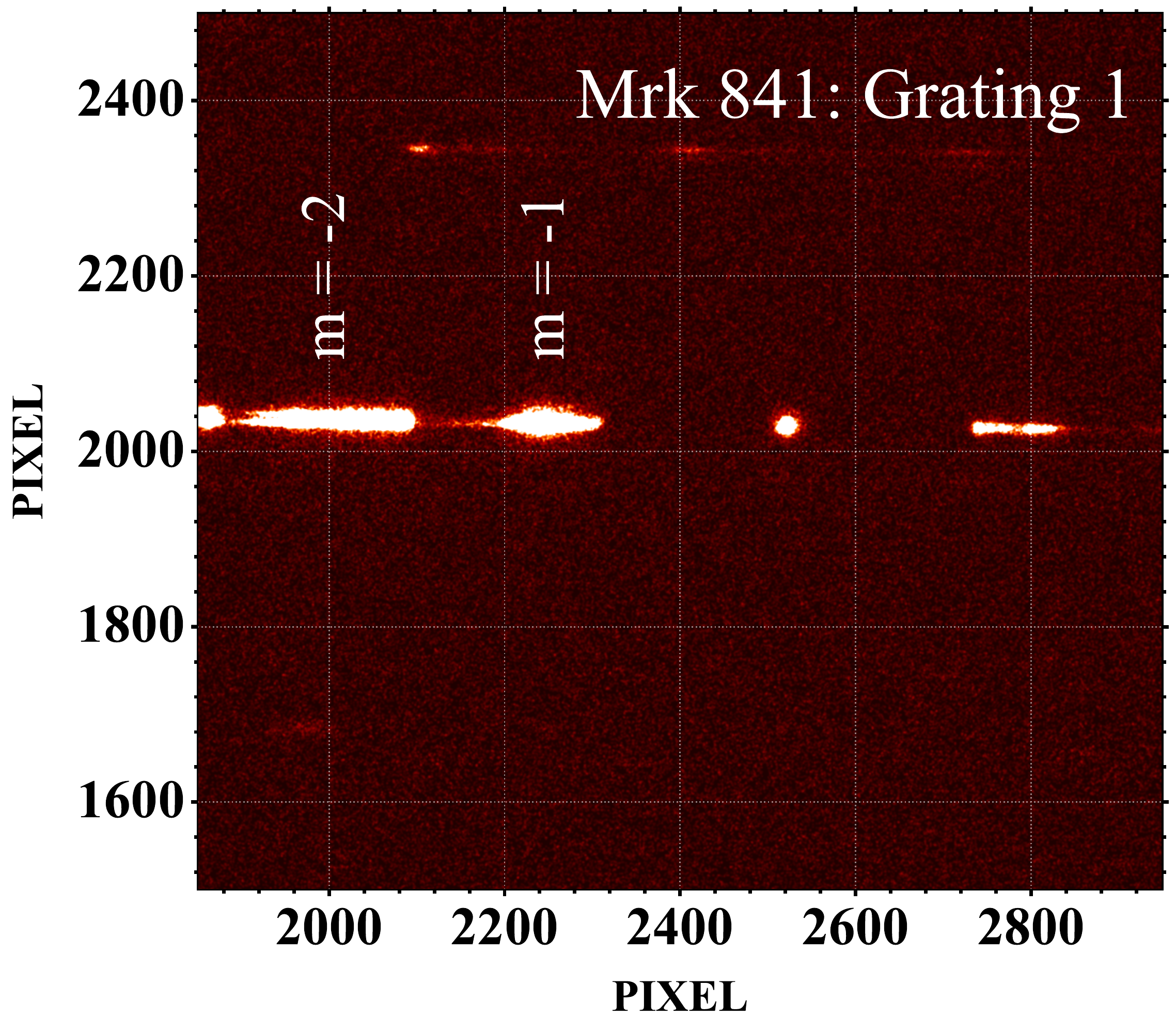}{0.3\textwidth}{(a)}
           \fig{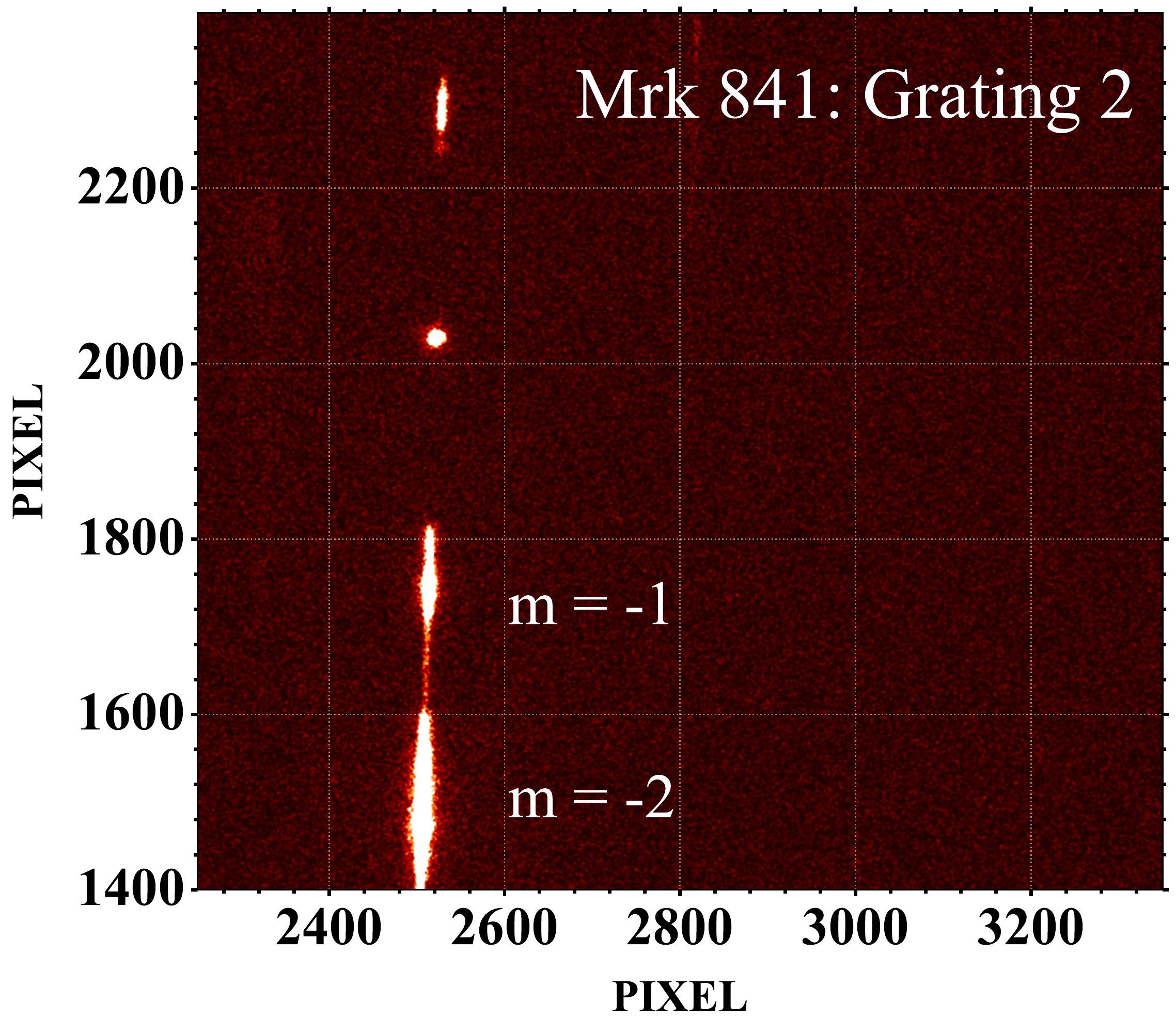}{0.3\textwidth}{(b)}
          \fig{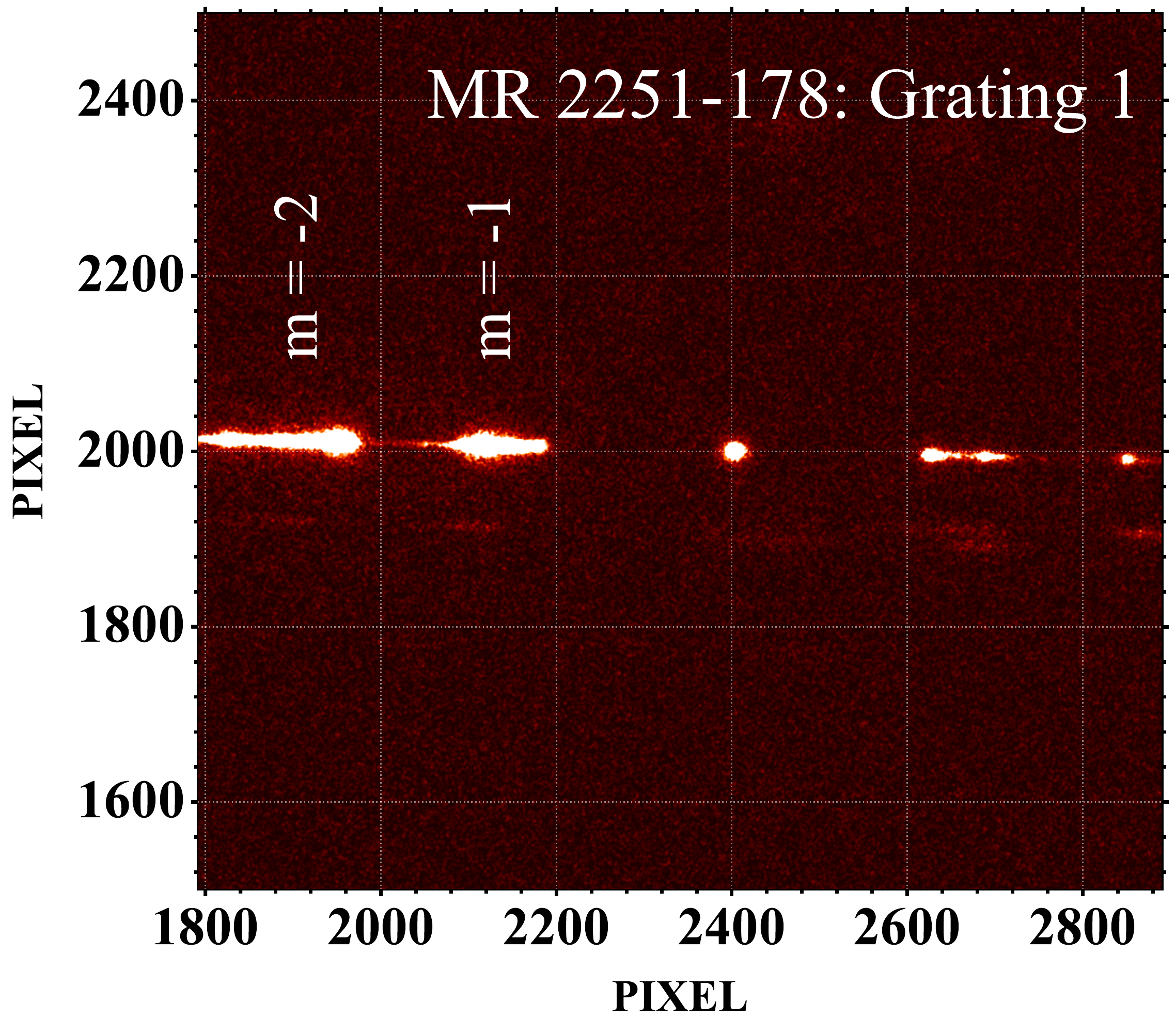}{0.3\textwidth}{(c)}
          }
\gridline{\fig{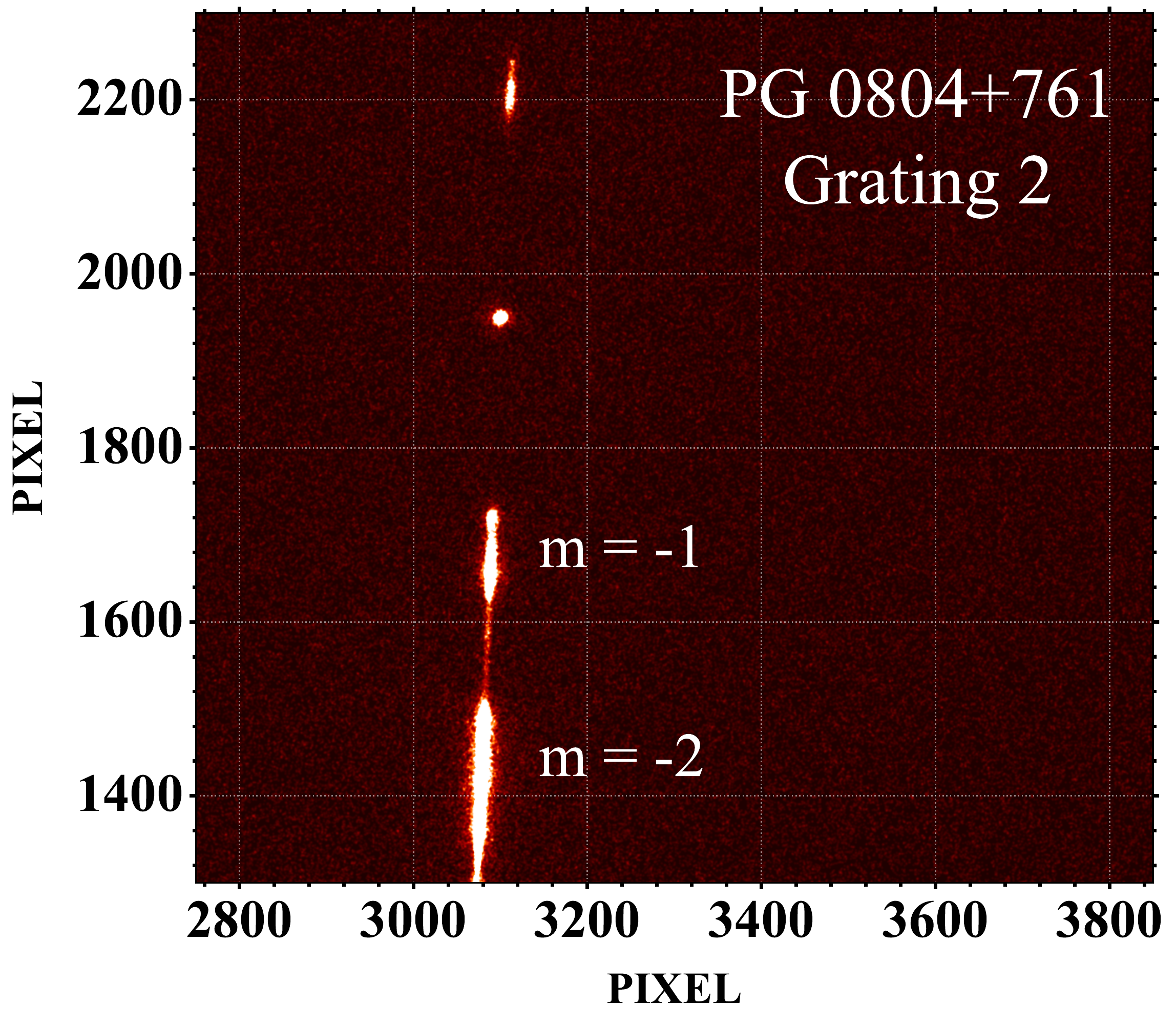}{0.3\textwidth}{(d)}
            \fig{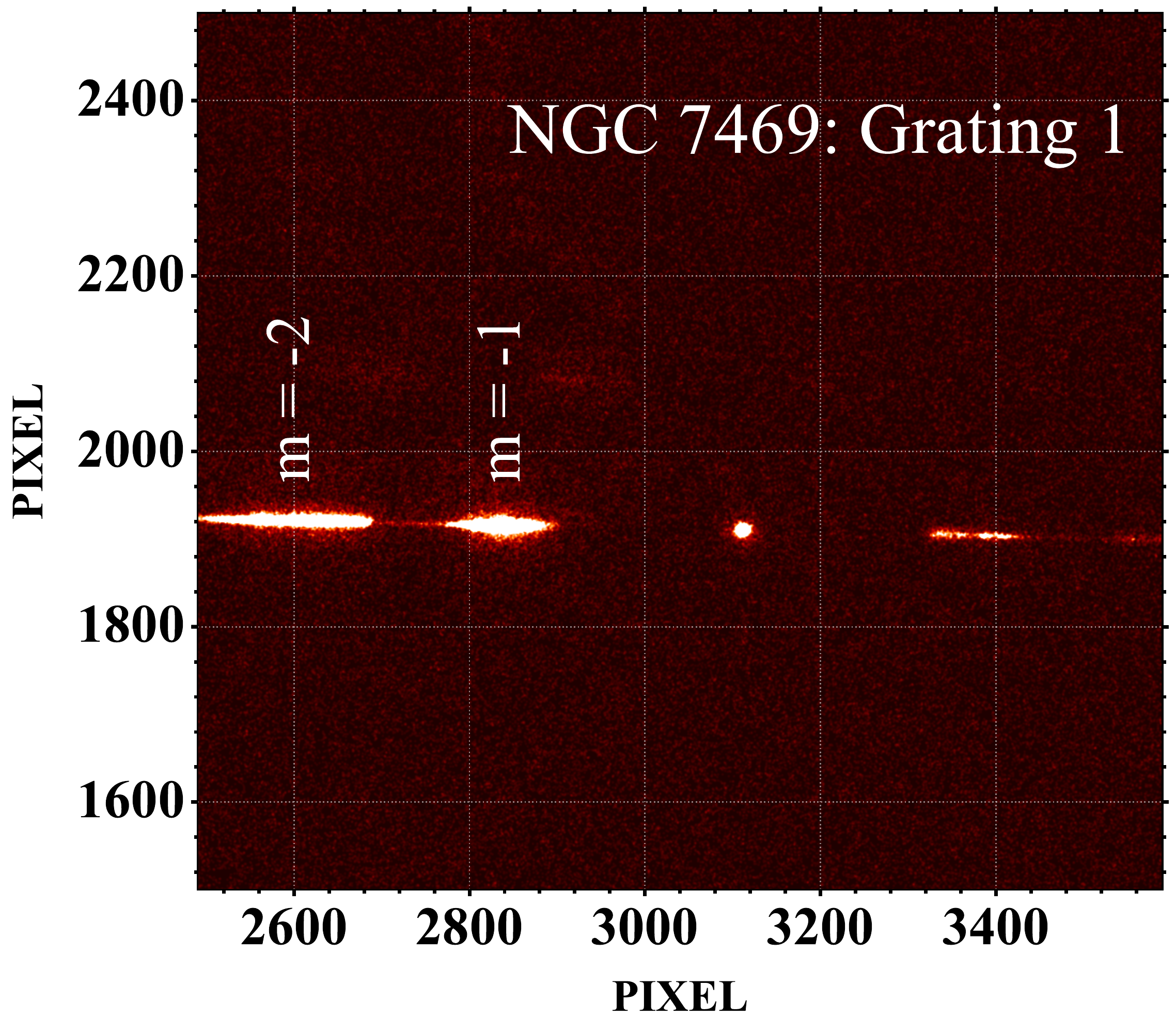}{0.3\textwidth}{(e)}
            \fig{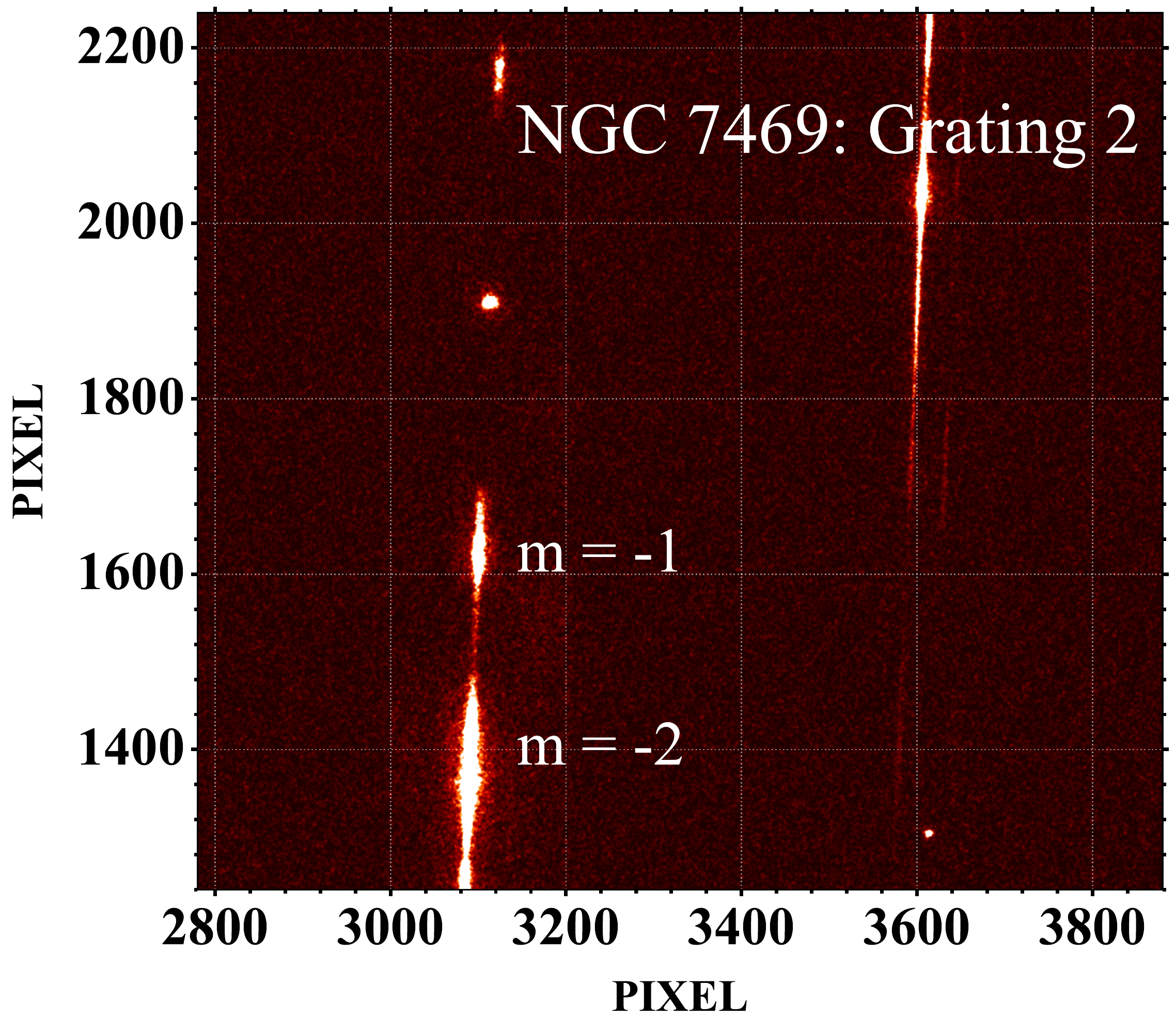}{0.3\textwidth}{(f)}
            }
\gridline{\fig{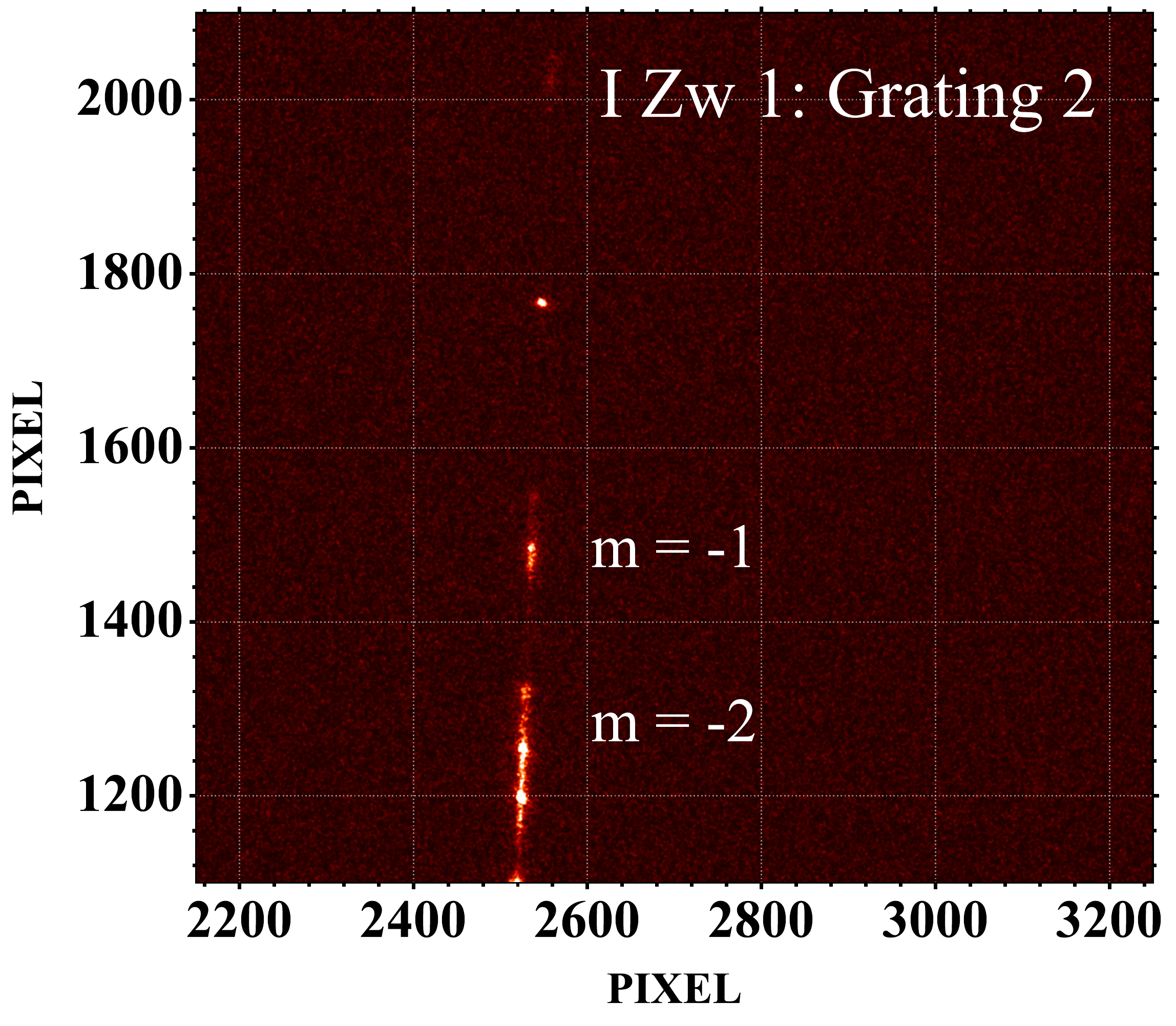}{0.3\textwidth}{(g)}
            \fig{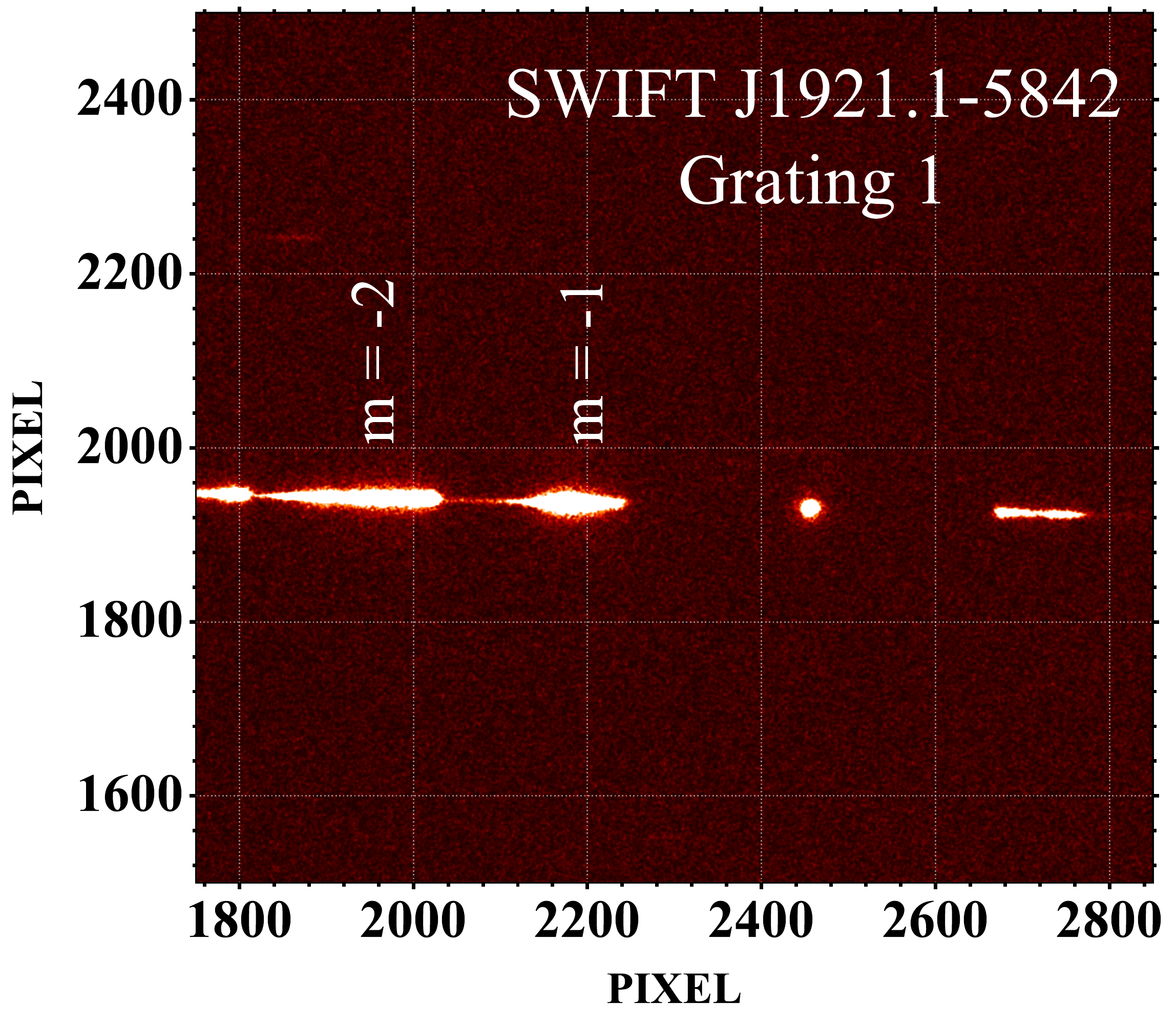}{0.3\textwidth}{(h)}
            \fig{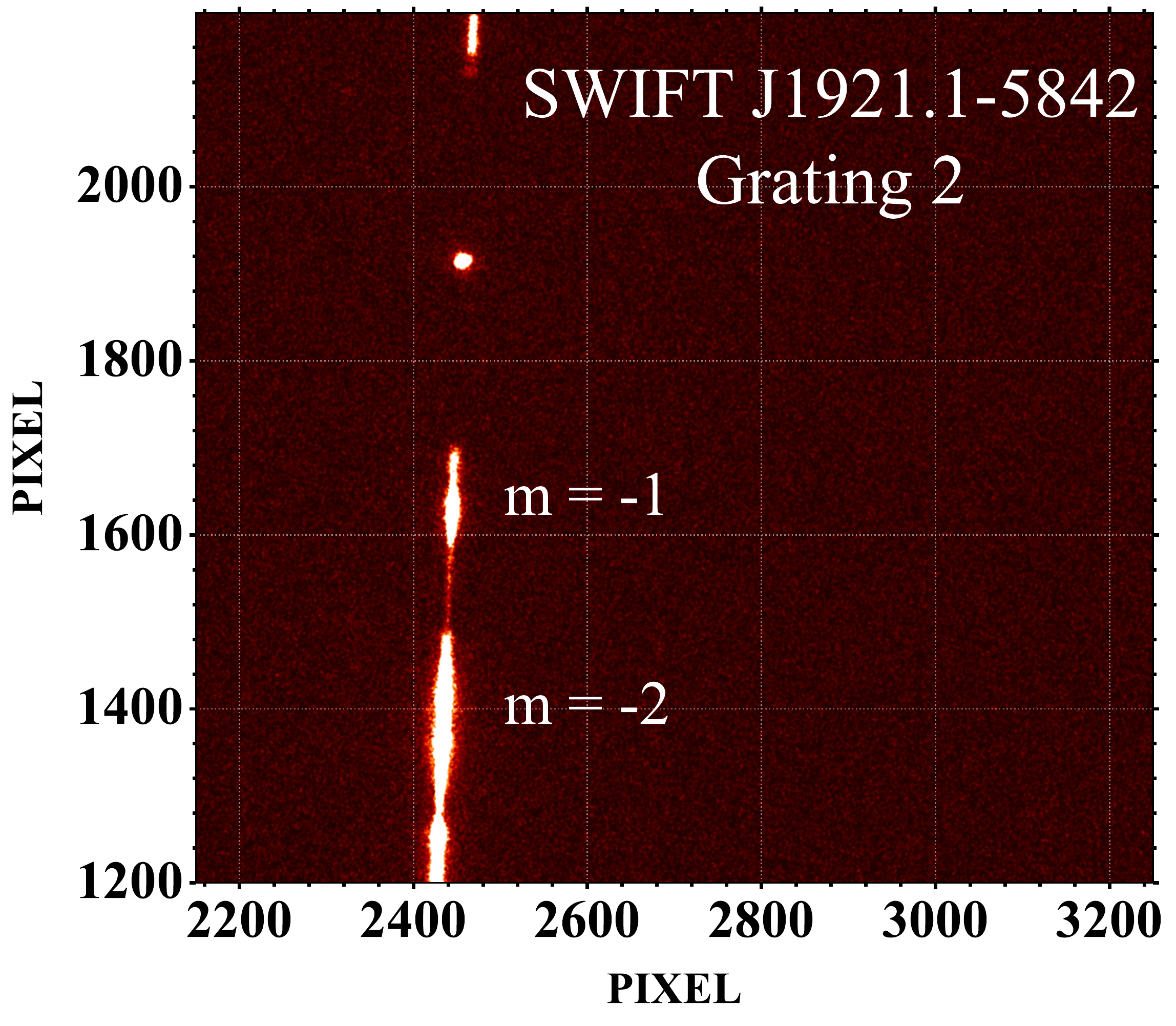}{0.3\textwidth}{(i)}
            }
\gridline{\fig{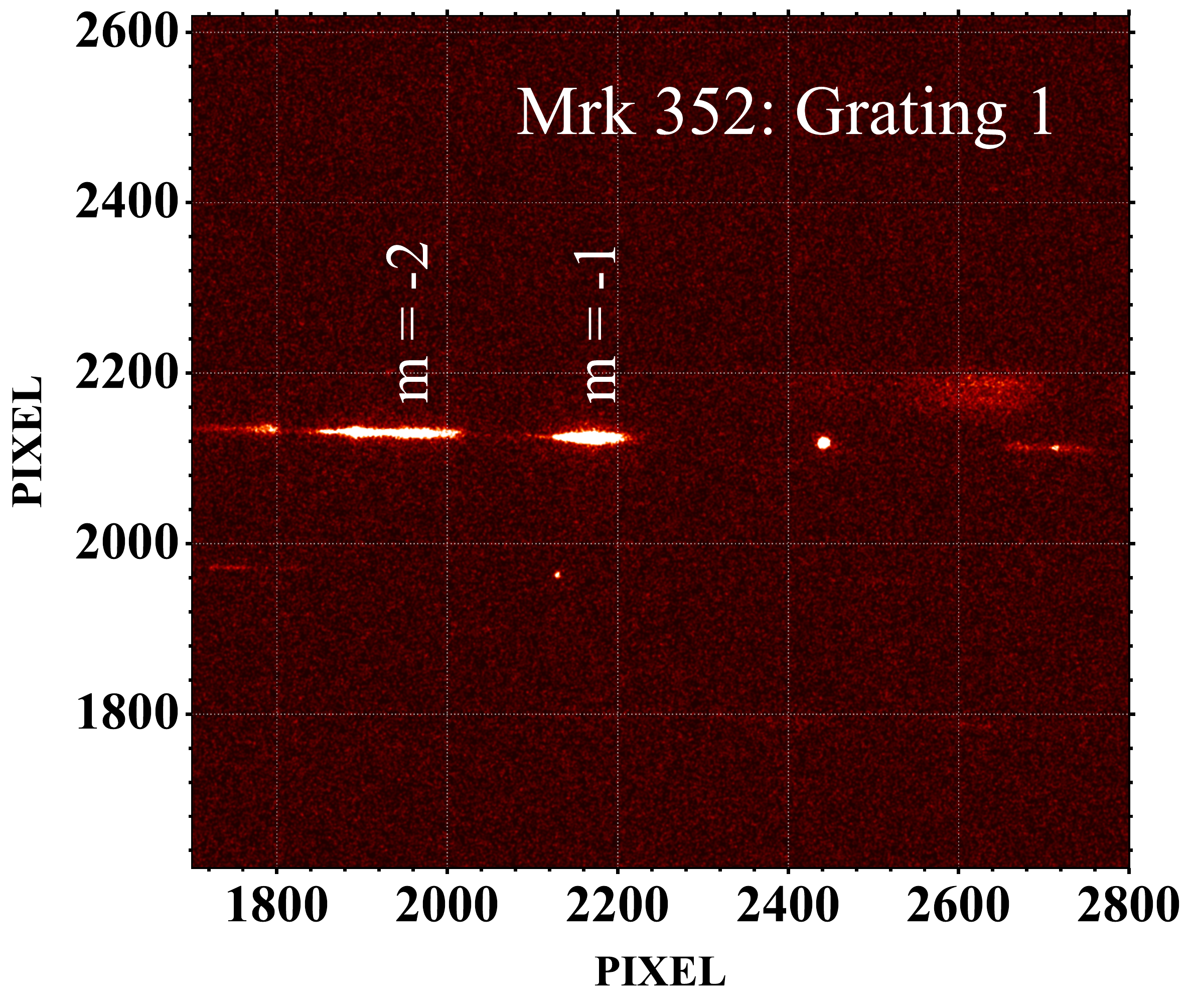}{0.3\textwidth}{(j)}
            \fig{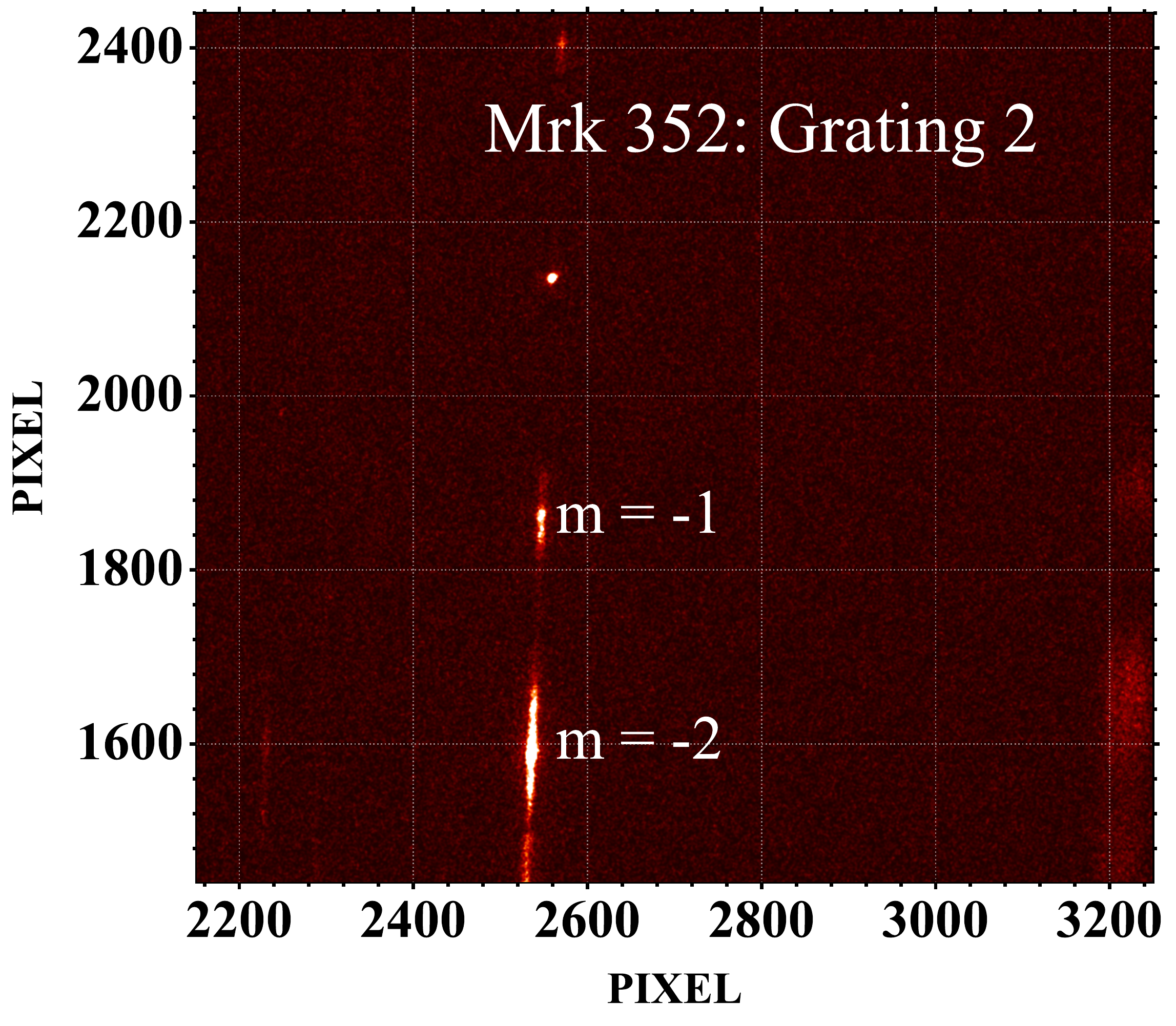}{0.3\textwidth}{(k)}
            \fig{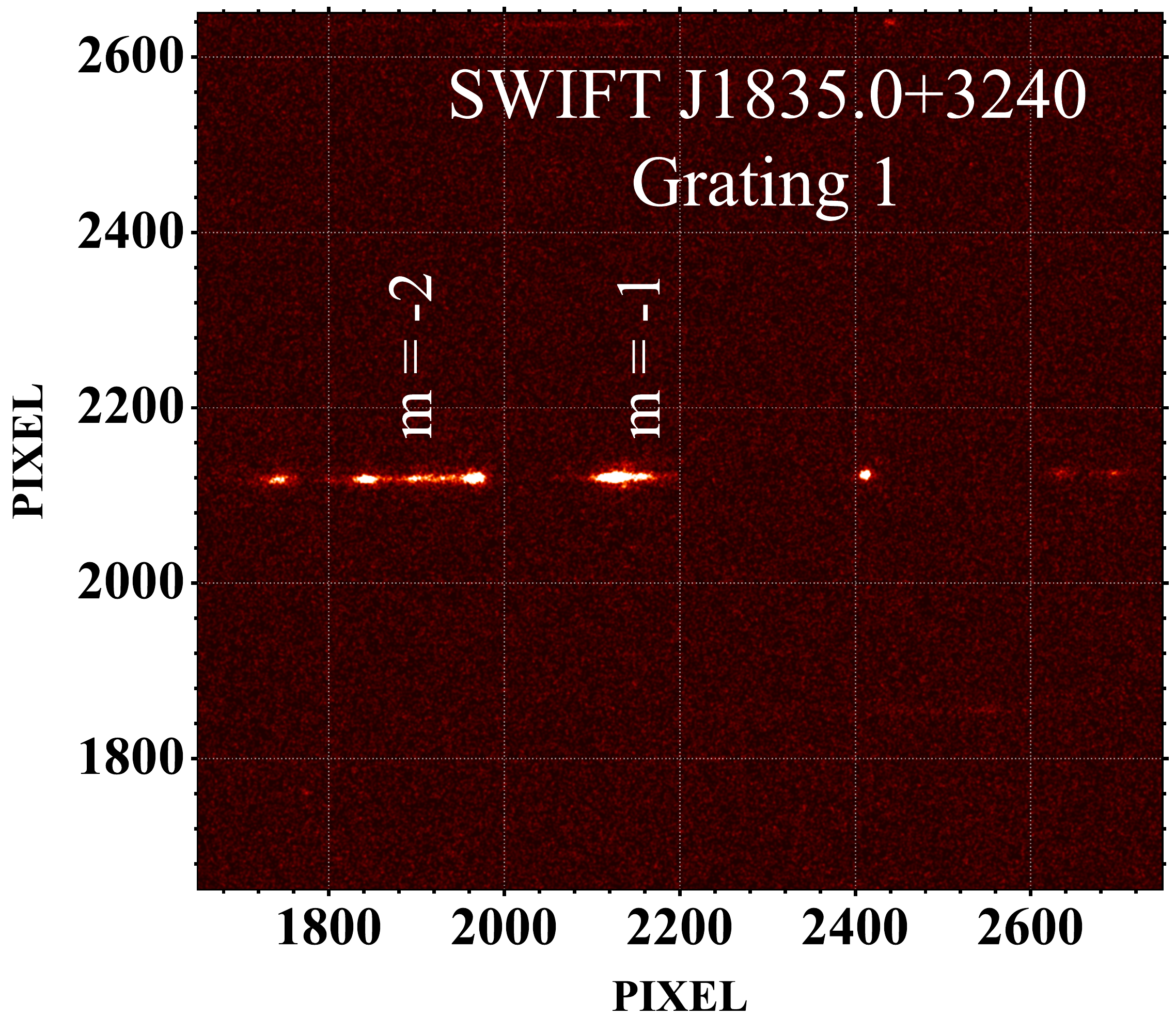}{0.3\textwidth}{(l)}
            }
\caption{The slit-less UVIT/FUV Grating images of the eight AGN -- (a,b) Mrk 841, (c) MR~2251-178, (d) PG0804, (e,f) NGC~7469, (g) I~Zw~1,  (h,i) SWIFT~J1921.1-5842, (j,k) Mrk~352, and (l) SWIFT~J1835.0+3240.}
\label{fig:grim}
\end{figure*}

We obtained level1 UVIT data for the eight sources: Mrk~841, MR~2251--178, PG~0804+761 (hereafter PG0804), NGC~7469, I~ZW~1, SWIFT~J1921.1--5842 (hereafter SWIFT1921), Mrk~352 and SWIFT~J1835.0+3240 (hereafter SWIFT1835) from the \textit{AstroSat} data archive \footnote{\url{https://astrobrowse.issdc.gov.in/astro_archive/archive/Home.jsp}} (see Table~\ref{tab:log}). 
%In the first column of Table~\ref{tab:log}, SWIFT J1921-5842 is abbreviated as SWIFT1921 and SWIFT J1835+3240 as SWIFT1835.
We list the available UVIT grating data sets in Table~\ref{tab:log}.   
Only FUV-G1 data are available for MR~2251--178 and SWIFT1835 while only FUV-G2 data are available for I~ZW~1 and PG0804. NUV grating data are available only for  PG0804.
We  processed the level1 UVIT data using the  CCDLAB \citep{Postma_2017} pipeline software. The CCDLAB extracts scientifically useful data for each of the orbits. To correct for the drift of the satellite pointing, we generated drift series (X, Y shifts relative to the reference (first) frame as a function of time) for bright point sources using images acquired with the VIS channel. We then applied the mean drift series (mean X, Y shifts as a function of time) to the FUV/NUV centroid lists.   
%bright point sources are tracked in the visible channel. The drift series (x and y shift in between frames assuming the 1st frame as reference) obtained using VIS channel are used to make drift correction for the FUV and NUV frames . 
The individual orbit-wise images and the centroid lists obtained after the drift corrections are not co-aligned.
We used the registration task available within the CCDLAB that uses two or more point sources to register the centroid lists in different frames into a common frame. Finally, we merged the co-aligned data to obtain a single science image with a greatly improved signal-to-noise ratio. Along the spatial direction in the dispersed images, one pixel corresponds to $0.41{\rm~arcsec}$ on the sky. The final grating images  of our sources are shown in Fig.~\ref{fig:grim}. 

We identified the zero-order positions of our target sources by comparing the patterns of zero-order positions of sources with the available \textit{ SWIFT}/UVOT or \textit{GALEX} images.
We then identified grating orders for our sources and then proceeded with spectral extraction. In addition to the grating images, we processed the available FUV broadband filter (FUV--Silica: $\lambda_{mean}=1717$\angs~with $\Delta \lambda \sim 125$ \AA, FUV--Sapphire: $\lambda_{mean}=1608$\angs~with $\Delta \lambda \sim 290$ \AA) images.  

 For the spectral extraction, we followed the procedures and tools described in \citet{Dewangan_2021}. We located the zeroth order of the source of interest in the grating image. The dispersion axes of the gratings are not aligned exactly with the X (FUV-G1) or Y (FUV-G2) axis in the grating images.  We calculated  the angle of the dispersion axis with respect to the X (FUV-G1, NUV-G) or Y (FUV-G2) image axes. This angle is used to find the centroids along the spatial direction at each pixel along the dispersion direction for the desired order ($-2$ for FUV gratings, $-1$ for NUV-G). 
 We used a 50-pixel width along the cross-dispersion direction and extracted the FUV and NUV grating PHA spectral data in the $-2$ and $-1$ orders, respectively, and the associated response matrices as described in \citet{Dewangan_2021}. We used a similar procedure and extracted background spectra from nearby source-free regions corresponding to each source spectrum. We list the background-subtracted net source count rates for FUV gratings in the $-2$ order or NUV-G in the $-1$ order in the last column of Table~\ref{tab:log}. The exposure time for gratings varies from $0.6-6.0{\rm~ks}$. The background-subtracted net count rate in FUV gratings varies from $\sim 1.0 - 9.7{\rm~counts~s^{-1}}$. 
 We used {\sc Sherpa} version $4.14.0$ \citep{2001SPIE.4477...76F,freeman2011sherpa} for the spectral analysis of all the sources. We also generated the fluxed spectra using the same extraction parameters mentioned above by applying the wavelength and flux calibration described in \cite{Dewangan_2021}. We show the fluxed FUV grating spectra of the eight AGN in Figure~\ref{fig:hstuv_slx}.
 
\section{The \hst{} Spectra } \label{sec:hst_spec_data}
We also used high-resolution UV spectra acquired with the  \textit{HST}'s Cosmic Origin Spectrograph (COS; \citealt{green2011cosmic}) and the Faint Object Spectrograph (FOS; \citealt{keyes1995faint}). We obtained the COS spectra for the sources Mrk~841, PG0804, I~Zw~1, SWIFT1921 and Mrk~352 from the archive (\url{https://archive.stsci.edu/hst/search.php}). We used FOS spectra  of MR~2251-178 and NGC~7469 from \citet{kuraszkiewicz2004emission} (\url{http://hea-www.harvard.edu/~pgreen/HRCULES.html}). We did not find \textit{HST} spectra for SWIFT1835. We  list the available \hst{} observations in Table~\ref{tab:log}. 

The \textit{HST}/FOS can cover a wide  wavelength range of $\sim 1150-8500$\angs~ with high and low-resolution gratings.
We used FOS spectra acquired with the high resolution gratings G130H ($1140  - 1606$\angs)  and G190H ($1573 - 2303$\angs) with spectral resolution $R = \lambda/\Delta\lambda \sim 1300$ for  MR~2251--178 and NGC~7469. The largest circular aperture size is $0.9^{\prime\prime}$ in diameter.
%G130H observes in $1140 - 1606$ \AA wavelength range and G190H covers $1573 - 2303$ \AA.  
We have used spectra acquired with  two COS medium resolution gratings, G130M ($\sim 900 - 1450 $\angs) and G160M ($1400 - 1775$\angs). The spectral resolution for these medium resolution gratings is $R \sim ( \lambda/\Delta\lambda)\sim 18,000$. The field of view for COS is $2.5^{\prime\prime}$ in diameter. For each of the COS gratings, there are two segments -- FUVA and FUVB, due to a gap in the active area. The gaps are about $14.3$\angs~ (G130M) and $18.1$\angs~ (G160M). 

Due to high spectral resolution, COS is widely used for absorption and emission line studies. \citet{tilton2013ultraviolet}  used COS data for 44 AGN including NGC~7469 and PG0804 to derive intrinsic absorption, broad emission/absorption line widths and inferred a rough estimate of black hole masses. In a sample of 159 AGN observed with the \textit{HST}/COS, \citet{stevans2014hst} reported the spectral indices ($\alpha$) of $-0.96$ (MR~2251--178), $-0.54$ (PG0804), $-0.89 $ (NGC~7469). They estimated the continuum slope in the line free narrow windows above $800 $\angs. 

\begin{figure*}
 \epsscale{1.25} 
\plotone{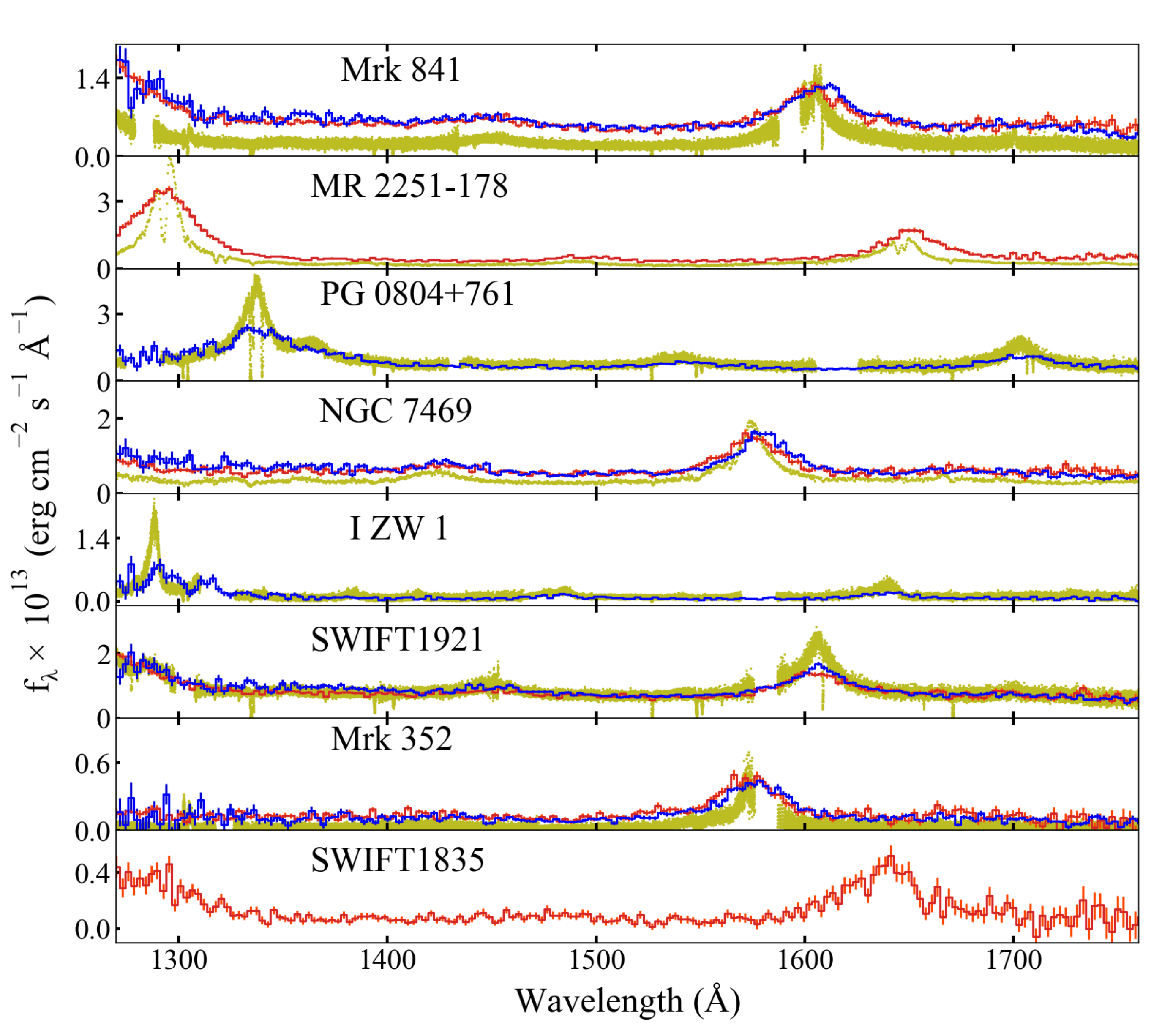}
   
    \caption{The UVIT FUV-G1 (red) and FUV-G2 (blue) fluxed spectra of the eight AGN in the observed frame. The \hst{} COS/FOS spectra (olive) are plotted for comparison. \hst{} spectra for SWIFT~J1835.0+3240 are not available.}
    \label{fig:hstuv_slx}
\end{figure*}

We compared the COS/FOS and UVIT spectra of our sources  in Figure~\ref{fig:hstuv_slx}. The differences in flux measured with UVIT gratings and COS/FOS (e.g., Mrk~841) are due to the variability of the AGN as the two spectrographs are well flux-calibrated.  Mrk~841 has been found to be highly variable in X-rays \citep{nandra1995soft,george1993broad} as well as in the UV \citep{miranda2021uv}. The emission lines and the continuum shapes appear to match well. We find  a slight discrepancy in the shapes of emission lines, particularly the \ion{C}{4}$\lambda1548.9$\angs~ line, possibly due to the differences in the spectral resolutions, presence of narrow  absorption lines not resolved in the UVIT spectra, and/or inaccuracies in the wavelength calibration. The main purpose of our study is to derive the far UV continuum shapes after accounting  for the emission lines in terms of Gaussian profiles, therefore, we ignore possible errors in the line positions. Using the \hst{} spectra, we measure the emission/absorption line parameters, and then we use these line parameters while performing UVIT spectral analysis. 
%{\color{blue}\textbf{Since we are interested in the continuum emission only, the choice in selecting the \hst{} spectrum will not affect the final result from UVIT spectral analysis. }}

%The emission lines appear to be shifted blue ward slightly. This is likely due to possible errors in the wavelength calibration of UVIT gratings. It is also possible that line asymmetry, presence of absorption lines, line variability, and possible contamination in the slit-less grating spectra  may also  alter the line peaks in the low resolution spectra such as the UVIT spectra.  
%The main purpose of our study is to derive the far UV continuum shapes after describing the emission lines in terms of Gaussian profiles, therefore we ignore possible errors in the line positions.

%\iffalse

%\iffalse
%\fi

\section{Spectral Complexities } \label{sec:obs_complx}
As mentioned in Section~\ref{sec:intro}, the observed UV spectra of AGN are affected  with  ($i$) Galactic and intrinsic reddening, ($ii$) host galaxy contribution, ($iii$) Balmer continuum, ($iv$) \ion{Fe}{2} emission  and ($v$) emission/absorption lines from the BLR and NLR. Below we describe these effects and the ways to correct them.

\subsection{Host Galaxy Contamination}
\begin{figure}[ht!]
 \epsscale{1.1}
\plotone{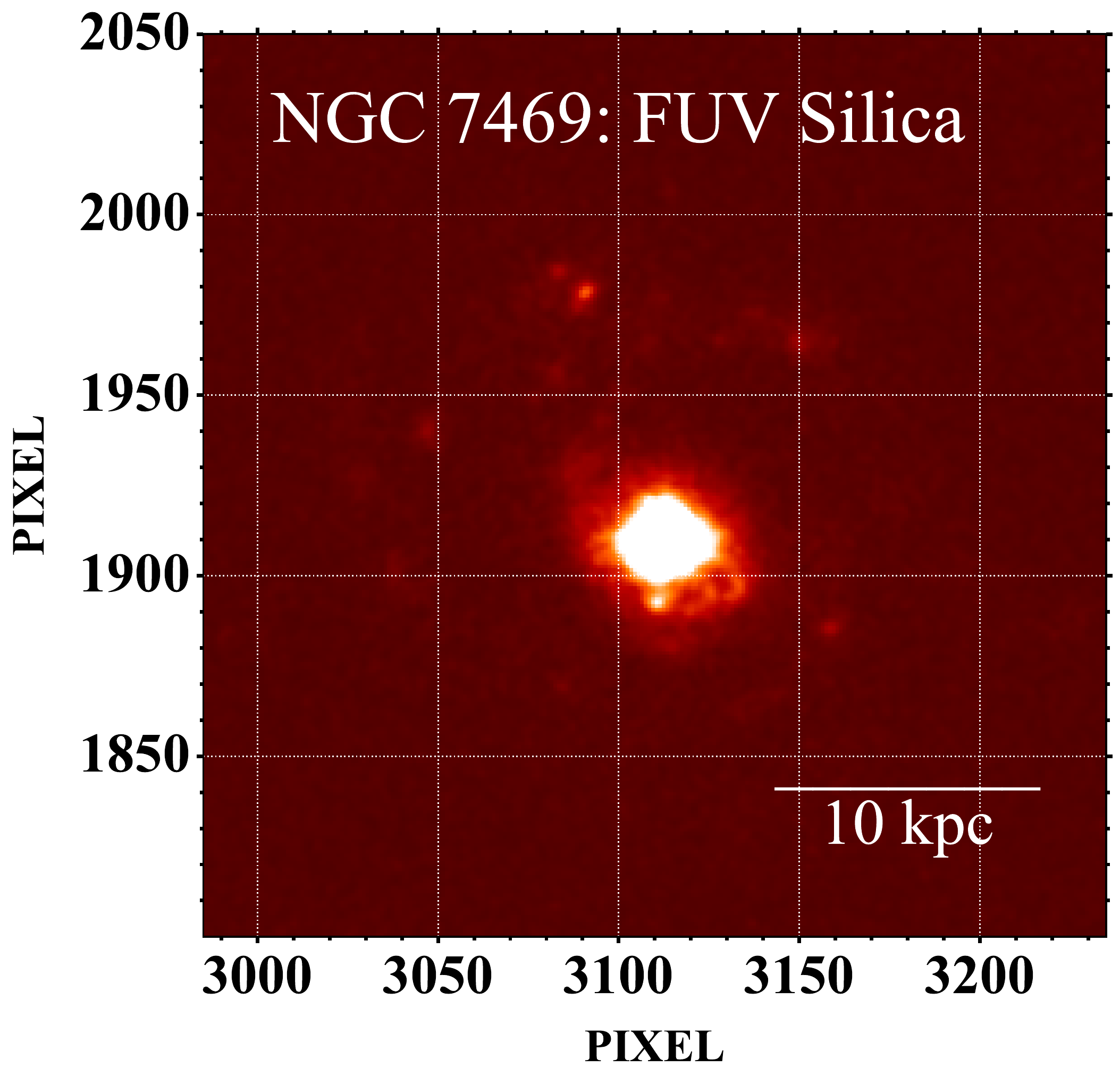}
\caption{FUV image of NGC~7469 in the Silica filter ($\lambda_{mean} \sim 1716.5 $\angs; $\Delta\lambda \sim 125$\angs). The image size is of 250x250 pixel along the axes (a single pixel corresponds to $0.416^{\prime\prime}$ or 139~pc). Faint diffuse emission connecting to spiral arms seen in the image may contaminate the AGN emission.  
%The spiral arms can have some contamination to the nucleus. Since the gratings are oriented orthogonal to each other, the contribution from the spiral arms will be different.
\label{fig:7469}}
\end{figure}

AGN are embedded into their host galaxies.   Naturally, the diffuse emission due to the presence of hot stars, H~II regions, supernovae remnants, etc. in  the central regions of host galaxies  can contribute to the observed UV emission. In our sample, all the sources except NGC~7469 are point-like objects. Figure~\ref{fig:7469} shows the UVIT image of NGC~7469 in the FUV-Silica filter. Clearly, the diffuse UV emission from the host galaxy in addition to the bright nuclear emission is detected.  To find out the host galaxy contribution, first, we measured the instrument's point spreading function (PSF). For this purpose, we generated  the radial surface brightness profile of a star BPS~CS~29521--0036 ($\alpha_{J2000}=23h2m52s$, $\delta_{J2000}=+08d59m27s$) which is located $9^{\prime}$ away from the source.  Following \citet{dewangan2021astrosat}, we fitted the radial profile  with a Moffat function, 

  %\iffalse
%\fi 

\begin{equation}
F(r) =  F_0\left[1+\left(\frac{r}{\sigma}\right)^{2}\right]^{-\beta}
\end{equation}
where the full width at half maximum (FWHM) is defined as $2\sigma\sqrt{2^{\frac{1}{\beta}}-1}$. We used a constant model to fit the background. The best-fit Moffat function with FWHM~$=1.18^{\prime\prime}$ provides the PSF of the instrument. Then, we used this  Moffat function i.e. the PSF, to fit the AGN radial profile. We kept the parameters $\sigma$ and $\beta$ fixed at the best-fit values derived for the star above. This component accounts for the AGN emission. Around $9^{\prime\prime}$ from the nucleus we noticed a large excess after using Moffat and a constant model for the background. This is due to one of the starburst regions observed at radius $\sim 8^{\prime\prime}-10^{\prime\prime}$. We used a Gaussian profile to account for this.  For the host galaxy contribution, we used an exponential profile 
\begin{equation}
I(r) = I_0 e^{-r/r_d}
\end{equation}
 The integrated galactic emission is $\sim 1 \%$ of the total emission (AGN+host galaxy) within $10^{\prime\prime}$ radius. We followed the same procedure for one of the point-like AGN, Mrk~841.  The integrated galactic emission within $10^{\prime\prime}$ radius is $2.9\%$ of the total emission. Therefore, we conclude that the host galaxy contribution is not substantial in the case of NGC~7469 or the point like AGN, and we do not make any corrections for such small contributions.

\subsection{Intrinsic reddening} 
The dust grains in the circumnuclear regions as well as in the ISM of the host galaxy can cause extinction to the AGN emission. The extinction generally rises steeply from the optical to UV bands, thus making the UV band much more susceptible to extinction.  A precise extinction correction is important to recover the intrinsic UV spectral  slope of  AGN \citep{koratkar1999ultraviolet}.  However,
the amount of intrinsic extinction is generally poorly known due to the lack of knowledge on the chemical composition and sizes of dust grains and molecules which are most likely dissimilar to that in our Galaxy  as the physical conditions in the AGN hosts are  quite different than that in our own galaxy. 

%\iffalse
\begin{figure}[ht!]
\epsscale{1.25}
\plotone{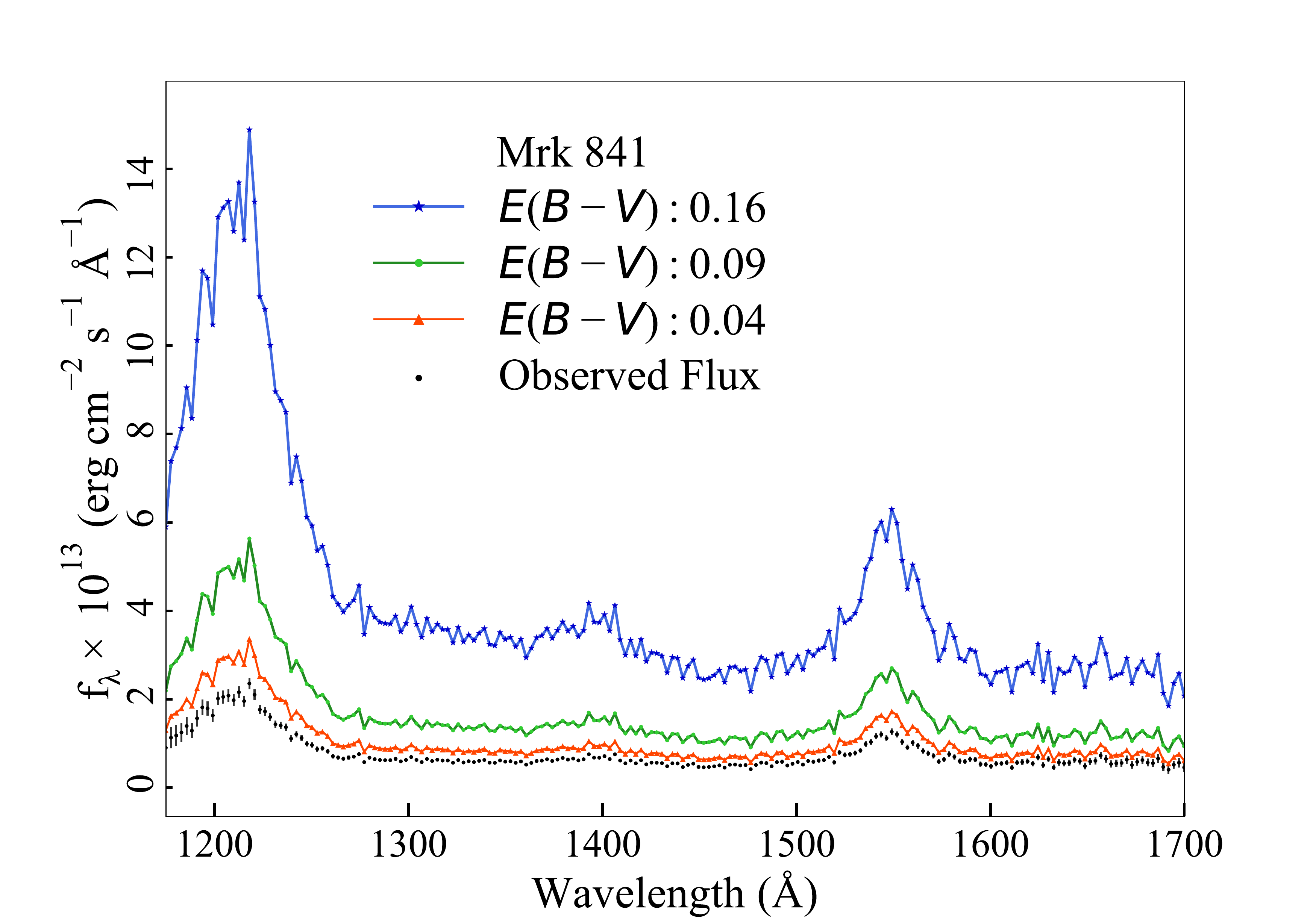}
\caption{Effect of de-reddening on an observed FUV spectrum. The FUV-G1 spectrum of Mrk~841 is de-reddened for different values of color excesses using the extinction law derived by \citet{czerny2004extinction}. The observed spectrum is shown in black, the de-reddened spectra appear in red, green and blue  for Balmer decrements of 3.5, 4 and 5  corresponding to $E(B-V) = 0.04$, 0.09 and 0.16, respectively.
%), green ( (Fuchsia ), 4 (Green) and 5 (Blue). Source name: Mrk 841, Model: 
\label{fig:red_ef}}
\end{figure}
%\fi
 %Accounting this intrinsic reddening is important because it can alter the spectral slope of intrinsic disk emission \citep{koratkar1999ultraviolet}. 
 Fortunately, type 1 AGN  are generally devoid of strong intrinsic extinction, and the wavelength dependence of the extinction can be characterized with empirical relations. 
 \citet{czerny2004extinction} used five composite SDSS spectra of quasars from \citet{richards2003red}, and derived the  extinction curve. They assumed, the bluest composite spectrum to be unaffected by dust while the other composite spectra differed only due to the dust reddening. Using these composite spectra, they obtained an empirical relation for the intrinsic extinction,
 %Using a large sample of quasars \citet{czerny2004extinction} obtained an empirical relation for AGN intrinsic extinction as 
\begin{equation}
    \frac{A_\lambda}{E(B-V)} = k_{\lambda} = -1.36 + 13\log{\frac{1}{\lambda}}
    \label{eq1:intred}
\end{equation}
where, $\frac{1}{\lambda}$ ranges from 1.5 to 8.5 $\mu m^{-1}$. The Balmer decrement, i.e., the ratio of  $H_\alpha$ and $H_\beta$ line fluxes has long been used as a reddening indicator. The color excess $E(B-V)$ can be written in terms of the Balmer decrement as follows \citep[see e.g.][]{dominguez2013dust}.

\begin{equation}
    E(B-V) = \frac{2.5}{k(\lambda_{H_\alpha}) - k(\lambda_{H_\beta})}\log_{10}\left[\frac{(H_\alpha/H_\beta)_{obs}}{(H_\alpha/H_\beta)_{int}} \right]
    \label{ebv_bd}
\end{equation}
where $k(\lambda_{H_\beta})$ and $k(\lambda_{H_\alpha})$ are the extinction curves evaluated
at $H_{\beta}$ and $H_{\alpha}$ wavelengths, respectively. 
As shown by \citet{gaskell2017case},  the broad line $H_\alpha/H_\beta$ ratio for the  optically blue AGN at low redshifts is $2.72$. We have used this value as the intrinsic ratio $(H_\alpha/H_\beta)_{int}$. We then used the extinction curve of \citet{czerny2004extinction} as given in Eq.~\ref{eq1:intred} and the color excess $E(B-V)$ based on the observed Balmer decrement $(H_\alpha/H_\beta)_{obs}$ (Eq.~\ref{ebv_bd}) and  created a spectral model for the spectral fitting package {\sc Sherpa}. The model parameters are redshift and Balmer decrement. This model calculates the extinction $A_{\lambda}$ which is then used to calculate intrinsic flux at the desired wavelength in the UV band
\begin{equation}
    f_{\lambda, int} = f_{\lambda, obs}10^{0.4A_{\lambda}}
\end{equation}

 In Fig.~\ref{fig:red_ef}, we show the effect of de-reddening for different amounts of extinction ($E(B-V) = 0.04, 0.09$ and $0.16$) on the observed spectrum. 
 In Table \ref{tab:balmDec} we list the  Balmer decrement obtained from previous study for the sources in our sample.
 
\subsection{\ion{Fe}{2} Emission}
\begin{figure}[ht!]
\epsscale{1.3}
\plotone{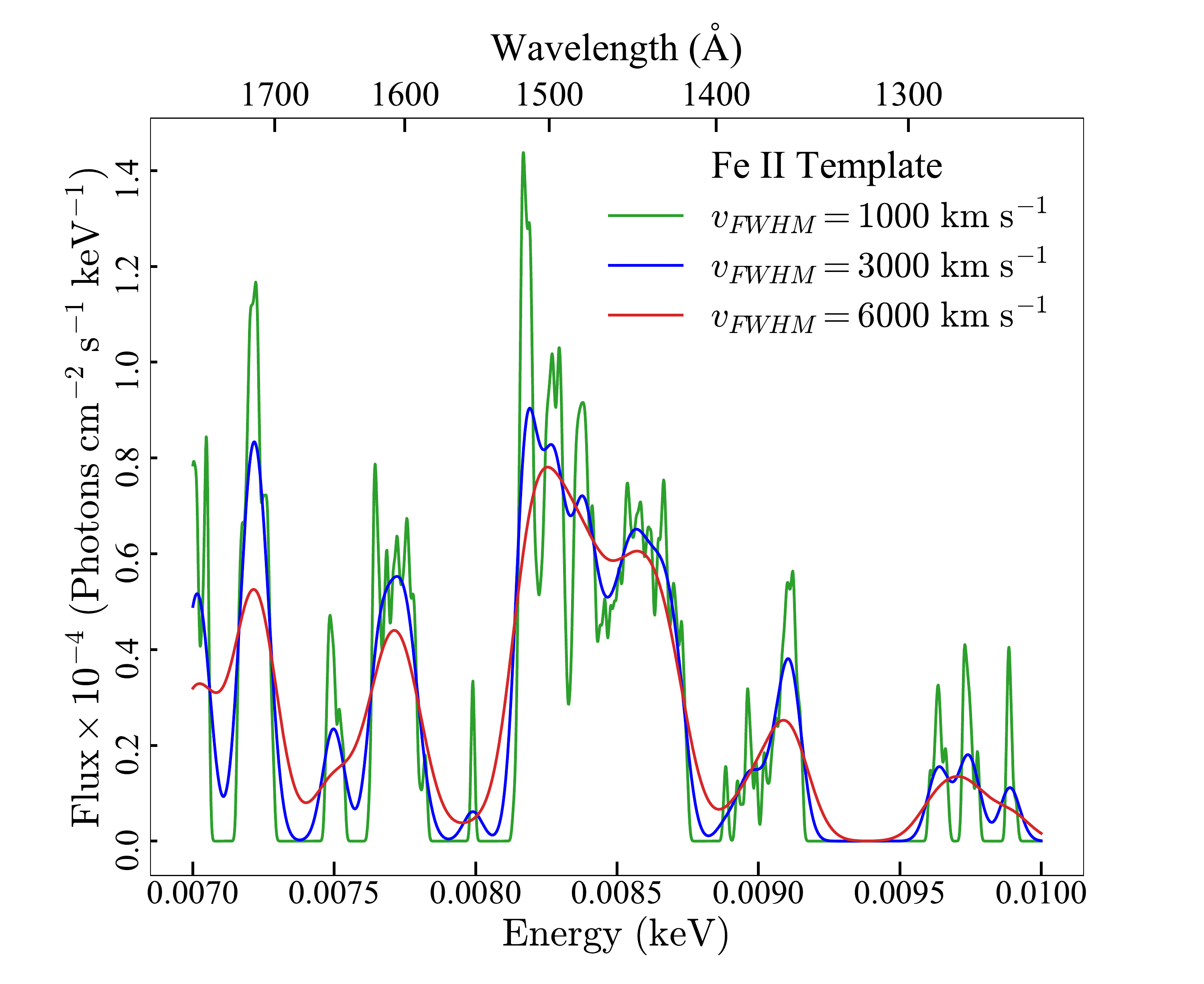}
\caption{Effect of broadening of \ion{Fe}{2} template spectrum for different velocities: $ v_{FWHM}=1000~\rm km~s^{-1}$ (green), $3000{\rm~km~s^{-1}}$ (blue), and $6000{\rm ~km~s^{-1}}$ (red) at $\lambda = 1500$\angs.
\label{fig:fe2}}
\end{figure}

Type 1 AGN,  particularly the  narrow-line Seyfert 1s, exhibit strong  \ion{Fe}{2} emission in the wavelength range of $1100-5000$\angs. The \ion{Fe}{2} complex is created by  thousands of electronic transitions of \ion{Fe}{2} ions  in the BLR. Due to the  Doppler broadening, these lines overlap with each other giving rise to a pseudo-continuum. To account for the \ion{Fe}{2} complex, we used the \ion{Fe}{2} template generated using  I~Zw~1   by \citet{Vestergaard_2001}  and, created a spectral model for the {\sc Sherpa} package. The parameters for this model are the redshift,  the Gaussian $\sigma$ for velocity broadening of individual lines, and a scaling factor to adjust the strength of the \ion{Fe}{2} complex. \citet{tilton2013ultraviolet} have used this approach and  studied the AGN continuum using \textit{HST}/COS observations. The template has an intrinsic width of FWHM $=900 {\rm~km~s^{-1}}$. The best-fitting $\sigma$ for an AGN spectrum is the width of the Gaussian used to broaden the \ion{Fe}{2} template that matches the  \ion{Fe}{2} emission in the AGN spectrum. Thus, 
 $\sigma = (\sqrt{FWHM_{AGN}^2 - FWHM_{intrinsic}^2})/{2.355}$,
 %
%The sigma after fitting is the convolved sigma given by, 
 %\begin{equation}
 %    \sigma _{conv} = \frac{\sqrt{QSO^2_{fwhm}-I~Zw~1^2_{fwhm}}}{2.533}
 %    \label{eq2:fesig}
 %\end{equation} 
 where $FWHM_{intrinsic} = 900{\rm~km~s^{-1}}$. 
 %Consequently, this puts a lower limit in the fitted template width.    
 In Fig.~\ref{fig:fe2}, we show the effect of different broadening ($\sigma$) on our model \ion{Fe}{2} template spectrum  with $z=0$.
 %\iffalse
%\fi
\subsection{Galactic Reddening} 
The dust grains and molecules in the ISM  of our own Galaxy contribute to the reddening of the optical/UV spectrum. We account for the Galactic reddening in the observed spectra of AGN by using the spectral model \texttt {REDDEN }\citep{cardelli1989relationship} in {\sc Sherpa}. This model has one parameter, the color excess $E_{B-V}$. We estimated the color excess from the observed linear relation between the Hydrogen column density  $N_{H}$ in $\rm cm^{-2}$  and  the extinction  $A_{V}$ in magnitude \citep{10.1111/j.1365-2966.2009.15598.x}:
\begin{equation} \label{eq3:nhgal}
  N_{H}[{\rm~cm^{-2}}] = (2.21 \pm 0.09) \times 10^{21} A_{V}[{\rm mag}]
\end{equation}
with $A_V = 3.1 \times E_{B-V}$.

\subsection{Balmer Continuum} 
The Balmer continuum emission in the wavelength range of $2000-4000~$\AA{}  arising from BLR  contributes to the intrinsic continuum.  For optically thin emitting region at a  fixed electron temperature, the Balmer continuum emission is given by \citep{1982ApJ...255...25G}  
\begin{equation} \label{eq4:balmcon}
 F_{\nu}^{BC} = F_{\nu}^{BE} \exp{[-h(\nu -\nu_{BE})/kT]}  ~~\rm{}  
 \end{equation}
 \noindent
 where, $F_{\nu}^{BE}$ is the flux at Balmer edge $\nu_{BE}$, $T$ is electron temperature and $h$ is Planck constant. $\nu_{BE}$ is theoretically at wavelength $3646$ \AA. We modeled this spectral component using the function given by Equation~\ref{eq4:balmcon}. We generated a spectral model for the Balmer continuum within the {\sc Sherpa} package. The parameters of this model are  redshift,  amplitude and electron temperature.  We used this model component convolved with the XSPEC \citep{1996ASPC..101...17A} model \texttt{GSMOOTH} (Gaussian smoothing) for the NUV-Grating spectrum available only for PG0804.
 
\section{Spectral analysis} \label{sec:spec_analys}
The slit-less UVIT gratings are useful for low-resolution spectroscopy (FWHM $\sim 2860~\rm km~ s^{-1}$) in a  limited wavelength band  (FUV: $1300-1800 $\angs, NUV: $2000-3000 $\angs). 
%We intend to derive the intrinsic continuum using \textit{AstroSat}/UVIT data. The  UVIT/FUV gratings have low spectral resolution (FWHM $\sim 2860~km s^{-1}$). 
The width of the emission lines from the NLR are typically $500{\rm~km~s^{-1}}$ whereas the BLR emission line widths  range from $\sim 1000{\rm~km~s^{-1}}$ to $\sim 10000{\rm~km~s^{-1}}$. The  absorption lines are generally narrow with typical FWHM  of the order of a $ \rm few~ hundred~km~s^{-1}$. Thus, the poor spectral resolution of UVIT gratings may not allow for  the separation of the emission and absorption lines. This may affect the measurement of continuum emission as our spectral window is populated with prominent broad and narrow emission/absorption lines. In order to separate the discrete features and to  measure the continuum shape reliably, we made use of the high resolution (FWHM(FOS)$\sim 230{\rm~km~s^{-1}}$;  FWHM(COS)$\sim 17{\rm~km~s^{-1}}$) \textit{HST} spectra. We first measure the emission/absorption lines in the \textit{HST} spectra and use the measured line widths and positions to fit the UVIT grating spectra.     
 For our spectral fitting, we  used the {\sc Sherpa} package developed for the \chandra{} X-ray observatory.  We  used $\chi^2$ statistics to find the best fit and calculated errors  on the best-fit parameters  at the  $1\sigma~ (68\%)$ confidence level.

\subsection{COS/FOS Spectral Analysis}\label{subsec:hst_spectral}
We first converted the fluxed \hst{} spectra into PHA spectral data and diagonal response using the {\sc ftflx2xsp} tool available within the HEASOFT package. This allowed us to treat the \hst{}  and the UVIT spectral data in the same fashion.
%We have converted the flux vs. wavelength spectrum into counts vs spectral channel (PHA) spectrum to analyze in {\sc Sherpa}.
For each source, we  used a common energy band  $0.007-0.0098\kev$  for the UVIT/FUV and the \textit{HST} data.  We used a simple  redshifted power-law (\texttt{ZPOWERLAW}) model for the underlying UV continuum in the observed spectrum. We reddened the continuum to account for the Galactic extinction using the  \texttt{REDDEN} model component with a single parameter -- the color excess ($E_{B-V}$). We kept this parameter fixed at the values listed  in Table~\ref{tab:genpar}. These values were calculated using Eq.~\ref{eq3:nhgal}.
%and the parameter is kept frozen at the quoted value.
We also tested the  model with the color excess calculated using the relation  $N_H = 5.8 \times 10^{21} E_{B-V}$ provided by \citet{1978ApJ...224..132B}. We found that these color excesses generally resulted in inferior fits compared to those calculated from Eq.~\ref{eq3:nhgal}, which we continued to use in the subsequent analysis.  Next, we added Gaussian components for the emission lines after inspecting the residuals. The most prominent lines observed are \ion{C}{4}~$\lambda 1449.5$\angs, \ion{Si}{4}/\ion{O}{4}]~$\lambda 1400$\angs, Ly$\alpha~\lambda 1215.7 $\angs~ and \ion{He}{2}~$\lambda 1640$\angs. 
For most of these broad emission lines, we  used two  Gaussian profiles -- broad and narrow components. The line widths are of the order of few thousand ${\rm~km~s^{-1}}$, typical of the BLR lines.  
A few more emission lines are added if required which are, \ion{O}{3}]/\ion{Al}{2}~$\lambda 1664.7$\angs, \ion{C}{2}~$\lambda 1335$\angs, \ion{O}{1}/\ion{Si}{2}~$\lambda 1305$\angs, \ion{Si}{2}~$\lambda 1262.6$\angs, and \ion{N}{5}~$\lambda 1241$\angs~(see Table \ref{tab:lines1} and \ref{tab:lines2}).  

In the case of MR~2251--178, we  used three Gaussian lines  for each of the \ion{C}{4} and Ly$\alpha$ emission lines. One of the \ion{C}{4} lines is blue-shifted by $13 $\angs~ in the source rest frame, indicating the possible presence of outflows. For Mrk~841 and SWIFT1921, we used a single component Gaussian emission line (\ion{N}{4}]~$\lambda 1485$\angs) at $1485$\angs ~$\rm{(FWHM\sim1704~km~s^{-1})}$ and $1486$\angs~ (FWHM$\sim 2200~ \rm km~s^{-1}$), respectively.  For I~Zw~1, some of the emission lines (\ion{C}{2}, \ion{N}{5} and \ion{O}{1}) are blue-shifted or red-shifted by $5-12$\angs{}. We list these additional weak emission lines of Mrk~841, MR~2251-178, SWIFT1921 and I~Zw~1 in Table~\ref{tab:lines3}. We added the absorption lines, which  further improved the fit significantly.  We list the rest wavelength, observed energy, width, and strength for the absorption lines in \hst{} spectra of six sources in Appendix Table \ref{tab:abs_lin1} and \ref{tab:abs_lin2}.

Although the iron emission is weak in the FUV band, nevertheless, we tested for the presence of the \ion{Fe}{2} complex.  For all the sources except for MR~2251-178, Mrk~352 and SWIFT1835, the addition of this component improved the $\chi^2$ by more than 5 per degree of freedom (dof). 

Next, we used the intrinsic reddening model, described in section 4.2, with the initial values of Balmer decrement set to those  found from the literature which ranges from $2.67 - 5.6$ (see Table~\ref{tab:balmDec}). If we fix the ratio at the values obtained from the literature, the fits worsened. Therefore, we left the parameter to vary and the fits improved significantly. We needed the intrinsic reddening  only for two AGN -- PG0804  and I~Zw~1. We listed the best-fit values in Table~\ref{tab:balmDec}. We calculated the errors on the best-fit Balmer decrements by fixing  the emission and absorption line centroids  at their best-fit values.  
In I~Zw~1, the color excess calculated from the Balmer decrement is $\sim 0.23_{-0.02}^{+0.02}$, which is slightly higher than that obtained using \ion{O}{1}~$\lambda1302$/\ion{O}{1}~$\lambda 8446$ line ratios ($E(B-V) \sim 0.19$, \citealt{rudy20001}).
 
 \begin{deluxetable}{ccc}
\tablenum{3}
\tablecaption{Comparison of 
Balmer decrements obtained from our best-fit values and those found in the literature. \label{tab:balmDec}}
\tablewidth{0pt}
\tablehead{
\colhead{Sources} & \colhead{  $\frac{H_\alpha}{H_\beta}$ (best-fit)  }  &\colhead{ $\frac{H_\alpha}{H_\beta}$}(Literature) }

\startdata
Mrk~841 & x  & $4.82^a$   \\
MR~2251--178   &  x  &$5.62^b$   \\
PG0804 & $2.749_{-0.002}^{+0.002}$ & $-$\\
NGC~7469 & x  &$5.4^a$   \\
I~Zw~1 & $3.9_{-0.1}^{+0.1}$ & $4.8^c$ \\
SWIFT1921 &  x &$3.6^c$  \\
Mrk~352  &  x  & $2.67^c$  \\
SWIFT1835 &  x & $4.43^d$\\
\enddata
\tablecomments{~~ $^a$~\cite{1983ApJ...273..489C}; $^b$~\cite{1978ApJ...226L...1C}; $^c$~\citet{1977ApJ...215..733O}; $^d$~\citep{popovic2003balmer}. `x': Not required in the modelling; `--': Not found in Literature. }
\end{deluxetable}

  %but this resulted in a spectral  index of $0.7_{-0.1}^{+0.6}$ which is higher than the predicted value. We discarded this component on this dataset. 

  %\iffalse
  \begin{figure*}
\gridline{\fig{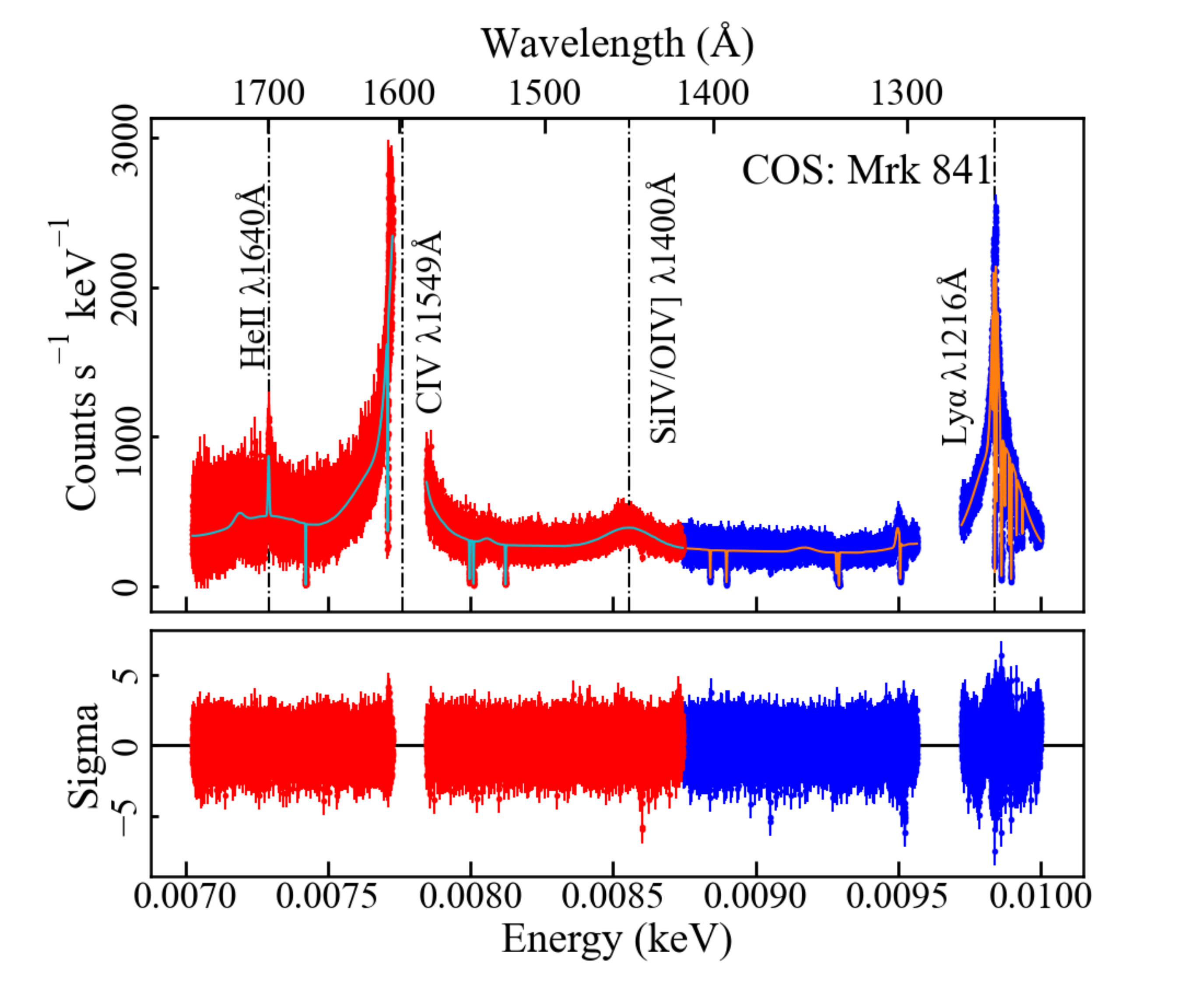}{0.55\textwidth}{(a)}
           \fig{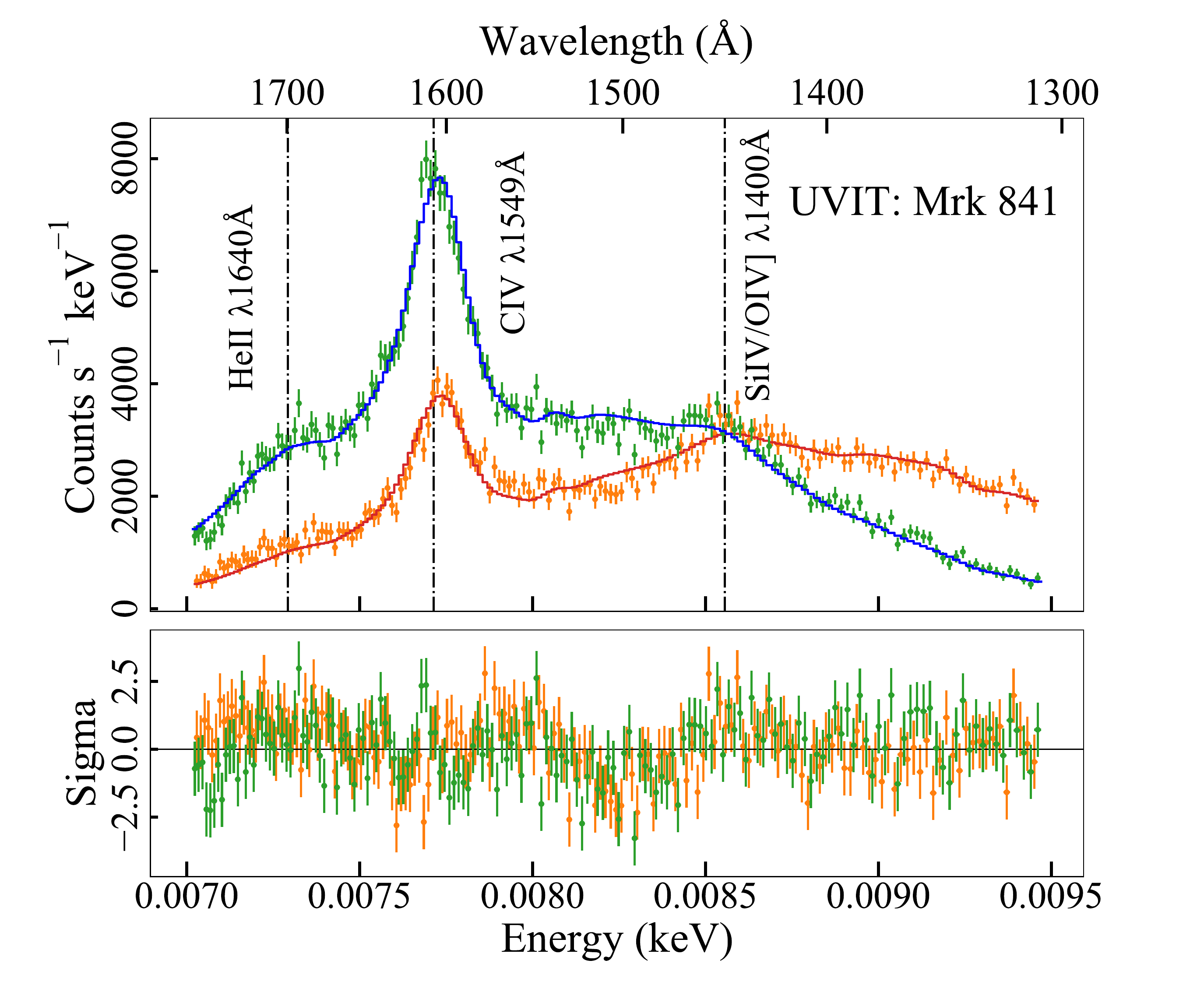}{0.55\textwidth}{(b)}}
           \caption{
 Top panels: (a) \hst{}/COS (red: 160M, blue: 130M) and (b) UVIT/FUV grating (orange: FUV-G1; green: FUV-G2) spectra of Mrk~841, and the best-fit models. Lower panels: fit-residuals in terms of the deviations ${\rm sigma=(data-model)/error}$.
 %Mrk~841: The top panels show the spectra and the best-fit models. The bottom panels  show fit-residuals in terms of the deviations of the data from the best-fit model in units of sigma  ((data--model)/error). 
 The prominent emission lines -- \ion{He}{2} $\lambda1640$\angs, \ion{C}{4} $\lambda1549$\angs, \ion{Si}{4}/\ion{O}{4}]$\lambda1400$\angs, \ion{O}{1}/\ion{Si}{2}$\lambda1305$\angs, Ly$\alpha \lambda1215.7$\angs~ are marked with  vertical dashed lines at their wavelengths in the observed frame. 
 %Here, (a) \hst{}/COS (Red: 160M; Blue: 130M) and  (b) UVIT/FUV  grating spectra (Orange: FUV-G1; Green: FUV-G2).
%for Mrk 841: (a)COS, (b) UVIT-FUV G1 and G2 
\label{fig:841spec}}
\end{figure*}
%\fi

 NGC~7469 is  known to host starburst regions around the AGN located at $1^{\prime\prime}-3^{\prime\prime}$ and $8^{\prime\prime}-10^{\prime\prime}$ from  the center \citep{mauder1994high,mehdipour2018multi}. For our wavelength range of interest and the spectral extraction region (both in the \textit{HST} and UVIT spectra), we account for the contribution from the starburst regions by using a template spectrum. \citet{kinney1996template} have generated six starburst template spectra suitable for different extinctions such as  $E(B-V)\le 0.1$ (SB1), $0.20-0.21$ (SB2), $0.25-0.35$ (SB3), etc. We generated starburst template models using the starburst template spectra by introducing a single parameter -- a normalization which is a multiplicative factor. We tested all the starburst template models from SB1 to SB6. Only SB3 resulted in a very marginal improvement with $\Delta\chi^2 =-3$ for an additional parameter. 
 %The color excess associated with the starburst template SB3 is $0.25<\rm E_{B-V}<0.35$ (Balmer decrement, $4.02-4.7$). 
 \citet{mehdipour2018multi} also found the suitability of the SB3 template for NGC~7469.
The spectral indices obtained in our analysis for MR~2251--178 and NGC~7469 are slightly different (see Table \ref{tab:index}) than that obtained by \citealt{kuraszkiewicz2004emission} ($\alpha = -0.82_{-0.03}^{+0.01}$ for MR~2251--178 and $ -1.1 \pm 0.01$ for NGC~7469). They derived the slope using line-free regions and for a wider band ($\sim 1250-3100$\angs), while we fitted the continuum and the lines simultaneously in the narrower $1300-1800{\rm~\AA}$ band.  We also used slightly different values of $E_{B-V}$ for the Galactic reddening based on the updated relation between the Galactic column density and the extinction (Eq.~\ref{eq3:nhgal}). These differences in the analyses likely resulted in different spectral indices.
%We fitted the continuum and the lines simultaneously which possibly resulted in a little different spectral slope. 
%Similarly, for MR~2251-178, \citealt{kuraszkiewicz2004emission} obtained slightly redder slope, $ \alpha = -0.82_{-0.03}^{+0.01}$. }}
 
 The spectrum of Mrk~352 showed the presence of extremely narrow emission lines in the observed frame  at $1302$\angs, $1305$\angs{} and $1306$\angs{}  with FWHM $\sim 150-200 {\rm~km~s^{-1}}$  in addition to the broad emission lines. These narrow emission lines observed at the exact position of \ion{O}{1} $\lambda\lambda\lambda 1302,1305,1306$\angs~ triplet are highly unlikely to be related to the AGN, and could possibly be of the Galactic origin. We obtained the best-fit reduced $\chi^2$ for \hst{} spectral fits in the range  of $0.6$ (Mrk~352) to $1.6$ (NGC~7469). The poor quality of the fit to the spectrum of NGC~7469 is most likely due to absorption-like broad features at $\sim 1250$\angs~ which are likely due to  multiple narrow absorption lines of \ion{N}{5}. We did not attempt to fit very weak features which are unlikely to affect continuum parameters significantly.  We show the \hst{} spectra and the corresponding best-fit models in the left panels of Fig.~\ref{fig:841spec}, \ref{fig:mr22_pg08_7469} and \ref{fig:zw1_1921_352}.
 %{\color{red} Any reference on the Galactic emission lines?}.

\subsection{UVIT Spectral Analysis} \label{subsec:uvit_spectral}
We started the fitting of  UVIT grating data using the same set of parameters  derived from  the \hst{} spectral analysis. Initially, we fixed all the parameters except the normalization and photon index of \texttt{ZPOWERLAW}. We used a multiplicative constant  factor for scaling the emission line fluxes. This allowed us to preserve the line ratios obtained from the \textit{HST} data. We also used another constant factor to account for any difference in cross-calibration between the UVIT/FUV-G1 and FUV-G2.  We fixed the color excess parameter for the Galactic reddening model \texttt{REDDEN} as in the case of \textit{HST} spectral fitting. We also fixed the Balmer decrement parameter in the  intrinsic reddening model for AGN unless mentioned otherwise. First, we varied the  emission line scaling factor, the constant factor between the two gratings and the power-law slope. 
 If we see strong residuals near emission lines or the fit is statistically not acceptable, we allowed to vary line centroids and/or widths  and check the improvement in the fit. 
%{\color{red} (Do you mean continuum spectral parameters here? If yes, mention "those continuum spectral parameters".)} which resulted in significant improvement in the fits,  otherwise, we kept the parameters fixed at the best-fit values obtained from the \hst{} spectral analysis.  
%
Thus we accounted  for any change in the power-law continuum and the emission lines which might have occurred since the \hst{} and UVIT observations were non-simultaneous, and the type 1 AGN are known to exhibit spectral variability in the UV band \citep{reichert1994steps,o1998steps,peterson1999keplerian,wang2011coexistence,miranda2021uv}.
%We fixed the color-excess parameter in the REDDEN component used for  the Galactic extinction and the Balmer decrement parameter in the intrinsic reddening model component to  the  \textit{HST} values for the rest of our spectral analysis. 

Once we finalized the spectral fits with the redshifted power-law (\texttt{ZPOWERLAW}) component, we used the following accretion disk models (in place of the \texttt{ZPOWERLAW} component) for the continuum emission from each AGN.
\begin{enumerate}[label=(\roman*)]
\item Simple multi-temperature accretion disk blackbody (\texttt{DISKBB}; \citealt{1984PASJ...36..741M}).

\item Fully-relativistic accretion disk around a Kerr black hole (\texttt{ZKERRBB}; \citealt{2005ApJS..157..335L}).

\item Accretion disk with Novikov and Thorne Temperature profile (\texttt{OPTXAGNF}; \citealt{done2012intrinsic}).

\end{enumerate}

Below we describe the spectral analysis for individual sources. 
%{\color{blue} For the continuum component the different disk models that we used  are -- \texttt{ZPOWERLAW} (ZPL), multi-temperature blackbody model \texttt{DISKBB}, fully relativistic model \texttt{ZKERRBB} and Comptonised disk model \texttt{OPTXAGNF}}. \\

%\par
%\noindent
%\underline{\textbf{\emph{MRK~841}}}: 
\subsubsection{Mrk~841} \label{subsec:mrk841}
We fitted both FUV-G1 and FUV-G2 spectra jointly. Part of the Ly$\alpha$ line falls around the edge of the spectrum at $\sim 9.7 \ev$. Therefore, we removed the entire Ly$\alpha$ line region from the spectrum, which resulted in our analysis region of  $7-9.5\ev$ (see Fig.~\ref{fig:841spec}(b)). We varied  the emission line scaling factor, redshifted power-law normalization and the photon index. This resulted in  $\chi^2~ \rm{of}~ 864$ for $323$ dof. Then we varied the widths and the centroids  of \ion{C}{4} emission lines. This improved the fit to $\chi^2/dof  = 410/320$.  We obtained the best-fit photon index,  $\Gamma \sim 1.36$ (Table~\ref{tab:index}) and the emission line scaling factor,   $f_{sc}\sim 0.69$ (Table~\ref{tab:lines1}). 

The UV continuum emission is thought to arise from the accretion disks in AGN. We first tested the simple multi-color accretion disk model, \texttt{DISKBB} \citep{1984PASJ...36..741M}. This model assumes a geometrically thin disk with each annulus emitting like a blackbody at the local temperature. The resulting spectrum consists of a superposition of multiple blackbody components with the highest temperature corresponding to the innermost disk. It does not consider the relativistic effects of the extreme gravity near the black hole. We  convolved the \texttt{DISKBB} component with the \texttt{ZASHIFT} to derive the emission in the source rest frame.   This model provided $\chi^2/dof =  414/320$  with the inner disk temperature ($\rm kT_{in}$) of  $\sim 5.8\ev$.

Next, we considered a fully relativistic accretion disk model \texttt{ZKERRBB} \citep{2005ApJS..157..335L}. This uses a ray tracing method for photon trajectory in each annulus. The disk is divided into a number of image elements. The orbit of the photon from each element is traced backward from the observer to the plane of the disk. Considering all the relativistic effects, individual flux density at each element is summed over all elements to obtain the observed flux density. For black holes with mass $\ge 10^6 \rm M_\odot$, the hardening factor is estimated to be $\sim 2.4$ \citep{done2012intrinsic}. We have considered isotropic emission from the disk, such that the limb darkening effect is not included. Also, we assumed zero torque at the inner boundary. We varied only the inclination, mass accretion rate, and normalization and performed the fit. We considered two cases of the spin parameter ($a^\star$): 0 and 0.998.  As the inclination angle is not well constrained, we tested the model for a fixed inclination angle, $i$.  \citet{nandra1997asca} estimated  inclination angles of $27_{-8}^{\circ+7}$ and $38_{-12}^{\circ+2}$ based on X-ray reflection spectroscopy for two sets of observations made in 1993 and 1994 using \textit{ASCA}. Keeping the inclination fixed at the lower end, i.e., 20$^\circ$ resulted in  $\chi^2/dof  = 461/321$ and mass accretion rate of $\sim 0.38~{\rm M_{\odot}~yr^{-1}}$  for $a^\star = 0.998$. The normalized mass accretion rate ($\dot{m} = \dot{M}/\dot{M_E}$ ) is $\sim 0.3$ with $a^\star = 0.998$ and $\sim 0.5$ with $a^\star = 0$.  The  best-fit parameters are listed in Table~\ref{tab:index}.

Many AGN are known to show excess emission in the soft X-ray band below $2\kev$ which could arise due to thermal Comptonization in the inner disk region. In such a scenario the standard disk is likely truncated at some radius ($\rm r_{cor}$) larger than the innermost stable circular orbit (ISCO). For $\rm r> r_{cor}$ disk emission is modified black body spectrum.  This is described by \texttt{OPTXAGNF} \citep{done2012intrinsic} available in XSPEC. We used this model to obtain the disk contribution only by making $\rm r_{cor}$ negative. This model utilizes the temperature profile of a relativistic accretion disk but unlike the \texttt{ZKERRBB} model, it does not consider the general relativistic effects on photon propagation. Again, we used two cases with spin parameter, $a^\star = 0.998$ and $0$. We obtained $\chi^2/dof  = 412/320$ for the $a^\star = 0.998$. The best-fit Eddington ratio ($L/L_{Edd}$) is  $\sim 1.3$. For the non-rotating case, the fit resulted in $\chi^2/dof = 416/320$. The $L/L_{Edd}$ in this case is $\sim 0.3$. Although the \texttt{OPTXAGNF} model generates the accretion disk model by setting the $\rm r_{cor}$ negative, some fraction of the accretion power also goes to the warm and hot comptonization component. Therefore we extrapolated the model from $0.0001-10\kev$ and calculated the disk luminosity ($L_{Bol}$) by integrating the model. The Eddington ratios are, $L_{Bol}/L_{Edd} = 1.7$ for  $a^\star = 0.998$ and $0.3$ for $a^\star = 0$. The difference in the $\chi^2$ between the two disk models \texttt{ZKERRBB} and \texttt{OPTXAGNF} arises largely due to the variation in the color-correction factor ($f_{col}$).   In the case of \texttt{ZKERRBB} model, this factor is fixed at 2.4 at all radii, whereas,  in the case of \texttt{OPTXAGNF} model,  it varies from 1  to a higher value ($\sim 2.7$)  depending on the local effective disk temperature ($T_{eff}$) as follows \citep[see][]{done2012intrinsic}, \\
\begin{equation}
    f_{col}=
    \begin{cases}
      1 \hspace{2.8cm}; ~T_{eff}<3\times10^4 {\rm~K} \\
      \Big(\frac{T_{eff}}{3\times10^4{\rm~K}}\Big)^{0.82} \hspace{0.91cm}; ~3\times10^4{\rm~K} \leq T_{eff} < 10^5\rm~K\\
      \Big(\frac{72\keV}{kT_{eff}(\kev)}\Big)^\frac{1}{9}~\hspace{0.58cm};~ T_{eff} \gtrsim 10^5{\rm~K}  \\
      
    \end{cases}
    \label{eq:f_col_eq}
  \end{equation}
   % old--As the effective disk temperature rises with decrease in radius and reaches to $\sim 10^5{\rm~K}$ or $\sim 8.6\ev$ at a certain radius, the FUV continuum slope changes sharply (see Fig.~\ref{fig:mrk841_zkopt}) due to the change in $f_{col}$ as a function of temperature (See Fig.~1 in \citet{Zdziarski_2022}). While in \texttt{ZKERRBB} model, the slope remains almost constant up to $\sim 10\ev$. As mentioned earlier, in \texttt{DISKBB} we obtained the peak disk temperature $\rm (kT_{in}) \sim 16\ev (1.8\times 10^5K)$ which is higher than the transition temperature, $10^5K$. Therefore, the flattening of spectral slope due to the change in the $f_{col}$ occurs in the UVIT band. This essentially changes the shape of the continuum, leading to different  $\chi^2$ for the \texttt{ZKERRBB} and \texttt{OPTXAGNF} models.

As described below, we followed a similar procedure for spectral analysis of the rest of the sources, and list the best-fit spectral parameters in Table~\ref{tab:lines1}, \ref{tab:lines2}, \ref{tab:lines3} and \ref{tab:index}. \\

%\noindent
%\underline{\emph{\textbf{MR~2251--178}}} : \\
\subsubsection{MR~2251--178} \label{subsec:mr22}
\begin{figure*}
\gridline{\fig{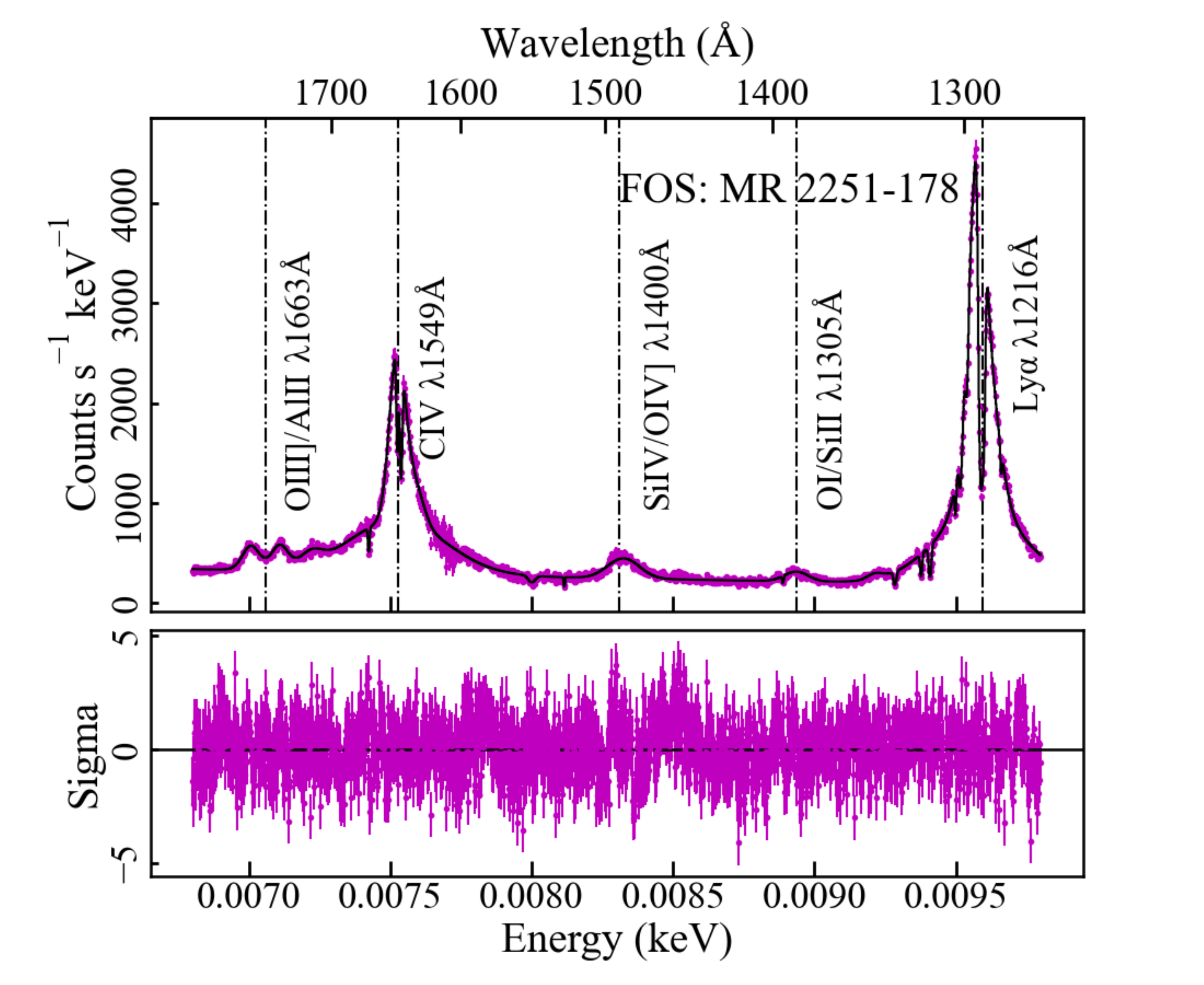}{0.47\textwidth}{(a)} 
           \fig{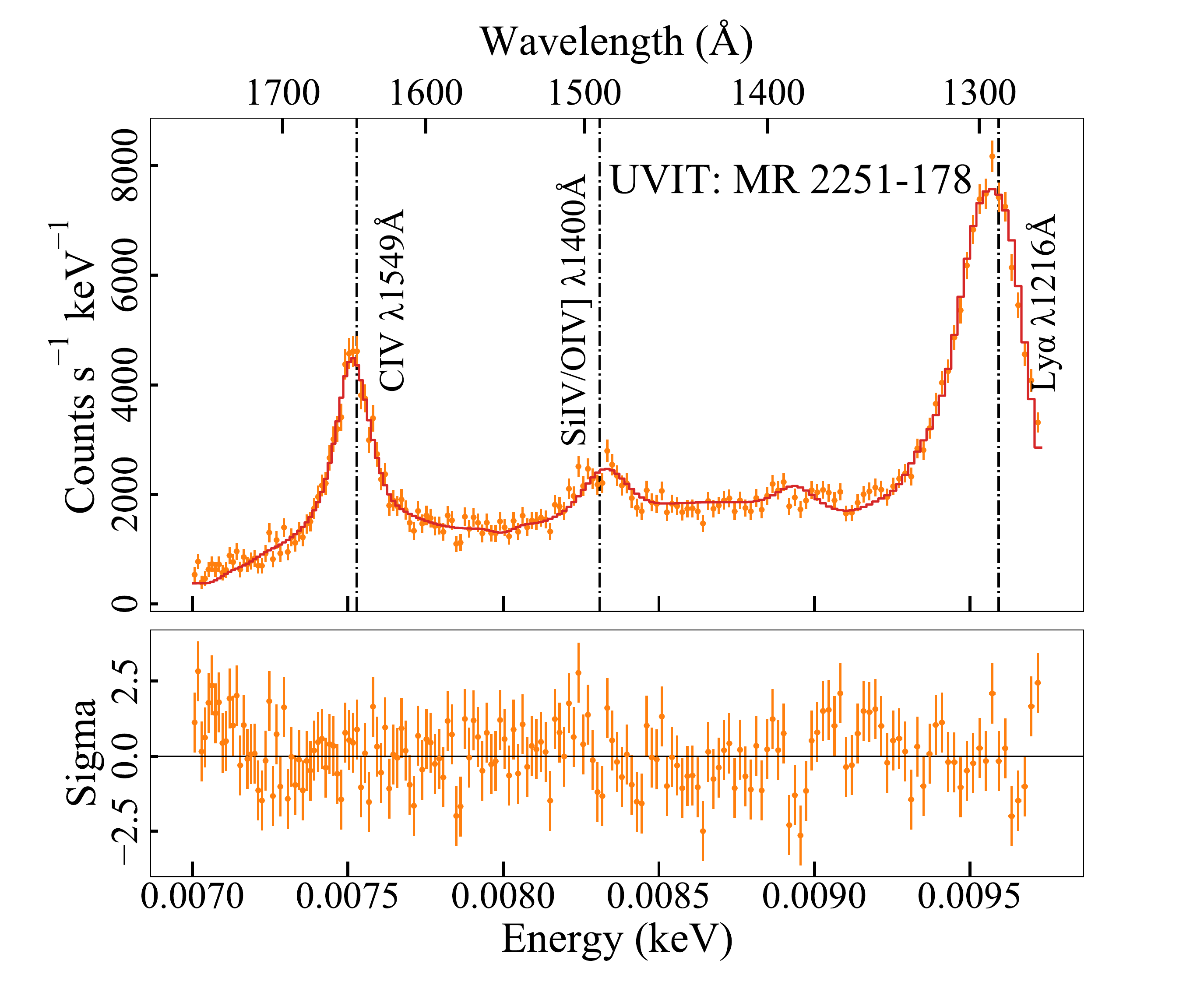}{0.47\textwidth}{(b)}}

\gridline{\fig{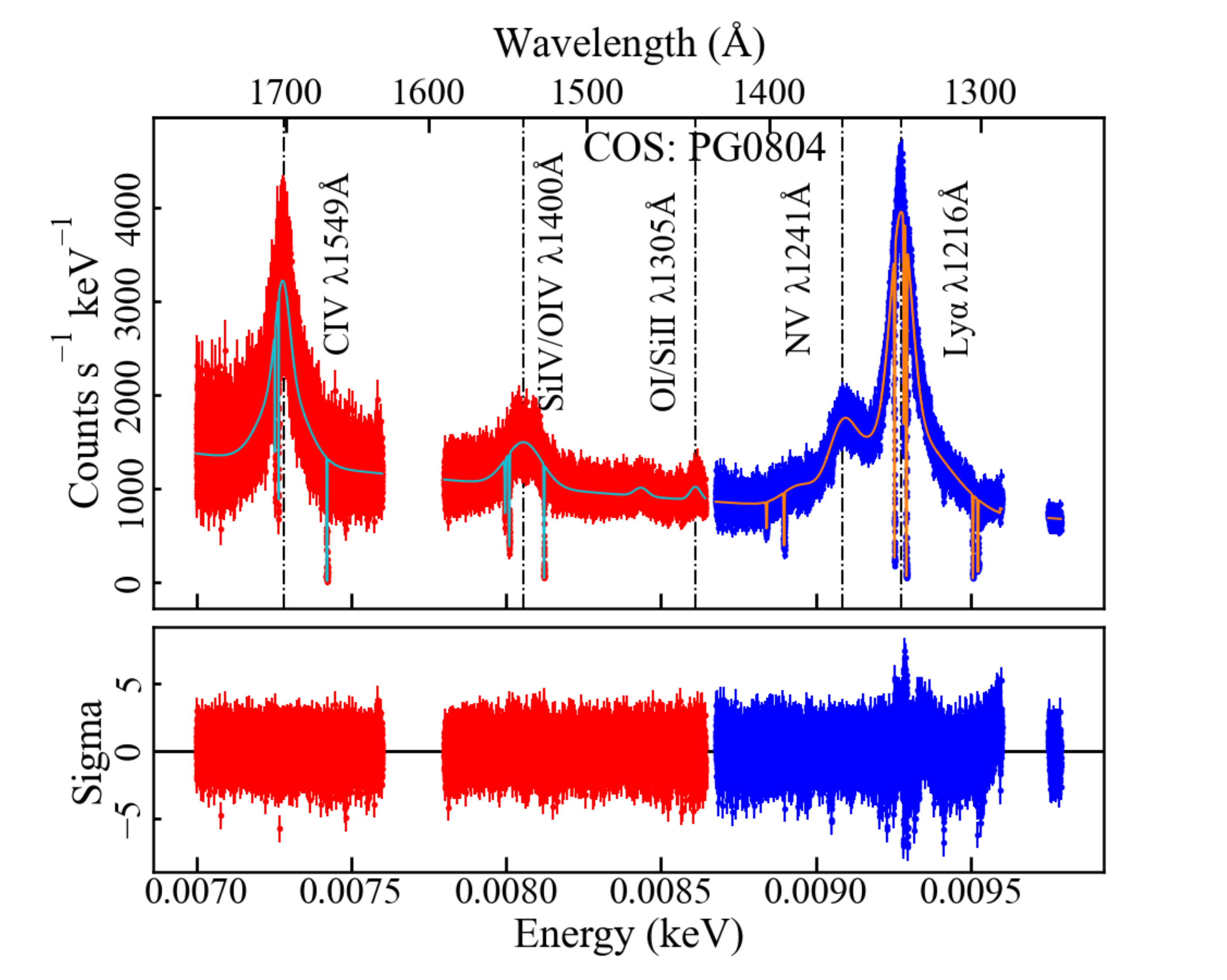}{0.47\textwidth}{(c)} 
           \fig{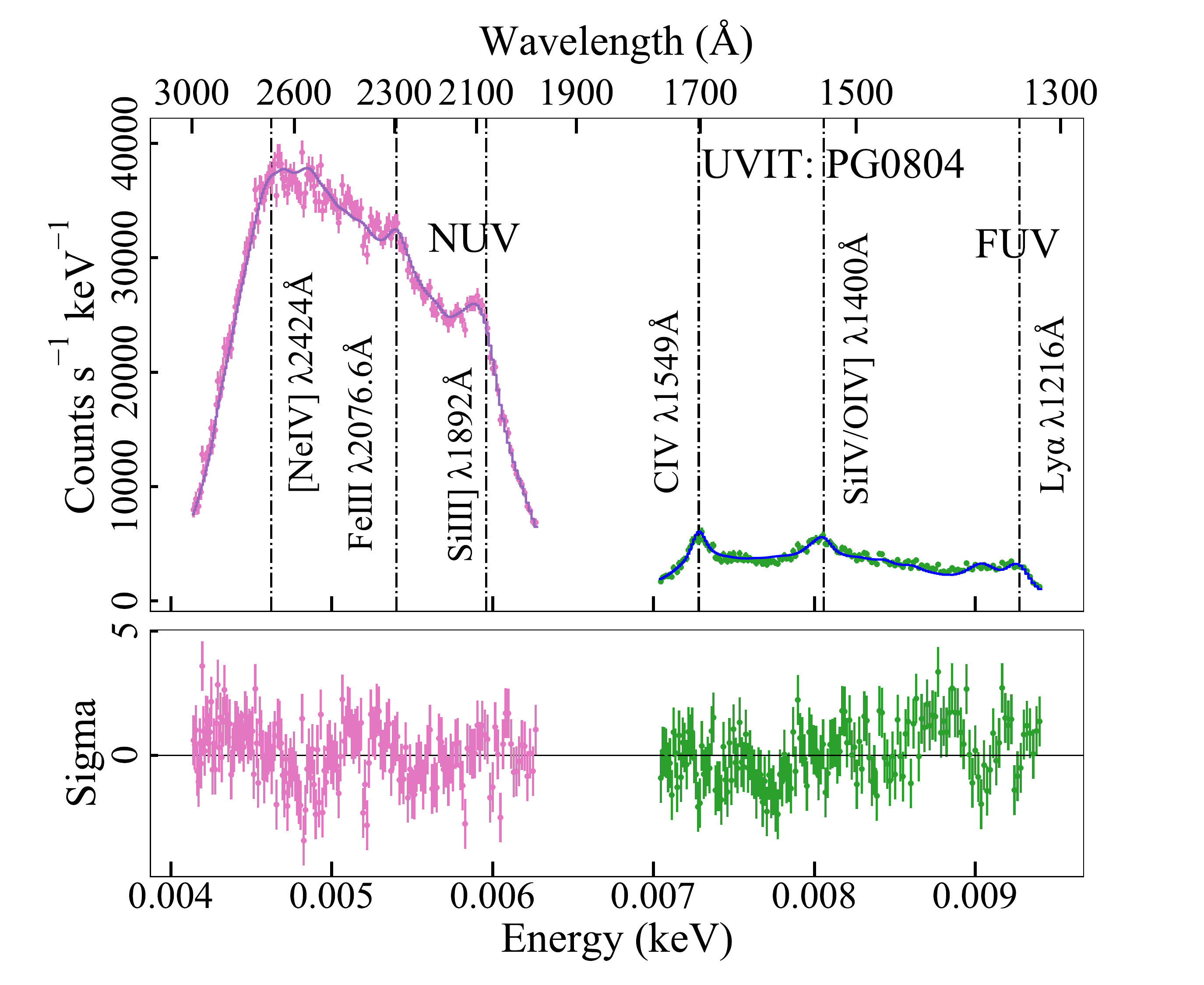}{0.47\textwidth}{(d)}}

\gridline{\fig{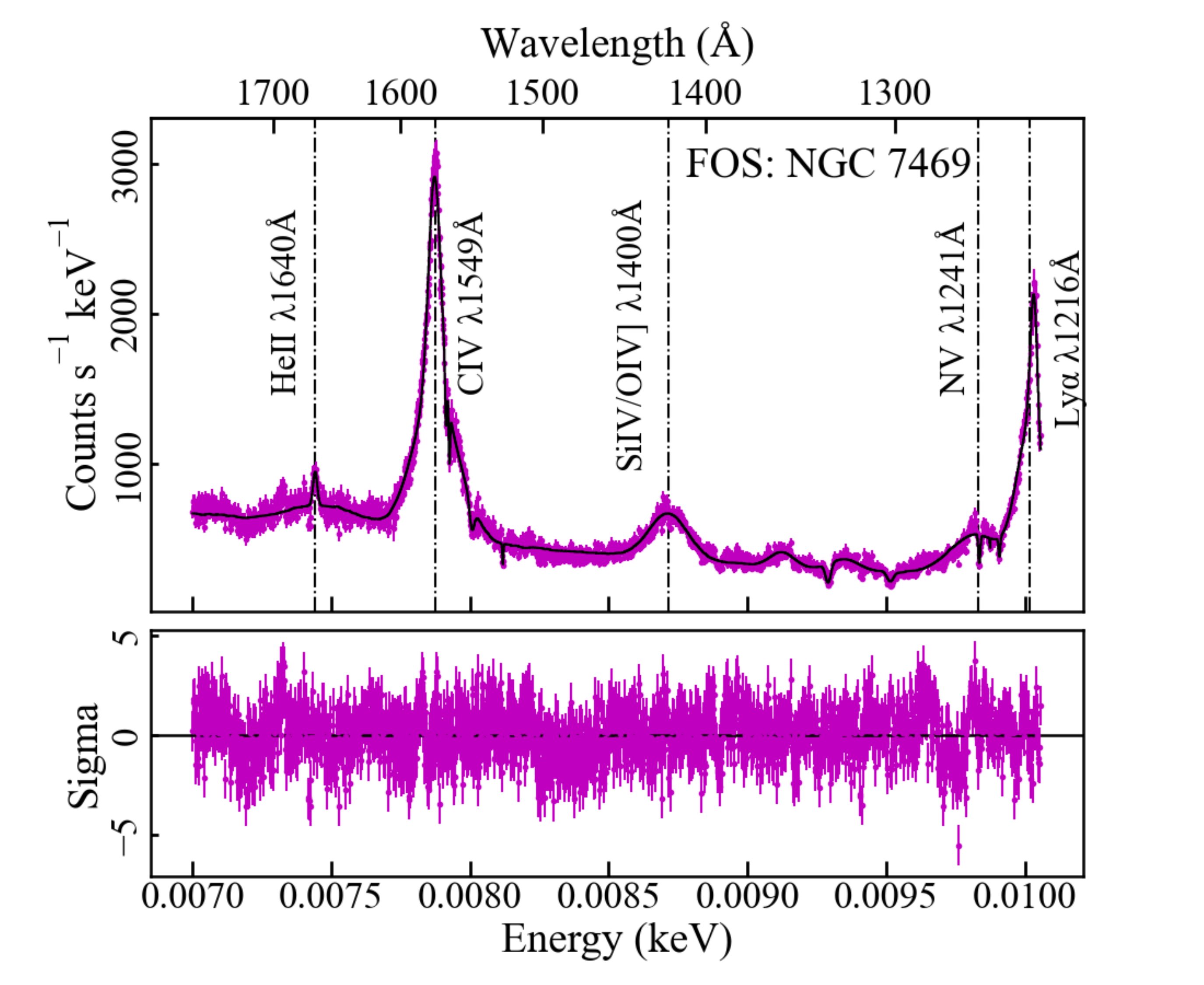}{0.47\textwidth}{(e)}
\fig{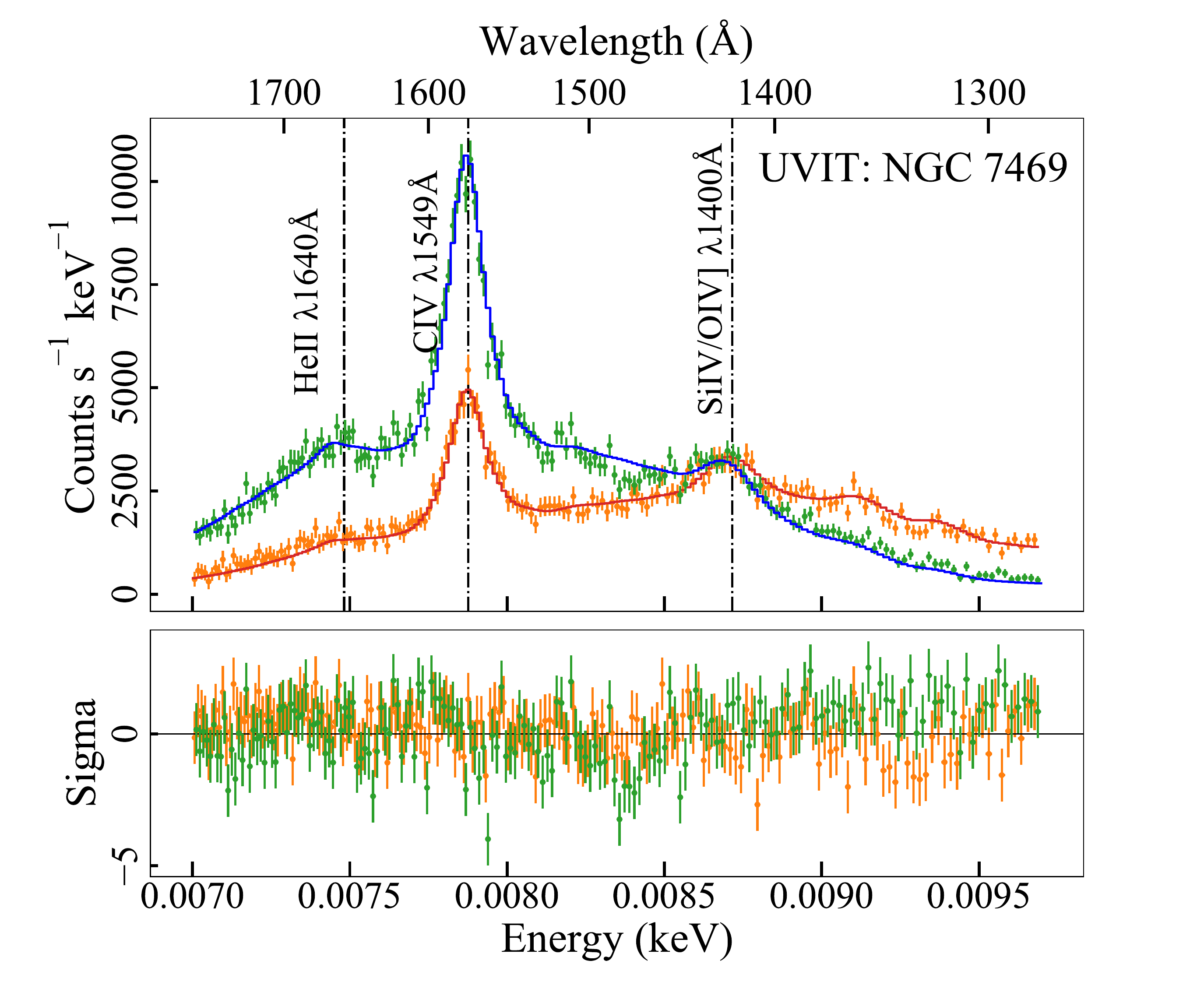}{0.47\textwidth}{(f)}}
            \caption{Same as Fig.~\ref{fig:841spec} but for MR~2251--178 (a,b), PG~0804+761 (c,d), and  NGC~7469 (e,f). \textbf{Left panels}: \hst{}/COS (red: 160M; blue: 130M),  \hst{}/FOS (magenta). \textbf{Right panels}: UVIT/FUV (orange: FUV-G1; green: FUV-G2) and UVIT/NUV-G (pink).}
             \label{fig:mr22_pg08_7469}           
\end{figure*}

After varying the simple power-law index, normalization, and scaling factor to the emission lines and emission line centroids, some residuals still remained which could be due to additional emission lines. To account for these residuals,  we added an emission line at $1243 $\angs~ (\ion{N}{5}$\lambda1241$\angs) which was not detected in the \hst{} spectrum (Table \ref{tab:lines1}). This improved the $\chi^2$ by 62 for three additional parameters. Further, we added one more line near the  \ion{C}{4} region  at $1561 $\angs~ (\ion{C}{4}$\lambda1549$\angs) which resulted in $\Delta \chi^2 = -29$ for three additional parameters. The FWHM and line normalization of \ion{C}{4} are 3861 $\rm km~s^{-1}$ and $0.06_{-0.01}^{+0.01}$, respectively.  The best-fit model consisting of redshifted power-law and emission lines  resulted in  $\chi^2/dof = 203/166$   (Fig.~\ref{fig:mr22_pg08_7469}(b)) with $\Gamma \sim 2.1$ and the scaling factor for the emission lines,  $f_{sc}\sim1.3$. The \texttt{DISKBB} model resulted in $\chi^2/dof= 205/166$   with the best-fit inner disk temperature of $\rm kT_{in}= 3.6_{-0.3}^{+0.4}\ev$. This is lower than  that derived ($\rm kT_{in} \sim 6.27\ev$) for the \textit{HST}/COS observation performed in 2020  \citep{mao2022multiwavelength}. The disk temperature can vary depending on the accretion rate and thereby the disk luminosity. As discussed below, we obtained  nearly five times lower accretion rate  than that derived by \citealt{mao2022multiwavelength} ($\sim 2.5 ~\rm M_\odot~yr^{-1}$). In the relativistic accretion disk model \texttt{ZKERRBB} with $i,  \dot{M}$ and normalization as free parameters, we could not constrain the $i$ and $\dot{M}$. \citet{brunner1997uv} estimated an inclination angle of ~70$^\circ$ by fitting the standard disk model to the \textit{IUE} and \textit{ROSAT} data. A recent study using X-ray reflection suggested that the inclination angle is  $\sim 25^\circ$ \citep{nardini14}.
We obtained the $\chi^2/dof= 242/167$ ($a^\star$= 0.998) and  $\dot{M}\sim 0.26~ \rm M_\odot ~yr^{-1}$ ($\dot{m}\sim  0.03$) for inclination angle, $i = 25^\circ$.  For $a^\star$= 0, we obtained $\chi^2/dof= 226/167$ and $\dot{M}\sim 0.46~ \rm M_\odot ~yr^{-1}$ ($\dot{m}\sim  0.06$). With  $i = 70^\circ$ the fit worsened by $\Delta \chi^2 = -13$. We obtained $\chi^2/dof  = 205/166$ with \texttt{OPTXAGNF} for both the spin cases. The model integrated Eddington ratios are, $L_{Bol}/L_{Edd} = 0.045$ ($a^\star = 0.998$ ) and $\sim 0.04~$ ($a^\star = 0$). We obtained the inner disk  to be truncated at a radius $\sim 16 r_g$, which  indicates the $f_{col}$ may be close to 1. This is apparent from the best-fit model with \texttt{ZKERRBB} as continuum component where the $\chi^2$  worsened by $-37$.  \\

%\iffalse

%\fi

%\iffalse
\begin{figure*}
          
 \gridline{\fig{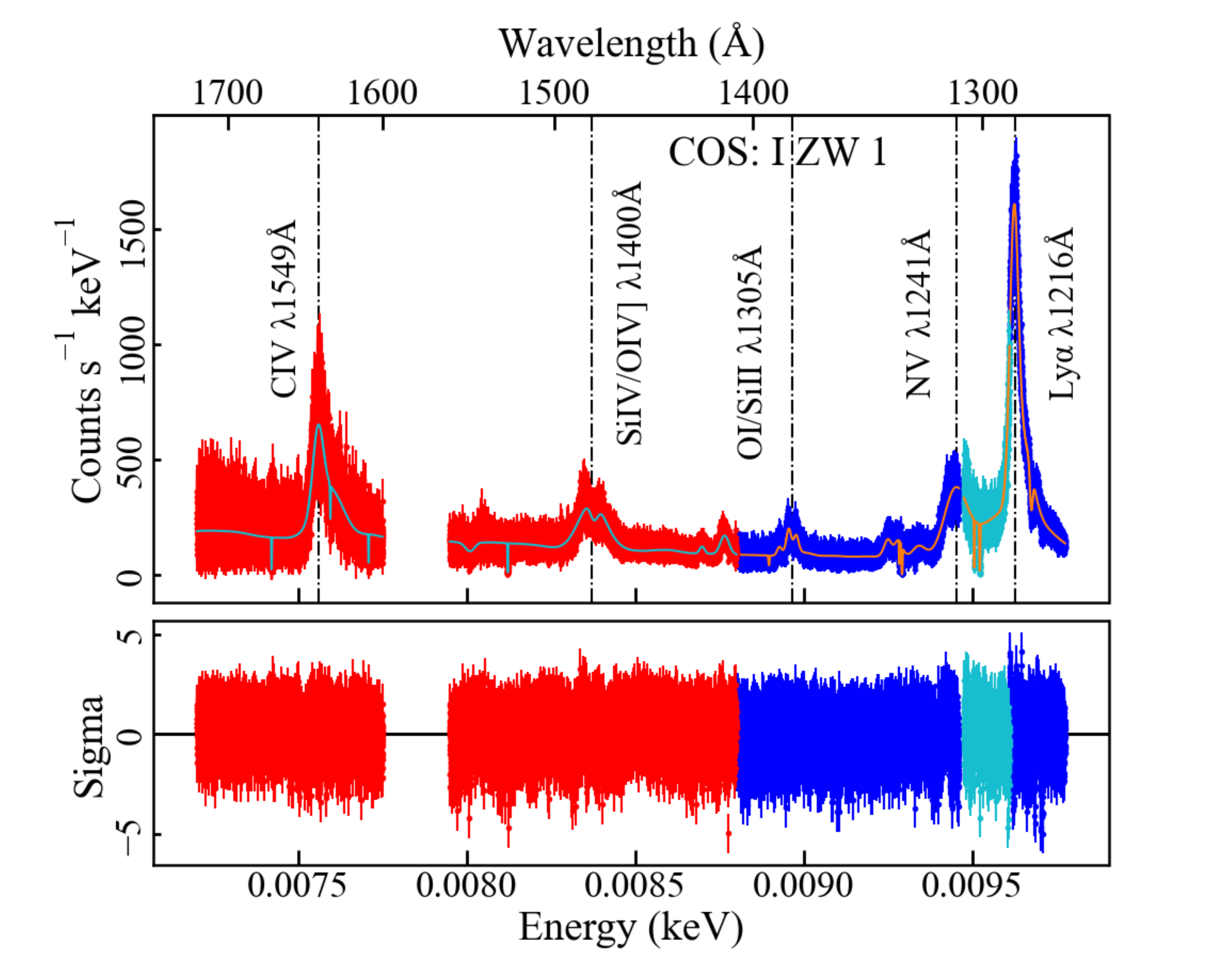}{0.47\textwidth}{(a)} 
           \fig{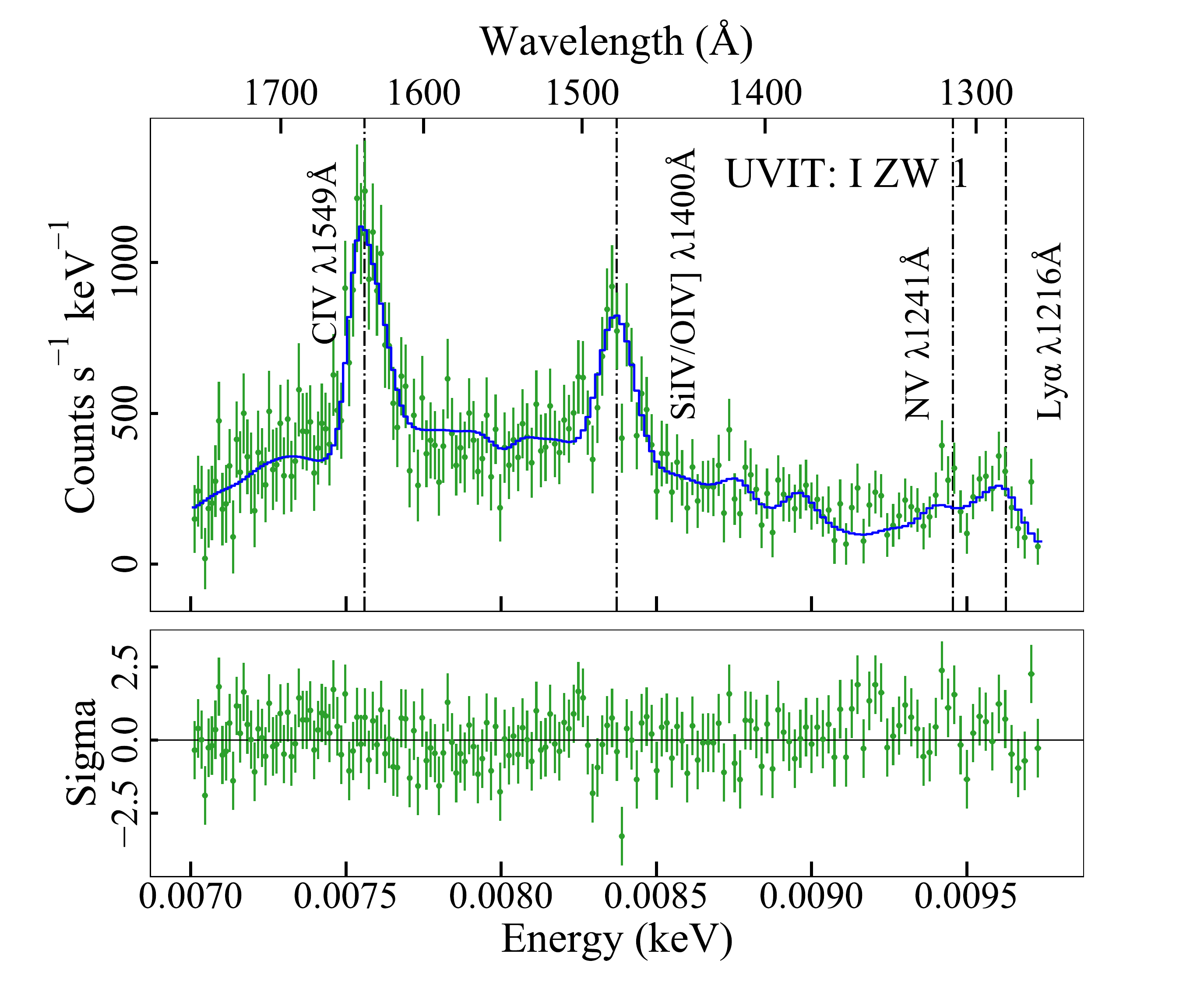}{0.47\textwidth}{(b)}}
            %\caption{Same as Fig.~\ref{fig:841spec} but for I~Zw~1: (a) COS and (b) UVIT/FUV-G2}
\gridline{\fig{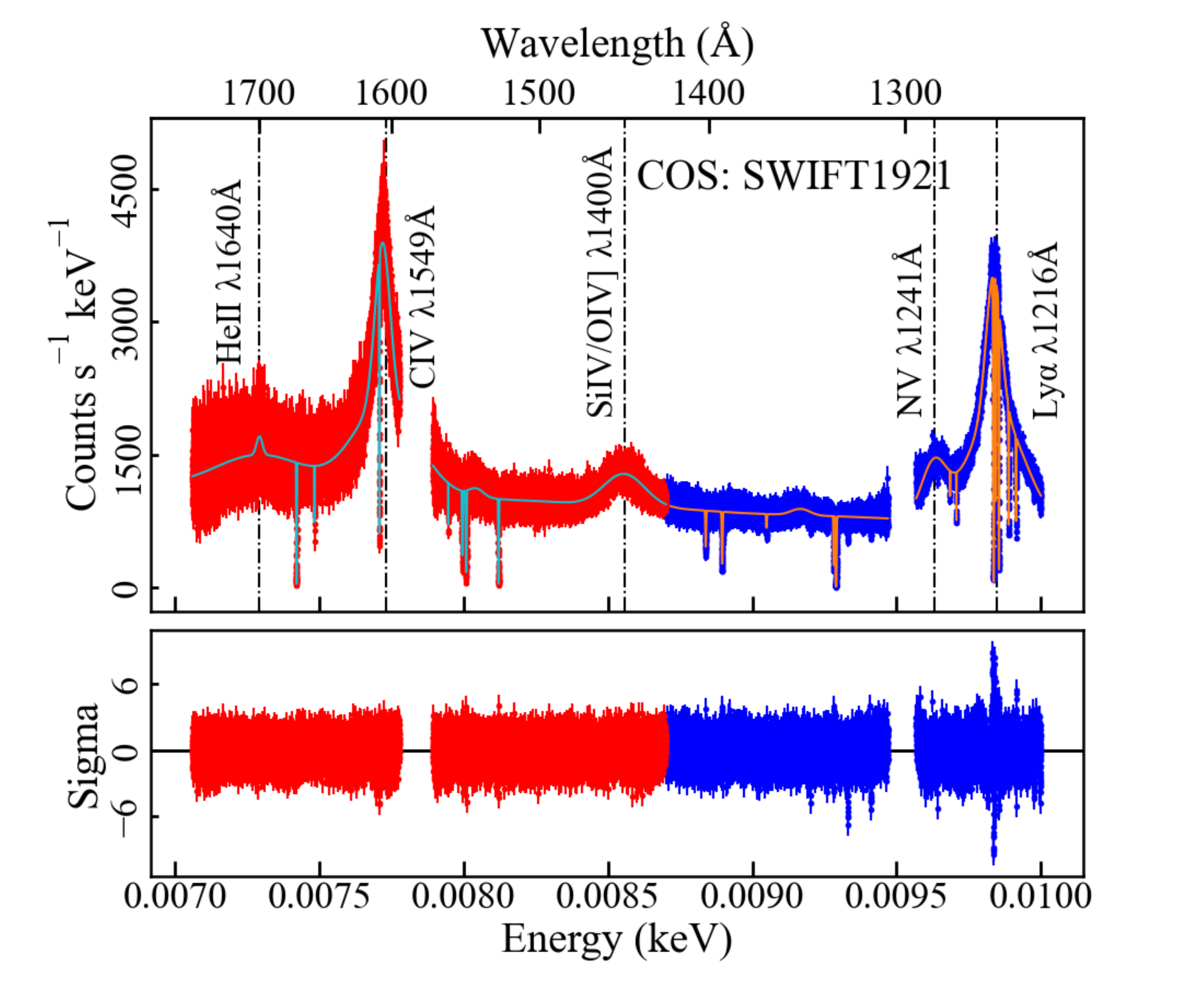}{0.47\textwidth}{(c)}
           \fig{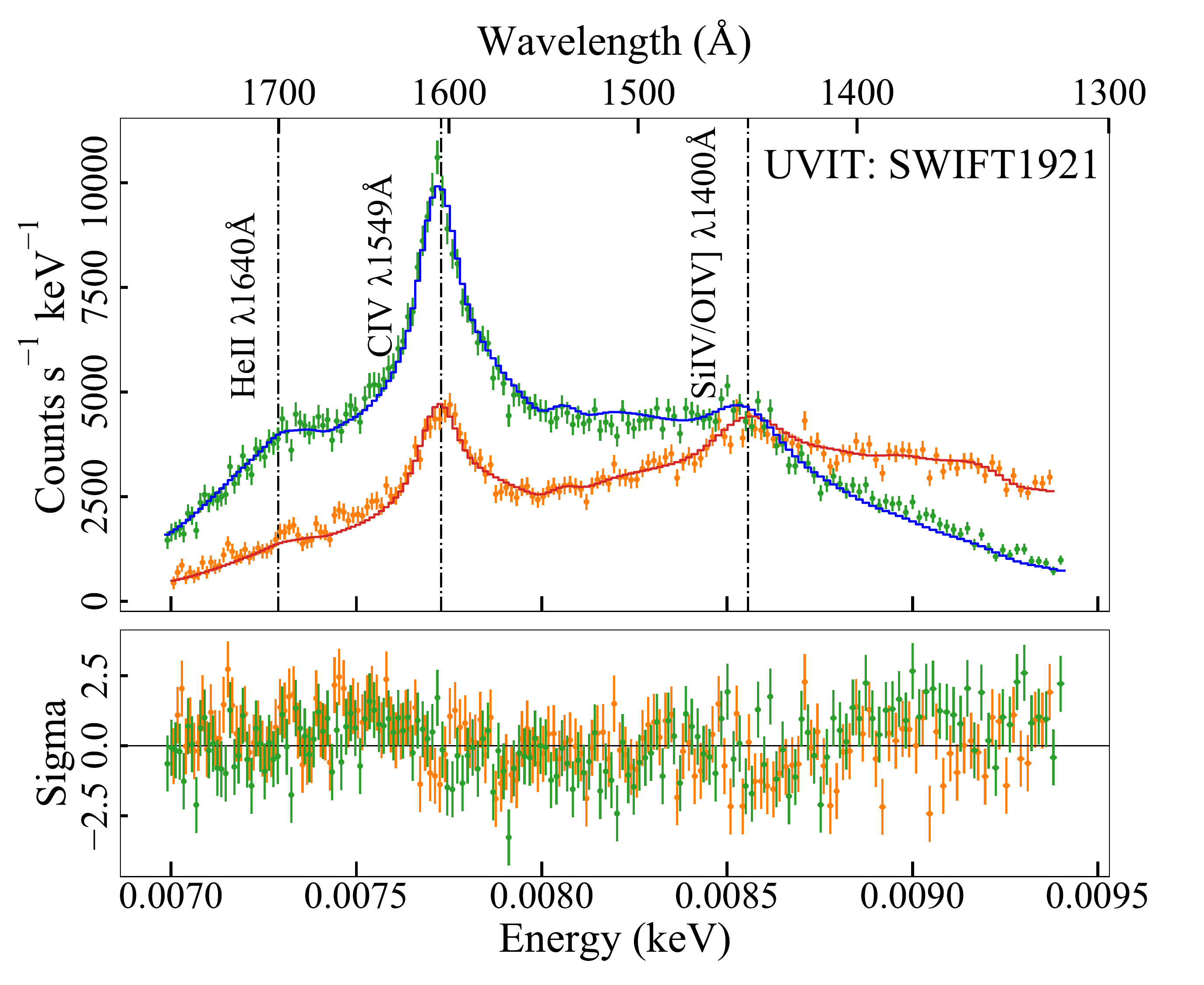}{0.47\textwidth}{(d)}}

\gridline{\fig{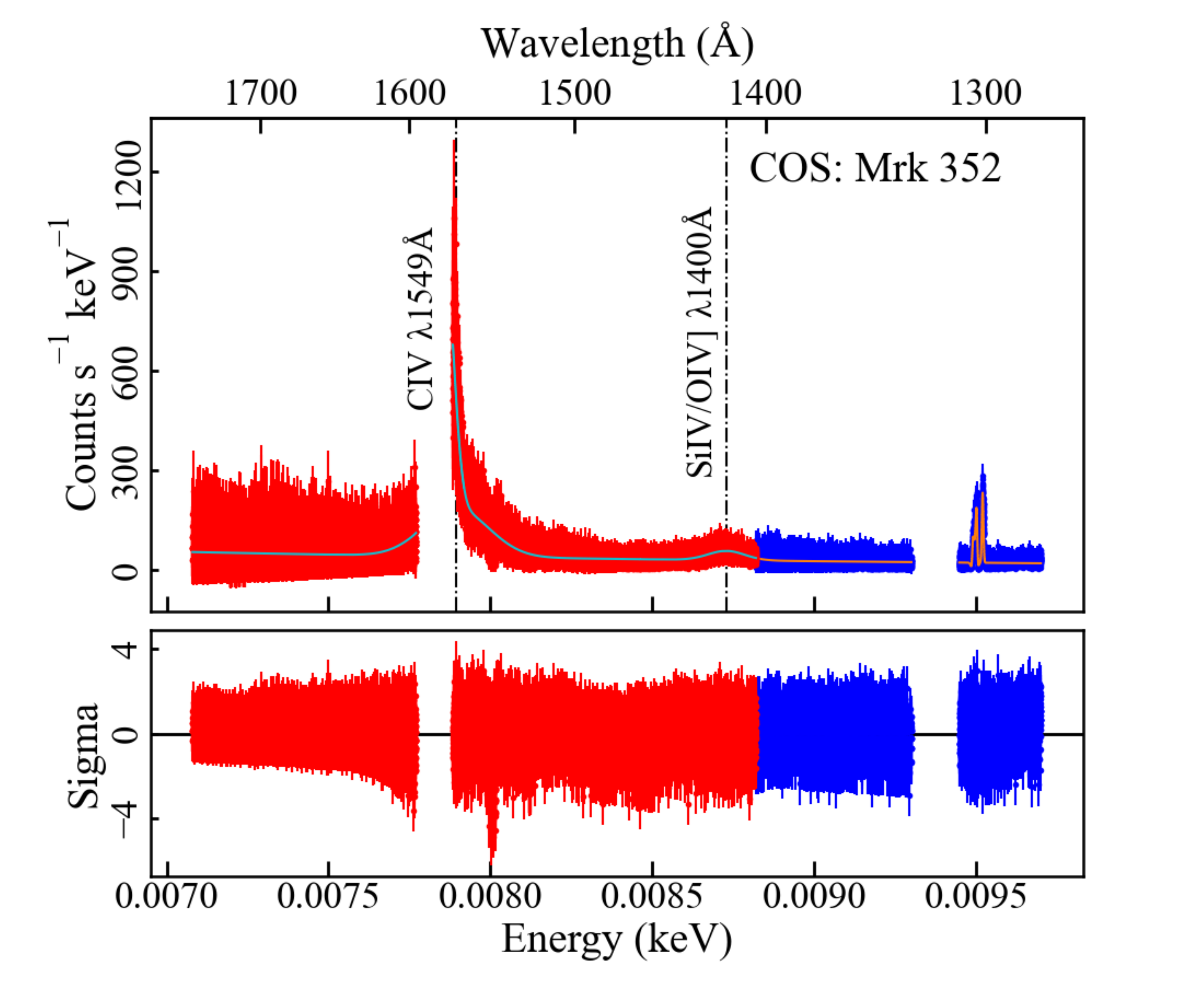}{0.47\textwidth}{(e)} 
           \fig{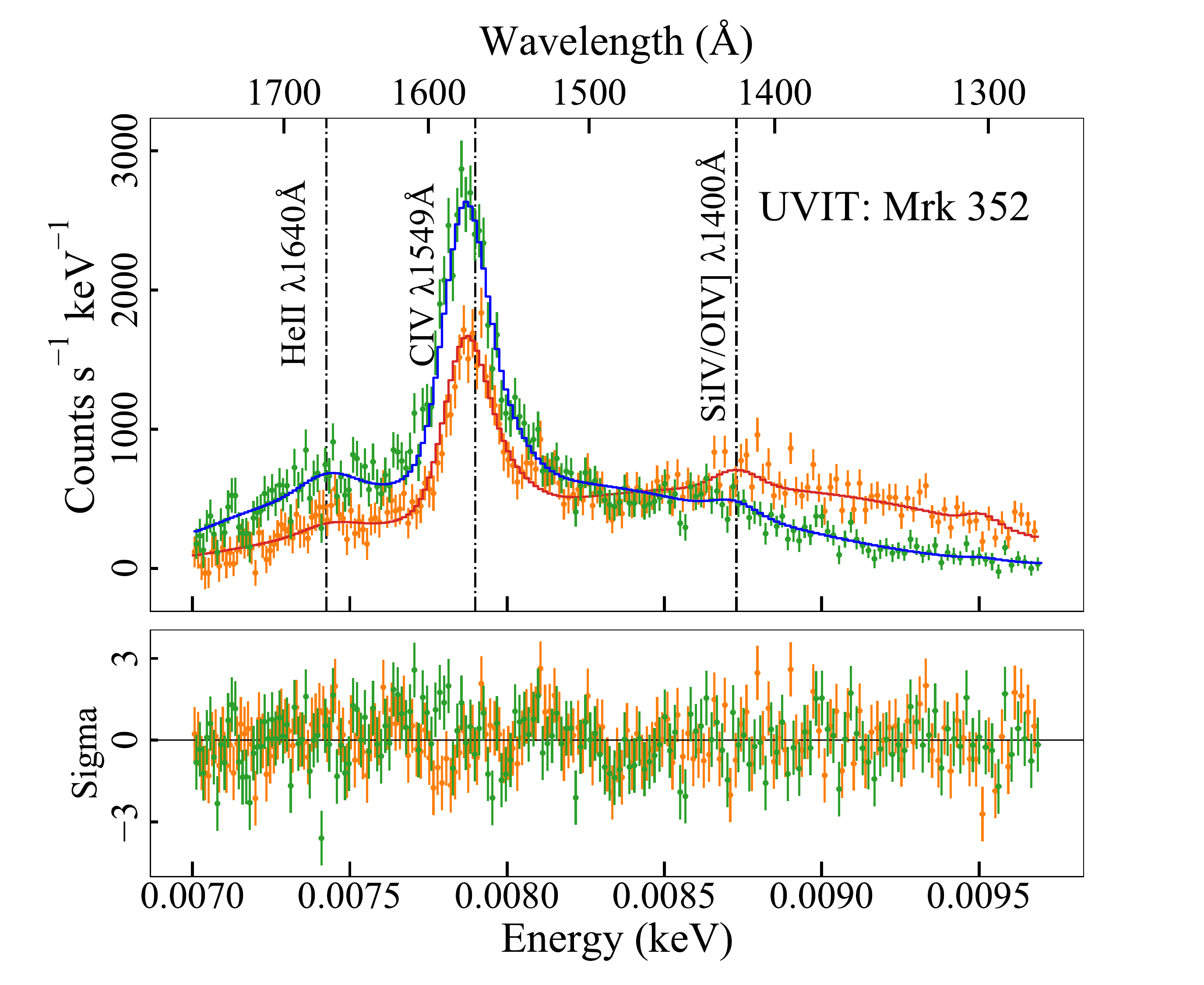}{0.47\textwidth}{(f)}}
            \caption{Same as Fig.~\ref{fig:841spec} but for I~Zw~1 (a,b),  SWIFT1921 (c,d) and Mrk~352 (e,f). \textbf{Left panels}: \hst{}/COS  (red: 160M; blue and cyan: 130M). \textbf{Right panels}:   UVIT/FUV (orange: FUV-G1; green: FUV-G2).}
            \label{fig:zw1_1921_352}
\end{figure*}
%\fi

\subsubsection{PG~0804+761} \label{subsec:pg08}
We fitted both FUV and NUV spectral data jointly. We used the same set of parameters derived from the COS data for the FUV grating and varied the power-law components, the $f_{sc}$   and line centroids of emission lines. As there are no \hst{} spectra available in the NUV band ($2200-2700$\angs), we added  emission lines in the NUV band as required by the data and using \citet{berk2001composite} for line identification.
First, we added \ion{Si}{3}]$\lambda1892$\angs~ which resulted in $\Delta \chi^{2}$ of $-1535$ for 3 parameters.
We  allowed the line centroid to vary within the range provided by \citet{berk2001composite} and fixed it to the best-fit value. The centroid (rest wavelength), FWHM and the normalization of \ion{Si}{3}] emission line  are 1896.5\angs{}, $10965_{-585}^{+613}~\rm km~s^{-1}$ and $0.27_{-0.01}^{+0.01}~\rm photons~cm^{-2}~s^{-1}$, respectively. Though the COS data did not show Fe~II emission,  we tested for the presence of \ion{Fe}{2} complex in  the entire FUV to NUV band. The addition of this component improved the statistic by $\Delta \chi^2=-49$ for one parameter. We fixed the sigma at $10^{-5}\ev$ due to the poor spectral resolution of UVIT. We further added an emission line for \ion{Si}{2}/[\ion{Ne}{3}]~$\lambda1817;1815 $\angs~ at 1818\angs{} that resulted in the $\Delta \chi^2$ of $-30$. The FWHM and the line normalization are $7961_{-1852}^{+2138}~\rm km~s^{-1}$  and $0.06_{-0.01}^{+0.01}~\rm photons~cm^{-2}~s^{-1}$, respectively. Addition of another emission line for [\ion{Ne}{4}]~$\lambda2423$\angs~ at $2438$\angs~ further improved the fit ($\Delta \chi^2$ of $-48$) for three additional parameters. The FWHM and the line normalization are $9115_{-1781}^{+1986}~\rm km~s^{-1}$ and $0.039_{-0.006}^{+0.006}~\rm photons~cm^{-2}~s^{-1}$, respectively. Finally, to account for an excess near $2086$\angs, the addition of an emission line due to \ion{Fe}{3}~$\lambda2076$\angs~ improved the statistic by $\Delta\chi^2=-33$ for three additional parameters. The FWHM and the line normalization are $3751_{-3733}^{+2867}~\rm km~s^{-1}$ and $0.018_{-0.003}^{+0.003}~\rm photons~cm^{-2}~s^{-1}$, respectively. Some of these lines may have broadened due to the   emission from \ion{Fe}{3} around this region. We fixed the widths and line centroids of NUV emission lines at the best-fit values in subsequent analysis.  We did not detect the emission from the Balmer continuum ($2000-4000$\angs) as the fit did not improve after adding this model. 
%We also tested for the presence of the Balmer continuum. We did not see any improvement in the statistic, therefore this emission component could be quite weak for this source. 
The best-fit parameters are shown in Table~ \ref{tab:lines1} and \ref{tab:index}.  We obtained the final $\chi^2/dof = 401/331$ with $\Gamma \sim 1.46$ (see Fig.~\ref{fig:mr22_pg08_7469}(d)). With \texttt{DISKBB} the fit worsened by $\Delta \chi^2= +68$ with the $\chi^2/dof = 469/331$ and inner disk temperature ($\rm kT_{in}$) of $\sim 4.3\ev$.
\cite{jiang2019high} estimated an inclination angle, $i = 65_{-9}^{\circ+14}$ based on  X-ray reflection spectroscopy. The \texttt{ZKERRBB} with $i = 65^\circ$ provided $\chi^2/dof$ = 616/332  ($a^\star = 0.998$) and 558/332 ($a^\star = 0$). We found $\dot{M} \sim 4.7 ~\rm M_\odot ~yr^{-1}$ ($\dot{m} \sim 0.4 $) for $a^\star = 0.998$ and 7.4 $~\rm M_\odot ~yr^{-1}$ ($\dot{m} \sim 0.6 $) for $a^\star = 0$.

%With the inclination angle fixed at $10^\circ$, the statistic improved by $-47$ and the accretion rate reduced to $\sim 1.6$ ($a^\star = 0.998$). 
As we have wider wavelength coverage for this source, we varied $i$ and $\dot{M}$ keeping the normalization fixed at 1. As the temperature estimated by \texttt{DISKBB} is somewhat low  for a maximally spinning black hole (see Fig.~\ref{fig:inDis_temp}), we expect the $f_{col}$ to be close to 1.  With $f_{col}$ fixed at 1, for $a^\star = 0.998$, we obtained $\chi^2/dof$ = 464/331 with $i <10^\circ$ and  $\dot{M}=0.62_{-0.01}^{+0.01}~\rm M_\odot~yr^{-1}$. For $a^\star = 0$, we obtained $\chi^2/dof$ = 455/331 with $i=60_{-5}^{\circ+5}$ and $\dot{M}=2.9_{-0.3}^{+0.4}~\rm M_\odot~yr^{-1}$. Fixing the $f_{col}$ to 2.4 affected the $\chi^2$ much more than for other sources (e.g., Mrk~841, NGC~7469 or Mrk~352) since the NUV emission region would be lying farther away from the SMBH and thereby,  at a much lower temperature than $10^5{\rm~K}$. Therefore, the worsening of fit with the \texttt{ZKERRBB} model as compared to the \texttt{DISKBB} model can be attributed to the high $f_{col}$. The \texttt{OPTXAGNF} model resulted in $\chi^2/dof$ = 432/331 ($a^\star$ = 0.998) and 440/331 ($a^\star$ = 0) and  the Eddington ratios are, $L_{Bol}/L_{Edd} = 0.6$ ($a^\star = $ 0.998) and $0.1$ ($a^\star = $ 0). The $a^\star = $ 0.998 resulted in a better fit for both the continuum models, \texttt{OPTXAGNF} and \texttt{ZKERRBB} which indicates the SMBH is likely to be associated with high spin \citep{piotrovich2020determination,jiang2019high}.

%\noindent
%\underline{\textbf{\emph{NGC~7469}}} : \\
\subsubsection{NGC~7469} \label{subsec:n7469}
  %{\color{red} We also tested for the stellar emission from the bulge using redial surface brightness profile.} 
  
  %Radial profile analysis showed $\sim 1\%$ galactic contribution to the nucleus within $\sim 10^{\prime\prime}~(25 \rm pixel)$ radius. This is verified in \citet{mehdipour2018multi} where, in the Far UV region i.e., above 0.007 keV has no contribution from the galactic bulge. 

  We varied the amplitude of the starburst template model as the UVIT spectral extraction region  will have  a contribution from both the starburst rings (see Fig.~\ref{fig:7469}) while the \hst{}/FOS aperture ($0.9^{\prime\prime}$ diameter) includes only the inner ring. Varying the amplitude of the starburst template model resulted in an improved fit with $\Delta\chi^2$ of $-8$ with a  flux contribution of $7.3 \times 10^{-12} \rm erg ~cm^2 ~s^{-1}$ (fractional contribution of $~0.2$ relative to the continuum) in the $1270-1770$\angs~ band. The best-fit model provided $\chi^2/dof  = 381/347$ with  $\Gamma \sim 1.64$ (see Fig.~\ref{fig:mr22_pg08_7469}(f)). The emission line scaling factor is $\sim 1.07$. The best-fit model with the \texttt{DISKBB} as the continuum component resulted in  $\chi^2/dof = 384/347$ with an inner disk temperature of $\rm kT_{in}\sim 4.4\ev$. 
 % {\color{blue} \textbf{The disk temperature has been observed to be two times lower ($\sim 1.4-2$~eV) in the broadband UV - X-ray SED modeling \citep{petrucci2004physical,mehdipour2018multi}.
  Using broadband UV/X-ray spectral data acquired with the IUE/RXTE in 1996, \citet{petrucci2004physical} obtained the disk temperature, $\rm kT_{in} \sim 2\ev$. They varied the Galactic extinction and obtained the $E_{B-V} \sim 0.12$ using the extinction curve of \citet{seaton1979interstellar}. \citet{mehdipour2018multi} found a disk temperature of $\sim 1.4\ev$ in their broadband SED modelling using \hst{}/COS, \swift{}/UVOT, and \chandra{} observations in 2015. They used the line-free windows of COS as in their broadband modelling. Both \citet{petrucci2004physical} and \citet{mehdipour2018multi} used  Comptonised accretion disk models where they approximated the disk emission as either the Wien tail or single temperature simple blackbody  instead of multi-temperature disk blackbody. The other emission components such as the soft X-ray excess and primary X-ray continuum being produced as the Comptonisation of the blackbody emission  may also be affecting the determination of the temperature. These differences in the accretion disk models and/or Galactic extinction, in addition to the higher flux ($\sim 60\%$) in the UVIT data as compared to the FOS data, may have resulted in different disk temperatures.

   Using different methods e.g., Fe K$_\alpha$ line, BLR orientation, the inclination angle has been estimated to be $\sim 11^\circ$, 24$^\circ$ to $45^\circ-50^\circ$ \citep{patrick2011iron,nguyen2021black}. The \texttt{ZKERRBB} model with  inclination angle fixed at $20^\circ$ resulted in  $\chi^2/dof = 428/348$ ($a^\star = 0.998$) and $424/348$ ($a^\star = 0$). The mass accretion rates are  $\sim 0.09{\rm~M_{\odot}~yr^{-1}}$ ($\dot{m} \sim 0.4$) and $\sim 0.1 {\rm~M_{\odot}~yr^{-1}}$ ($\dot{m} \sim 0.45$) for $a^\star = 0.998$ and $0$, respectively. The \texttt{OPTXAGNF} model resulted in $\chi^2/dof = 383/347$ and $L_{Bol}/L_{Edd} = 0.15$ for both the spin cases. This source also indicates truncation of disk at a radius ($\sim 90r_g$) larger than ISCO. 

\subsubsection{I~Zw~1} \label{subsec:zw1}
We obtained  $\chi^2/dof = 141/173$ for the best-fitting power-law continuum with $\Gamma \sim 1.65$, and the emission lines scaling factor of $\sim 0.82$ (see Fig.~\ref{fig:zw1_1921_352}(b)). The \texttt{DISKBB} model provided the  $\chi^2/dof = 148/173$  with $\rm kT_{in} \sim 4.2\ev$.   The \texttt{ZKERRBB} model with an inclination angle of 42$^\circ$\citep{porquet2004xmm} resulted in  $\chi^2/dof = 145/174$ for both the spin cases. We obtained  high relative accretion rates, $\dot{m} \sim 6.6$ for $a^\star = 0.998$ and  8.5 for $a^\star = 0$. We obtained, $L_{Bol}/L_{Edd} = 0.9$ ($a^\star = 0.998$) and $1.06$ ($a^\star = 0$), in the case of \texttt{OPTXAGNF} model.
%Similar $\chi^2$ between the best-fit models with different continuum model component in I~Zw~1 and SWIFT1835 could be due to low count rate for these two sources.  \\

 \begin{figure}
    \epsscale{1.2}
\plotone{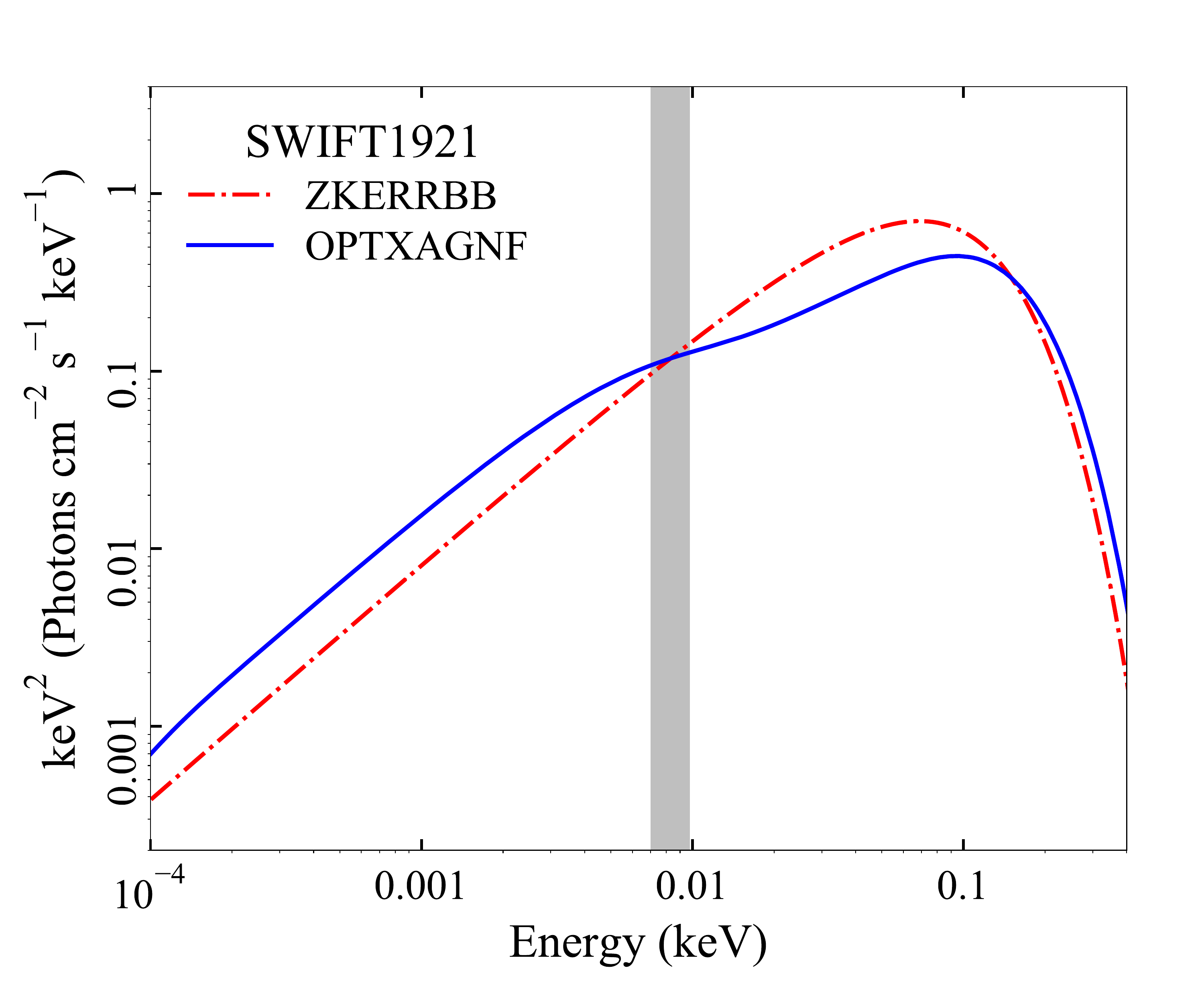}
   \caption{ 
   %Best-fit model for \texttt{ZKERRBB} ($i=30^\circ$ and $a^\star = 0$) and \texttt{OPTXAGNF} ($a^\star = 0$). The gray shaded region corresponds to UVIT/FUV energy band. The two model differing in shape particularly around this band has lead to different $\chi^2$ values.
   The best-fit \texttt{ZKERRBB} ($i=30^\circ$ and $a^\star = 0$) and \texttt{OPTXAGNF} ($a^\star = 0$) models for SWIFT1921 based on the UVIT/FUV grating spectra. The gray-shaded region marks the FUV energy band. The two models differ in shape particularly around this band resulting in different values of the $\chi^2$ fit-statistic.
   }
   \label{fig:s1921_zkopt} 
\end{figure}

%\noindent
%\underline{\textbf{\emph{SWIFT~J1921--5842}}} :\\
\subsubsection{SWIFT~J1921.1--5842} \label{subsec:s1921}
We obtained $\chi^2/dof = 377/321$ for a simple power-law model as the continuum with  $\Gamma \sim 0.7$  and $f_{sc} \sim 0.75$ (Fig.~\ref{fig:zw1_1921_352}(d)). The addition of an intrinsic reddening model did not improve the fit, and, we did not include this component. 
The \texttt{DISKBB} model resulted in $\chi^2/dof = 378/321$ with best-fit $\rm kT_{in} > 23\ev$, the upper limit could not be constrained as the BBB peak lies far away from our FUV band and in the unobservable part of the SED.

With the \texttt{ZKERRBB} as continuum component, we obtained $\chi^2/dof = 378/322$  for $a^\star = 0.998$  and  $\chi^2/dof = 380/322$  for $a^\star = 0$ with inclination angle set to  30$^\circ $ estimated previously by \cite{gondoin2003xmm}. The normalized mass accretion rate, $\dot{m} \sim 1.3$ for $a^\star = 0.998$ and $\sim 1.8$ for $a^\star = 0$. We obtained $\chi^2/dof = 475/321$  for $a^\star = 0.998$  with \texttt{OPTXAGNF}  and $\chi^2/dof= 492/321$  for $a^\star = 0$ and the Eddington ratios are, $L_{Bol}/L_{Edd} = 5.9$ ($a^\star = 0.998$) and $1.1$ ($a^\star = 0$). 

%{\color{blue}
We obtained a large difference in $\chi^2$ between the \texttt{OPTXAGNF} and \texttt{ZKERRBB} models which is largely due to the variation in color-correction factor, $f_{col}$.  
%old --This is a similar situation as  noticed in the case of Mrk~841 (see Fig.~\ref{fig:mrk841_zkopt}). 
As the effective disk temperature rises with the decrease in radius and reaches to $\sim 10^5{\rm~K}$ ($8.6\ev$) at a certain radius, the FUV continuum slope changes sharply. This change occurring in the FUV band is shown in Fig.~\ref{fig:s1921_zkopt}. It occurs due to the variation in $f_{col}$ as a function of local effective temperature as given by Eq.~\ref{eq:f_col_eq} (see also Figure~1 in \citealt{Zdziarski_2022}). While in the \texttt{ZKERRBB} model, the slope remains almost constant up to $\sim 20\ev$ due to the constant $f_{col}$. As mentioned earlier, with the \texttt{DISKBB} model we obtained the peak disk temperature $\rm (kT_{in}) >23\ev$  ($2.6\times 10^5{\rm~K})$,  which is higher than the transition temperature, $10^5{\rm~K}$ (see Eq.~\ref{eq:f_col_eq}). Thus, the flattening of spectral slope due to the change in the $f_{col}$ % occurs in the UVIT band. This essentially 
alters the shape of the continuum in the UVIT band, leading to different values of $\chi^2$ for the \texttt{ZKERRBB} and \texttt{OPTXAGNF} models.

  %\iffalse
\begin{figure}
    \epsscale{1.3}
\plotone{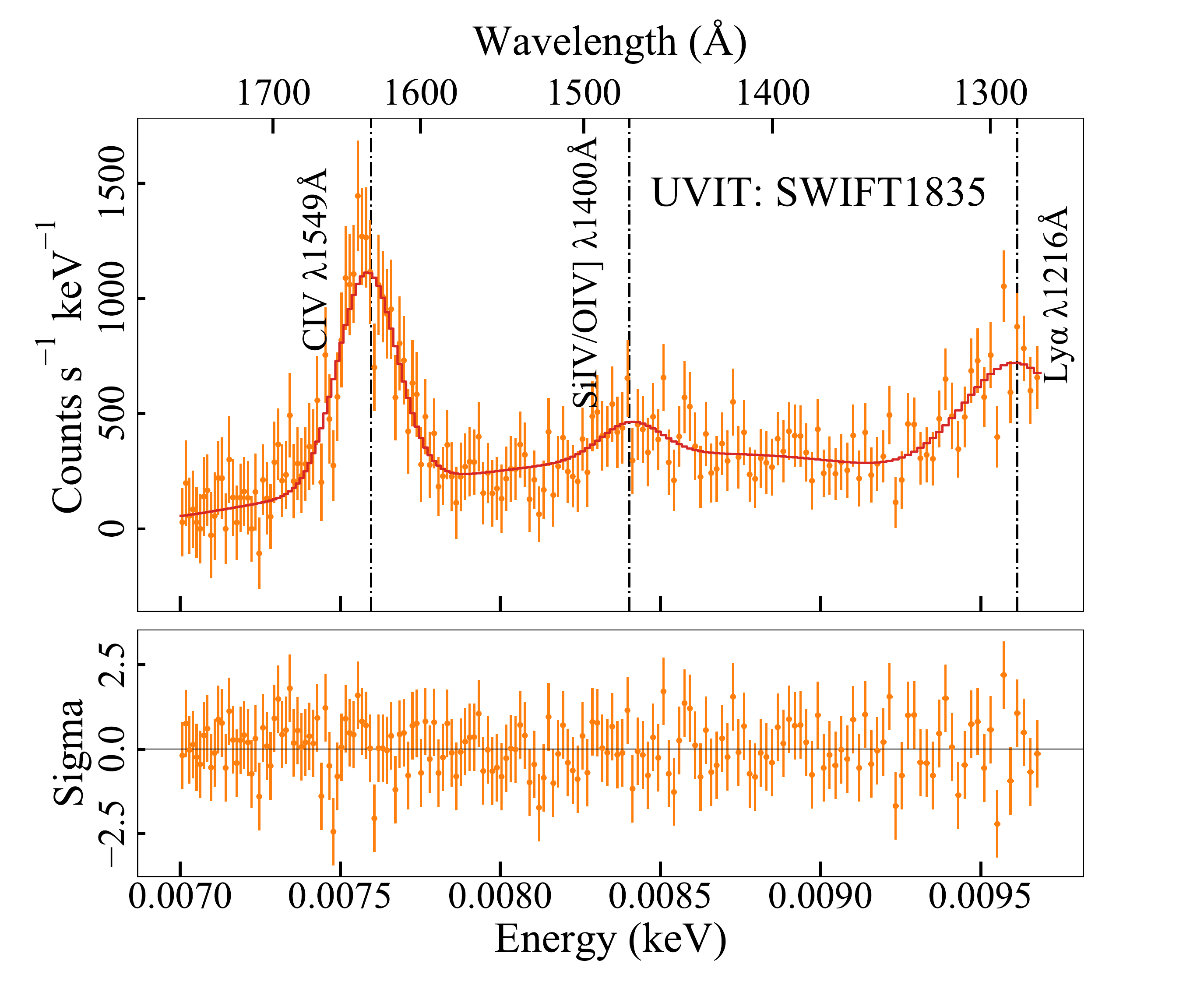}
   \caption{Same as Fig.~\ref{fig:841spec}(b) but for SWIFT1835 (UVIT/FUV-G1).}
    \label{fig:1835spec}
\end{figure}
%\fi

%\noindent
%\underline{\textbf{\emph{Mrk~352 }}} :\\
\subsubsection{Mrk~352} \label{subsec:m352}
For the simple power-law continuum, we obtained the final $\chi^2/dof = 349/346$ after varying the line widths and centroids of \ion{C}{4} line components in comparison to that obtained for the \textit{HST} data (see Table~\ref{tab:lines2} and Fig.~\ref{fig:zw1_1921_352}(f)). After inspecting the residuals, we added a Gaussian line to account for the \ion{He}{2}~$\lambda 1640$\angs~ emission line. This resulted in $\chi^2/dof = 335/345$ with $\Gamma \sim 1.79$.   The \texttt{DISKBB} model resulted in $\chi^2/dof$ = 335/345 with $\rm kT_{in} \sim 3.9\ev$. With \texttt{ZKERRBB} model, for $i=10^\circ$ (chosen arbitrarily) resulted in $\chi^2/dof$ = 355/346   ($a^\star = 0.998$) and 353/346 ($a^\star = 0$) and  $\dot{m} \sim 0.14$ ($a^\star = 0.998$) and 0.21 ($a^\star = 0$).   We obtained $\chi^2/dof$ = 335/345 with \texttt{OPTXAGNF} for both the spin values and the Eddington ratios are, $L_{Bol}/L_{Edd} = 0.04$ ($a^\star = $ 0.998) and $0.004$ ($a^\star = $ 0). \\

%\noindent
%\underline{\emph{\textbf{SWIFT~J1835+3240}}} : \\
 \subsubsection{SWIFT~J1835.0+3240} \label{subsec:1835}
 \textit{HST} data are not available for this source.  We fitted  three emission lines and a redshifted power-law continuum. This provided a satisfactory fit (Fig.~\ref{fig:1835spec}) with $\chi^2/dof$ = 111/169 and  $\Gamma \sim 2.1$.  Using the intrinsic reddening model with the Balmer decrement fixed at 4.43 \citep{popovic2003balmer} did not improve the fit, therefore we did not use this component. The \texttt{DISKBB} model resulted  in a similar $\chi^2$ as the simple  power-law model. We obtained the best-fit inner disk temperature of $\sim 3.7\ev$. \texttt{ZKERRBB} resulted in $\chi^2/dof$ = 112/170 with $i = 30^\circ$ \citep{sambruna2011suzaku}.  We obtained the Eddington ratios, $L_{Bol}/L_{Edd} = 0.003$ ($a^\star = $ 0.998) and $0.006$ ($a^\star = $ 0) with the \texttt{OPTXAGNF} model. 
 %These ratios are the lowest among all the sources.

%\movetabledown=7.5cm
%\begin{rotatetable*}
\begin{deluxetable*}{llcccccccc}
\small
\tablenum{4}
\tablecaption{The best-fit emission line parameters derived from the \hst{} and UVIT spectra of the AGN. The symbols used are as follows. $f_{sc}$: multiplicative constants used for the strengths of  emission lines in the UVIT grating spectra relative to those in the \hst{} spectra, $\lambda_{o}$: line centroid in \angs{} in the source rest frame, $ v_{FWHM}$: FWHM of lines in ${\rm km~s^{-1}}$, $ f_{line}$: line flux in units of $\rm photons~cm^{-2}~s^{-1}$. \label{tab:lines1}
%are the multiplicative constant used to the flux of emission lines of UVIT gratings spectra to preserve the line ratio. 
%$\lambda_{o}$, the observed line centroid in the source rest frame, is in \angs. $v$, the FWHM of emission lines, is in $\rm km~s^{-1}$.  The flux is in the unit of $\rm photons~cm^{-2}~s^{-1}$. \label{tab:lines1} 
}
\tablewidth{700pt}
\tablehead{
\multirow{2}{*}{Objects}& &\multicolumn2c{ Mrk~841} &  \multicolumn2c{MR~2251--178}  & \multicolumn2c{PG0804} & \multicolumn2c{NGC~7469}    \\
& & \colhead{HST} &  \colhead{UVIT} & \colhead{HST} &  \colhead{UVIT} & \colhead{HST} &  \colhead{UVIT} & \colhead{HST} &  \colhead{UVIT}
}
\startdata
$f_{sc}$ && 1 & $0.69_{-0.02}^{+0.02}$ &1 &  $1.32_{-0.02}^{+0.02}$& 1& $0.99_{-0.03}^{+0.03}$ & 1 & $1.07_{-0.02}^{+0.02}$ \\
&&&&&&&&&\\
Ly$\alpha~ \lambda 1215.7$\angs & $\lambda_{o}$(\AA)  & 1214.9  & * &1214.8 & 1220 &1214.8& 1222 &1218.1  & * \\
 & $v_{FWHM}$ &  5383   & * & 11682 & * & 10681& *   &3367 & *  \\
 & $f_{line}$ & $0.196_{-0.001}^{+0.001}$ & * & $0.40_{-0.01}^{+0.01}$ & *& $0.538_{-0.004}^{+0.004}$&*  & $0.215_{-0.004}^{+0.004}$ & *   \\
  & $\lambda_{o}$(\AA)  & 1216.3  & * & 1215.1 & 1210& 1215.2 & *  & 1216.3 & * \\
  & $v_{FWHM}$&716   & * & 3779 &  * & 2542& * & 901 & *   \\
  & $f_{line}$&$0.0420_{-0.0003}^{+0.0003}$ & * & $0.32_{-0.01}^{+0.01}$& * & $0.306_{-0.002}^{+0.002}$& * & $0.057_{-0.003}^{+0.003}$& *  \\
&&&&&&&&&\\
\ion{N}{5} $\lambda 1241$\angs &$\lambda_{o}$(\AA)  &  1249 & *  & x  & $1242$ & 1240.6 &1245& 1241.2 & * \\
& $v_{FWHM}$& 6652  & * &x & 3864 &3308 &* & 7120& * \\
&$f_{line}$&$0.021_{-0.001}^{+0.001}$& * & x & $0.04_{-0.01}^{+0.01}$ & $0.100_{-0.001}^{+0.001}$& * & $0.131_{-0.004}^{+0.004}$& *  \\
&&&&&&&&&\\
\ion{Si}{2} $\lambda 1262.6$\angs
 &$\lambda_{o}$(\AA)  
 & 1260  & *  &1264.5  & * &1260.3 & *& x &x   \\
& $ v_{FWHM}$ & 440&* &1994&*& 4197 &* & x& x \\
& $ f_{line}$&$0.0024_{-0.0001}^{+0.0001}$& * & $0.005_{-0.001}^{+0.001}$& * & $0.038_{-0.001}^{+0.001}$& * & x & x  \\
&&&&&&&&&\\
\ion{O}{1}/\ion{Si}{2} $\lambda 1305$\angs & $\lambda_{o}$(\AA)  &  1305 & *   & 1304   & * & 1309&* & 1304.6 & *   \\
& $v_{FWHM}$& 2435 & * & 3321 & * & 1435 & *   & 3876 & *   \\
&$ f_{line}$ & $0.0032_{-0.0002}^{+0.0002}$ & * & $0.014_{-0.001}^{+0.001}$& *& $0.0098_{-0.0003}^{+0.0003}$& * & $0.017_{-0.002}^{+0.002}$ & *  \\
&&&&&&&&&\\
\ion{C}{2} $\lambda 1335$\angs &$\lambda_{o}$(\AA)  
 &  x & x  & x & x& x&x & 1337 & *  \\
& $ v_{FWHM}$ & x  &x   &x  &  x & x&x& 3373 & * \\
&$ f_{line}$ & x & x & x & x & x&x & $0.02_{-0.001}^{+0.001}$ & *  \\
&&&&&&&&&\\
\ion{Si}{4} $\lambda1400$\angs & $\lambda_{o}$(\AA)  &1398.8 & * & 1400  & * & 1399& 1403 & 1400.1  & *  \\
& $v_{FWHM}$ & 6788  & * & 4758 & * & 5060& * & 5724 & *  \\
&$f_{line}$ & $0.0368_{-0.0004}^{+0.0004}$ & * & $0.037_{-0.001}^{+0.001}$ & * & $0.104_{-0.001}^{+0.001}$& * & $0.092_{-0.002}^{+0.002}$ &  *\\
&&&&&&&&&\\
\ion{C}{4} $\lambda1549$\angs &$\lambda_{o}$(\AA)  &1551.5  &1561 & 1548  &1559.6 & 1548&* & 1549.2 & *  \\
& $v_{FWHM}$&10229  & * & 15925 & * &6648& 10569  & 7180 & *  \\
&$f_{line}$&$0.204_{-0.002}^{+0.002}$& *& $0.300_{-0.002}^{+0.002}$& * & $0.223_{-0.004}^{+0.004}$& * & $0.323_{-0.004}^{+0.004}$& * \\
&$\lambda_{o}$(\AA)  & 1542 & 1548 & 1548 & 1550 &1548&* & 1550.0& * \\
& $v_{FWHM}$& 3213 & 5282 & 2585& * &2202&* &1934 & * \\
&$f_{line}$&$0.267_{-0.003}^{+0.003}$&*& $0.154_{-0.002}^{+0.002}$& *& $0.092_{-0.004}^{+0.004}$ & * & $0.125_{-0.004}^{+0.004}$& *  \\
&&&&&&&&&\\
\ion{He}{2} $\lambda1640$\angs &$\lambda_{o}$(\AA)  & 1640  & * & 1613.5 &* &x&x & 1630.1& * \\
&$ v_{FWHM}$&  11332 & * & 3644 & * &x&x &13804 &* \\
&$ f_{line}$&$0.058_{-0.002}^{+0.002}$ &*& $0.014_{-0.001}^{+0.001}$ & * &x&x& $0.093_{-0.005}^{+0.005}$& * \\
&$\lambda_{o}$(\AA)  & 1641  & * & x & x &x&x & 1639.7 & * \\
&$ v_{FWHM}$& 250  & * &  x&  x&x&x  & 725 & *   \\
&$ f_{line}$&$0.0032_{-0.0002}^{+0.0002}$& * & x & x &x&x & $0.007_{-0.001}^{+0.001}$& *\\
&&&&&&&&&\\
\ion{O}{3}]/\ion{Al}{2}$\lambda 1664.7$\angs &$\lambda_{o}$(\AA)  
 &  1665 & *  & 1651 & *& x&x & x & x  \\
& $ v_{FWHM}$ & 1612  &*   & 4229 &  * & x&x& x & x \\
&$ f_{line}$ & $0.0036_{-0.0004}^{+0.0004}$ & * & $0.18_{-0.05}^{+0.05}$ & * & x&x & x & x  \\
\enddata
\tablecomments{~ ~`*' $\equiv$ same as HST; `x'  $\equiv$ not detected.}
\end{deluxetable*}
%\end{rotatetable*}

%next half of the table
\begin{deluxetable*}{llccccccccc}
\tablenum{5}
\tablecaption{Same as Table~\ref{tab:lines1}, but for  I~Zw~1, SWIFT1921, Mrk~352 and SWIFT1835. Since, for these sources, the emission line \ion{O}{3}]/\ion{Al}{2}$\lambda 1664.7$\angs is not detected, this row is omitted from this table.  \label{tab:lines2} 
}
\tablewidth{700pt}
\tablehead{
\multirow{2}{*}{Objects}& &\multicolumn2c{I~Zw~1} & \multicolumn2c{SWIFT1921}&\multicolumn2c{Mrk~352}& \multicolumn1c{SWIFT1835}   \\
& &  \colhead{HST} &  \colhead{UVIT} & \colhead{HST} & \colhead{UVIT} & \colhead{HST} & \colhead{UVIT} &  \colhead{UVIT}
}
\startdata
$f_{sc}$& & 1 & $0.82_{-0.05}^{+0.05}$ &1 & $0.75_{-0.02}^{+0.02}$ &1 & $1.18_{-0.01}^{+0.01}$  &  --    \\
&&&&&&&&&\\
Ly$\alpha~ \lambda 1215.7$\angs & $\lambda_{o}$(\AA) &1212.8 & * &1214.2 & *  &x& x &1213.5  \\
 & $ v_{FWHM}$& 4964 & *&6823 & *& x&x&12193   \\
 &$ f_{line}$ & $1.08_{-0.17}^{+0.12}$& *& $0.591_{-0.002}^{+0.002}$ & *& x& x& $0.27_{-0.02}^{+0.03}$  \\
  & $\lambda_{o}$(\AA) & 1215 &*  & 1215.5 & *  &x & x& x  \\
  & $ v_{FWHM}$ & 1024 & *  & 1820 & *& x & x& x  \\
  & $ f_{line}$ & $0.72_{-0.17}^{+0.08}$& * & $0.193_{-0.001}^{+0.001}$& * &x &x & x \\
&&&&&&&&&\\
\ion{N}{5} $\lambda 1241$\angs &$\lambda_{o}$(\AA) & 1237 &* & 1241.6& * &x&x& x  \\
& $v_{FWHM}$ & 2922&* &3189  & *&  x& x  & x  \\
&$ f_{line}$ & $0.52_{-0.11}^{+0.05}$& * & $0.127_{-0.001}^{+0.001}$ & *&x &x & x  \\
&&&&&&&&&\\
\ion{Si}{2} $\lambda 1262.6$\angs
 &$\lambda_{o}$(\AA)  
& 1262.4 &*& x & x &  x & x &x  \\
& $v_{FWHM}$ & 1461 & *& x & x &x &x & x \\
& $ f_{line}$ & $0.11_{-0.02}^{+0.01}$& *& x & x&x &x & x  \\
&&&&&&&&&\\
\ion{O}{1}/\ion{Si}{2} $\lambda1305$\angs& $\lambda_{o}$(\AA) & 1305 & * &1304& * &  x &x &x  \\
& $v_{FWHM}$ & 905& * &2483& *&x & x& x \\
&$ f_{line}$  & $0.10_{-0.02}^{+0.05}$& *&  $0.0110_{-0.0004}^{+0.0004}$& * &x &x & x    \\
&&&&&&&&&\\
\ion{C}{2} $\lambda 1335$\angs &$\lambda_{o}$(\AA) & 1334 & *& x & x & x&x &x \\
& $v_{FWHM}$ & 1038 & * & x & x &  x&x &x  \\
&$ f_{line}$ &$0.039_{-0.008}^{+0.003}$&  *  & x & x & x & x &  x \\
&&&&&&&&&\\
\ion{Si}{4} $\lambda1400$\angs & $\lambda_{o}$(\AA) & 1397 & *  & 1397.3 & * &1399.4&* &1394.6   \\
& $ v_{FWHM}$ & 3332 &*  & 5969 & * & 4592& * & 6757  \\
&$ f_{line}$ & $0.28_{-0.09}^{+0.02}$& * & $0.121_{-0.001}^{+0.001}$&*& $0.0073_{-0.0003}^{+0.0003}$& * & $0.02_{-0.01}^{+0.01}$ \\
&&&&&&&&&\\
\ion{C}{4} $\lambda1549$\angs &$\lambda_{o}$(\AA) & 1539 & * &1547.0 &* & 1547 & 1543 &1542.7   \\
& $ v_{FWHM}$ & 3477& *  &10148 & * & 9044 &* &7474   \\
&$f_{line}$ & $0.22_{-0.04}^{+0.02}$& *  & $0.50_{-0.01}^{+0.01}$ & *& $0.064_{-0.001}^{+0.001}$ & *& $0.26_{-0.02}^{+0.02}$  \\
&$\lambda_{o}$(\AA) & 1547&* & 1549 &* &1553.7 &*&x  \\
& $v_{FWHM}$ & 1500&* &2381 &* &2402& 4821 & x \\
&$\rm f_{line}$ & $0.17_{-0.03}^{+0.02}$& * & $0.198_{-0.003}^{+0.003}$& *   &$0.08_{-0.02}^{+0.02}$ & * & x  \\
&&&&&&&&&\\
\ion{He}{2} $\lambda1640$\angs &$\lambda_{o}$(\AA) & x&x &1640 & *  &x & $1644.6$ &x   \\
&$ v_{FWHM}$ & x&x & 16131 & * & x&7609& x \\
&$ f_{line}$ & x&x & $0.29_{-0.01}^{+0.01}$& * &x &$0.018_{-0.005}^{+0.005}$ & x  \\
&$\lambda_{o}$(\AA) & x&x & 1639& x  &x &x & x   \\
&$ v_{FWHM}$ & x&x & 981& x &x & x& x \\
&$f_{line}$ & x&x & $0.008_{-0.001}^{+0.001}$ & x & x& x& x  \\
\enddata
\tablecomments{~ ~`*' $\equiv$ same as HST, `x'  $\equiv$ not detected.}
\end{deluxetable*}

\begin{deluxetable*}{llcccccccccc}
\tablenum{6}
\tablecaption{Same as Table~\ref{tab:lines1}, but for  Mrk~841, MR~2251-178, I~Zw~1 and SWIFT1921 for the additional weak emission lines which are not mentioned in Table \ref{tab:lines1} and \ref{tab:lines2}.  \label{tab:lines3} 
}
\tablewidth{700pt}
\tablehead{
\multirow{2}{*}{Objects}& &\multicolumn2c{Mrk 841} & \multicolumn2c{MR 2251-178}&\multicolumn2c{I Zw 1}& \multicolumn2c{SWIFT1921}   \\
& &  \colhead{HST} &  \colhead{UVIT} & \colhead{HST} & \colhead{UVIT} & \colhead{HST} & \colhead{UVIT} & \colhead{HST}&  \colhead{UVIT}
}
\startdata
Ly$\alpha~ \lambda 1215.7$\angs & $\lambda_{o}$(\AA) &x & x &1217 & *   &x&x&x& x  \\
 & $ v_{FWHM}$& x& x&1856 & *&x&x  &x& x \\
 &$ f_{line}$ & x& x& $0.17_{-0.01}^{+0.01}$ & *& x& x &x& x \\
&&&&&&&&&\\
\ion{N}{5} $\lambda 1241$\angs &$\lambda_{o}$(\AA) & x &x & x&x & 1252&* &x& x \\
& $v_{FWHM}$ & x&x &x  & x  & 1383&* &x& x \\
&$ f_{line}$ & x& x &x & x& $0.042_{-0.004}^{+0.012}$ &*&x& x \\
&&&&&&&&&\\
\ion{O}{1}/\ion{Si}{2} $\lambda1305$\angs& $\lambda_{o}$(\AA) & x& x &x &x  &1298 &*  &x& x\\
& $v_{FWHM}$ & x& x &x & x & 1784& * &x& x\\
&$ f_{line}$  & x& x &x & x  & $0.015_{-0.007}^{+0.003}$ & *&x& x    \\
&$\lambda_{o}$(\AA) & x& x &x & x  &1310 &* &x& x \\
& $v_{FWHM}$ & x& x &x & x& 541& *&x& x\\
&$ f_{line}$  & x& x &x & x  & $0.010_{-0.001}^{+0.001}$ & *   &x& x \\
&&&&&&&&&\\
\ion{C}{2} $\lambda 1335$\angs &$\lambda_{o}$(\AA) & x& x &x & x  &1344 &* &x& x\\
& $v_{FWHM}$ & x& x &x & x& 611& * &x& x\\
&$ f_{line}$  & x& x &x & x & $0.008_{-0.001}^{+0.001}$ & * &x& x   \\
&&&&&&&&&\\
\ion{N}{4}] $\lambda1486$\angs &$\lambda_{o}$(\AA) & 1485&* &x & x   &x &x &1486& * \\
&$ v_{FWHM}$ & 1704&* &x & x  &x &x &2200& *  \\
&$ f_{line}$ & $0.0022_{-0.0002}^{+0.0002}$&* &x & x   &x &x&$0.0095_{-0.0006}^{+0.0006}$& *   \\
&&&&&&&&&\\
\ion{C}{4} $\lambda1549$\angs &$\lambda_{o}$(\AA) & x & x &1536 &* & x & x &x&x \\
& $ v_{FWHM}$ & x & x &2932 &* & x & x &x&x   \\
&$f_{line}$ & x & x & $0.039_{-0.002}^{+0.002}$ &* & x & x &x&x  \\
\enddata
\tablecomments{~ ~`*' $\equiv$ same as HST, `x'  $\equiv$ not detected.}
\end{deluxetable*}

\begin{deluxetable*}{clcccccccc}
\tablenum{7}
\tablecaption{Best-fit parameters of \texttt{ZPOWERLAW} model (abbreviated as ZPL) fitted to the \hst{} and UVIT/Grating data and the \texttt{DISKBB}, \texttt{ZKERRBB} and the \texttt{OPTXAGNF} models fitted to the  UVIT grating data. }
\label{tab:index}
\tablewidth{650pt}
\tablehead{
\colhead{Model}&\colhead{Parameters} &\colhead{ Mrk 841} & \colhead{MR~2251--178} & \colhead{PG0804} &  \colhead{NGC~7469} & \colhead{I~Zw~1} &\colhead{SWIFT1921}   & \colhead{Mrk~352}& \colhead{SWIFT1835}
}
\startdata
%\multirow{4}{*}{ZPOWERLAW}
\multirow{3}{*}{\begin{tabular}{c} ZPL (HST)
                   \end{tabular}}
&$\alpha$& $-0.17^{+0.04}_{-0.04}$ & $-0.52^{+0.03}_{-0.03}$ & $-0.96^{+0.01}_{-0.01}$& $-1.59^{+0.05}_{-0.05}$&$-1.68^{+0.05}_{-0.55}$ & $0.28^{+0.03}_{-0.03}$ & $ -1.53^{+0.16}_{-0.10}$ & --  \\
 & $F_{UV} $&  $1.2$  & $1.1$ & $5.2$& $2.4$ & $6.0$ & $6.5$ & $0.2$ & --  \\
   & $\chi^{2}/dof$  & $\frac{43506}{42044}$  & $\frac{2582}{1905}$ & $\frac{52319}{38087}$& $\frac{3010}{1891}$ & $\frac{31979}{36087}$ & $\frac{51231}{42302}$  & $\frac{20427}{35833}$ & --   \\ \hline
\multirow{3}{*}{\begin{tabular}{c}ZPL
                   \end{tabular}}
& $\alpha$ & $-0.36^{+0.08}_{-0.08}$  &$-1.1^{+0.2}_{-0.2} ~$ & $-0.45^{+0.05}_{-0.05}$ & $-0.64^{+0.11}_{-0.11} ~$ &  $-0.65^{+0.50}_{-0.50}$ & $0.30^{+0.07}_{-0.07} ~$ & $-0.79^{+0.23}_{-0.22} ~$&  $-1.1^{+1.0}_{-1.0}$\\
 & $F_{UV} $ &  $3.5$   & $2.4$ & $4.8$ & $3.8$  & $3.9$ &$6.1$ & $0.9$ & $0.6$ \\
  & $\chi^{2}/dof$  & 410/320  & 203/166 & 401/331 & 381/347 & 141/173 & 377/321   &335/345 & 111/169  \\
  \hline
\multirow{3}{*}{\begin{tabular}{c}DISKBB
                   \end{tabular}}
&$\rm kT_{in}~(eV)$  &  $5.8_{-0.4}^{+0.5}$ & $3.6^{+0.4}_{-0.3}$ &  $4.3^{+2.8}_{-1.6}$& $4.4^{+0.4}_{-0.3}$ & $4.2_{-1.0}^{+2.4}$ & $>23$ & $3.9^{+0.6}_{-0.5}$ & $3.7^{+12.2}_{-1.3}$\\
 & Norm$\times 10^{10}$&  $1.1_{-0.3}^{+0.3}$  & $4.7_{-1.6}^{+2}$ & $ 4.2_{-0.2}^{+0.2}$& $2.8_{-0.8}^{+0.9}$ & $4.1_{-3.2}^{+8.9}$ & $0.006_{-0.006}^{+0.03}$  & $1.0_{-0.5}^{+0.7}$ &  $1.0_{-1.0}^{+6.5}$ \\
 &$\chi^{2}/dof$  & 414/320 &205/166& 469/331  & 384/347& 148/173 & 378/321  & 335/345 &111/169  \\
 \hline
%\multirow{4}{*}{ZKERRBB\\a= 0.998} 
\multirow{4}{*}{\begin{tabular}{c}ZKERRBB\\ $a^\star = 0.998$
                \end{tabular}}
 & $i$ & $20^\circ(f)$  & $25^\circ(f)$& $65^\circ(f)$ & $20^\circ(f)$  & $42^\circ(f)$ & $30^\circ(f)$  & $10^\circ(f)$ &$30^\circ(f)$  \\
 &$\dot{M}$&$0.379_{-0.004}^{+0.004}$ & $0.263_{-0.003}^{+0.004}$ & $4.7_{-0.1}^{+0.1}$ & $0.09_{-0.01}^{+0.01}$ & $4.6_{-0.2}^{+0.2}$ & $1.22_{-0.01}^{+0.01}$   & $0.021_{-0.001}^{+0.001}$ & $0.023_{-0.001}^{+0.001}$  \\
 &Norm& $1$ &$1$  & $1$ & $1$  & $1$ & $1$ & $1$ & $1$ \\
 &$\chi^{2}/dof$ & 461/321  & 242/167 & 616/332  & 428/348& 145/174 &378/322 & 355/346& 112/170 \\
 \hline
 %\multirowcell{4}{ZKERRBB\\a=0}
 \multirow{4}{*}{\begin{tabular}{c}ZKERRBB\\$a^\star = 0$
                  \end{tabular}}
  &$i$ &$20^\circ(f)$  & $25^\circ(f)$ & $65^\circ(f)$ & $20^\circ(f)$ & $42^\circ(f)$& $30^\circ(f)$  & $10^\circ(f)$ & $30^\circ(f)$ \\
 &$\dot{M}$&$0.525_{-0.005}^{+0.005}$ &$0.464_{-0.006}^{+0.006}$ & $7.4_{-0.1}^{+0.1}$  & $0.13_{-0.01}^{+0.01}$ &$5.9_{-0.3}^{+0.3}$ & $1.63_{-0.01}^{+0.01}$   & $0.028_{-0.001}^{+0.001}$ & $0.082_{-0.004}^{+0.004}$   \\
 &Norm & $1$ & $1$  & $1$ & $1$   & $1$  & $1$ &$1$ & $1$ \\
&$\chi^{2}/dof$ &447/321  & 226/167 & 558/332 & 424/348 &145/174 &380/322 & 353/346 & 112/170  \\
\hline
%\multirowcell{3}{OPTXAGNF\\a=0.998}
\multirow{3}{*}{\begin{tabular}{c}OPTXAGNF\\$a^\star = 0.998$ 
                  \end{tabular}}
& $\log\big(\frac{L}{L_{Edd}}\big)$  & $0.120_{-0.004}^{+0.007}$ & $-0.7^{+0.1}_{-0.1}$ & $-0.27_{-0.01}^{+0.01}$& $0.48_{-0.06}^{+0.06}$ &$1.24_{-0.05}^{+0.16}$& $0.677_{-0.002}^{+0.005}$ & $-0.07_{-0.07}^{+0.07}$ & $-2.68_{-0.04}^{+0.39}$  \\
&$\rm r_{cor}$ & $<9$ &$16.5_{-4}^{+3.5}$ & $<6$ &$99_{-40}^{+27}$ &$<241$ & $<4$& $101_{-40}^{+33}$ & $<5$ \\
&$\chi^{2}/dof$ &412/320 &205/166 & 432/331 &383/347 &141/173 & 475/321& 335/345 & 111/169  \\
\hline
%\multirowcell{3}{OPTXAGNF\\a=0.0}
\multirow{3}{*}{\begin{tabular}{c}OPTXAGNF\\$a^\star = 0$
                  \end{tabular}}
& $\log\big(\frac{L}{L_{Edd}}\big)$  & $-0.562_{-0.004}^{+0.005}$ &$-1.41_{-0.03}^{+0.06}$& $-0.86_{-0.01}^{+0.01}$  & $-0.24_{-0.06}^{+0.06}$ & $0.52_{-0.04}^{+0.16}$ & $-0.022_{-0.004}^{+0.004}$ &  $-1.79_{-0.07}^{+0.07}$ &$-2.3_{-0.03}^{+0.12}$  \\
&$\rm r_{cor}$ &$<14$ & $<14$ & $<8$ & $85_{-50}^{+28}$ & $<230$& $<10$ & $92_{-43}^{+33}$ & $<13$  \\
&$\chi^{2}/dof$ &416/320 &205/166 & 440/331 & 383/347 & 141/173& 492/321 & 335/345 & 117/169 \\
\enddata
\tablecomments{The FUV flux ($F_{UV}$) is in the units of $10^{-11}~\rm erg~cm^{-2}~s^{-1}$ in the $0.007$ to $0.0097\kev$ ($1280-1770$\angs) band . $\dot{M}$ is in the unit of $\rm M_{\odot} ~yr^{-1}$. }
\end{deluxetable*}
%\end{rotatetable*}

%\newpage
\section{discussion} \label{sec:discuss}

We analyzed \textit{AstroSat}/UVIT grating observations of eight nearby AGN with black hole masses ranging from $10^6 - 10^9 M_{\odot}$, and studied the shape of the intrinsic continuum.
We confirm the presence of strong emission lines due to \ion{C}{4}, \ion{N}{5}, \ion{He}{2}, Ly$\alpha$ and \ion{Si}{4}/\ion{O}{4}]. The equivalent width (EW) of these broad emission lines fall in the range  $12 - 218$\angs~ (Ly$\alpha \lambda 1216$\angs), $2 - 56 $\angs~ (\ion{N}{5}$\lambda 1241$\angs), $4 - 32 $\angs~ (\ion{Si}{4}/\ion{O}{4}])$\lambda 1400$\angs, $5 - 263 $\angs~ (\ion{C}{4}$\lambda1549$\angs) and $0.4 - 26 $\angs~ (\ion{He}{2}$\lambda 1640$\angs). We found emission line variations as large as $\sim 30\%$ compared to the line strengths found in the \hst{} observations.  
%We also observed possible variations in the line profiles. 
 Though the \ion{Fe}{2} line complex is weak below $1800 $\angs, we detected this complex in five sources with fractional contributions of $0.002$ (Mrk~841), $0.008$ (PG0804), $0.006$ (NGC~7469), $0.05$ (I~ZW~1) and $0.01$ (SWIFT1921)  in the $1278-1770$\angs~ band relative to the continuum. 

We accounted for the intrinsic and the Galactic reddening, host galaxy contamination, and emission lines from the BLR and NLR, and derived the intrinsic continua  from the observed UVIT grating spectra.  We measured the spectral slopes, $\alpha \sim -1.1$ to $0.3$, of the intrinsic power-law continua. These slopes are comparable to those obtained from earlier studies. \citet{stevans2014hst} observed the spectral index $\alpha$ to range from $-0.36$ to $-2.76$ using line-free spectral regions for 159 AGN using \textit{HST}/COS ($1200-2000$\angs). \citet{shull2012hst} observed  slope $\alpha$ ranging from  $-0.25$ to $-1.72$ in $1200-2000$\angs~ for their sample of 22 AGN using \textit{HST}/COS. These studies did not account for the contamination of \ion{Fe}{2} emission or the intrinsic reddening. 
%After using these corrections, we observed that the spectral slope $\alpha$ for most sources is redder than that predicted from standard disk model ($\alpha = 0.33$).\\

The UV slopes of observed spectra are strongly affected by intrinsic reddening due to the dust and gas in the host galaxies \citep{koratkar1999ultraviolet}. This is one of the major uncertainty in deriving the accretion disk continuum. We used the empirical relation given by \citet{czerny2004extinction} where we used  Balmer decrement as  a reddening indicator. This relation is thought to describe the intrinsic extinction of AGN appropriately. For many Seyfert 1s, \citet{lu2019reddening} found that the Balmer decrement  is distributed as  a  Gaussian  with a peak value of $\sim$ 3.16.
%They showed that the reddening in the NLR is higher than that in BLR and are not correlated. 
For our sources, the best-fit Balmer decrement is in the range of $\sim 2.75 -3.9$ which we obtained using the \hst{} spectra. We observed negligible intrinsic reddening for all the sources except for I~Zw~1 (see Table~\ref{tab:balmDec}).  It is possible that the reddening suffered by the far UV spectra is somewhat different than that in the optical band, and the  Balmer lines in the optical band may not be the true indicators of extinction in the FUV band. For example, as noted by \citet{pottasch1960balmer}, a higher optical depth of the higher order lines in the Balmer series can affect the line ratio. In this case, $H_\alpha$ radiation is scattered less while some fraction of $H_\beta$ radiation is re-absorbed and emitted in Paschen $\alpha$ and $H_\alpha$ and thereby further steepening the ratio. Apart from hydrogen lines, other line ratios such as \ion {He}{2} $\lambda1640/\lambda4686$, \ion{O}{1} $\lambda1302/\lambda8446$ can be used as reddening indicator. As pointed out by  \citet{1983grandiRedd}, these line ratios also suffer from re-absorption or scattering in such a way that may be seen as reddening. Another possibility is that the continuum may not suffer the same reddening as the broad emission lines. As we derived the internal extinction by fitting  the high-resolution \hst{} spectra (see section~\ref{subsec:hst_spectral}), some of the above issues are unlikely to affect our results on the continuum shape.

\subsection{Intrinsic far UV continuum and the non-relativistic accretion disk model}
\begin{figure*}
    \epsscale{0.55}
\plotone{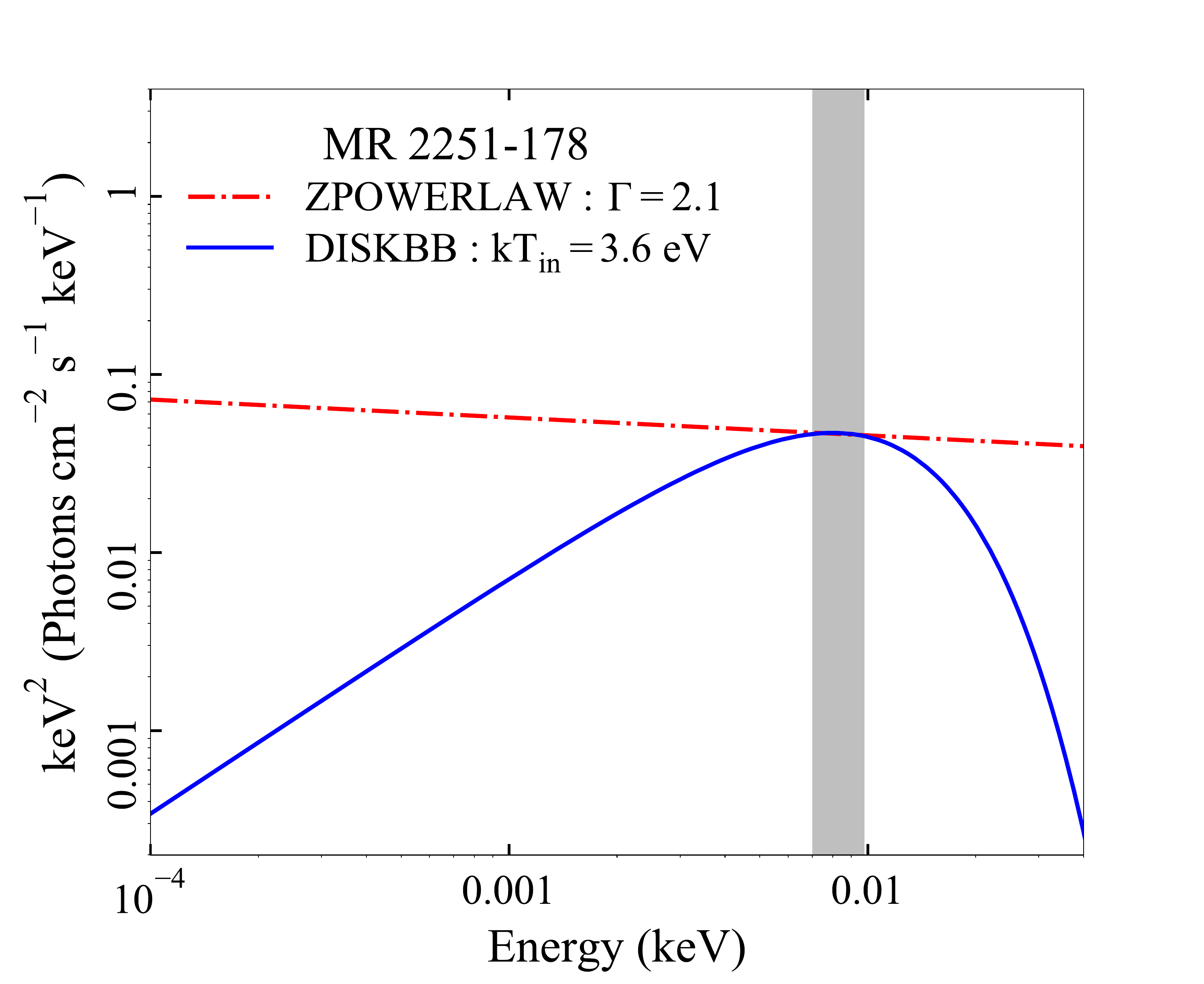}
%\end{figure}
 %\begin{figure}
    \epsscale{0.55}
\plotone{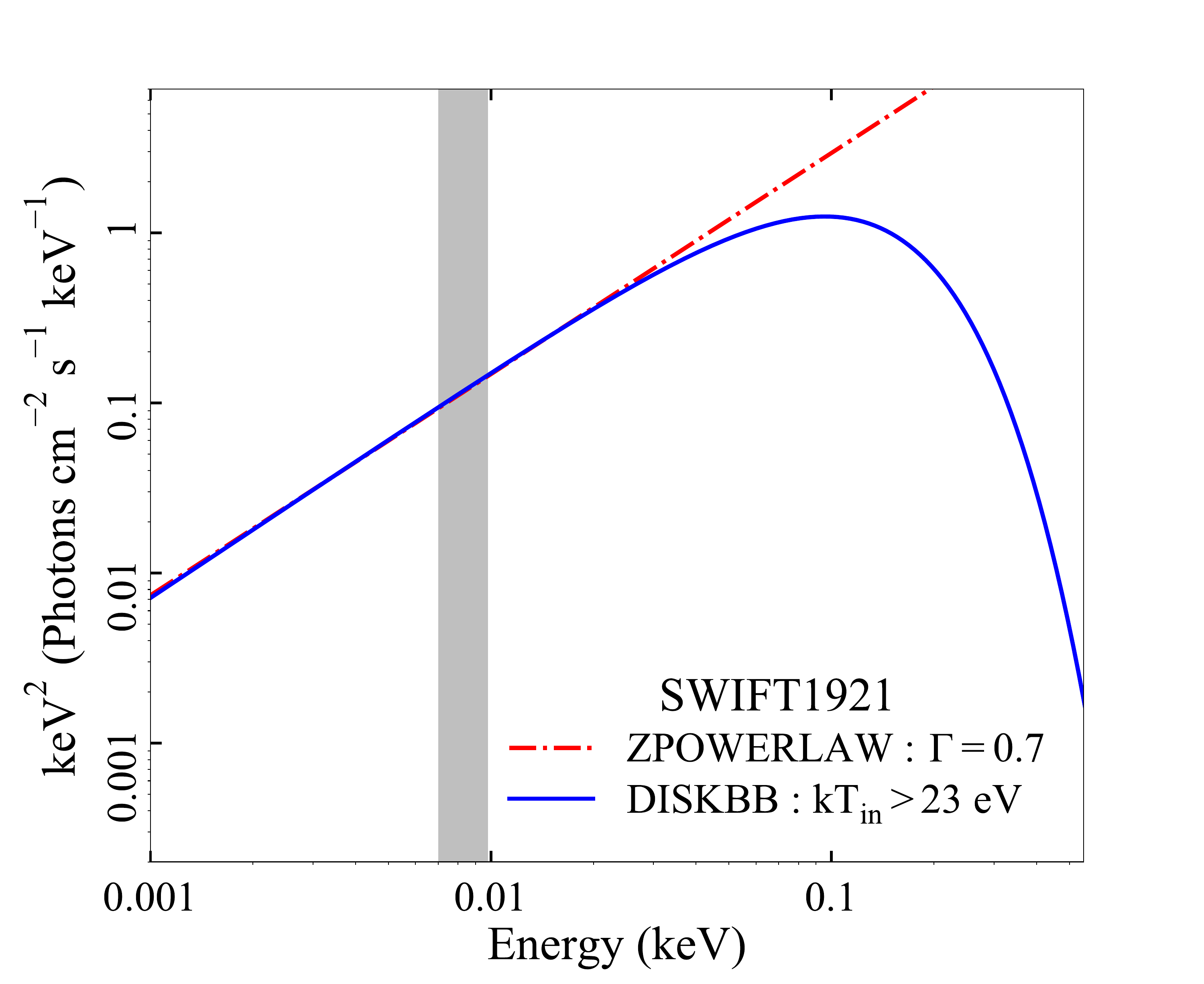}
\caption{The best-fitting \texttt{ZPOWERLAW} (red dot dashed line) and the \texttt{DISKBB} (blue solid line) spectral components derived for  MR~2251--178 and SWIFT1921 (see Section~\ref{subsec:uvit_spectral} and Table~\ref{tab:index}). The power-law slope ($\alpha$) is redder ($\sim -1.1$) for MR~2251--178 (left panel) whereas it is similar ($\sim 0.3$) for SWIFT1921 (right panel) compared to that predicted by the standard accretion disk model. The gray-shaded regions mark the FUV band of $0.007-0.0097\kev$.}
\label{fig:pow_bb_n74_mr22}
\end{figure*}

The far UV spectra when fitted with the multi-color accretion disk model (\texttt{DISKBB}), describing a non-relativistic accretion disk around a non-rotating black hole, provided  inner disk temperatures in the range of $3.6-5.8\ev$ (see Table~\ref{tab:index}).
While both the accretion disk and the  simple power-law models resulted in similar fit-statistic,  the apparent discrepancy between the redder power-law slopes ($\alpha$) compared to that  predicted by the standard disk model for most of our AGN is due to the fact that the FUV band does not always lie in the power-law part of disk emission for AGN with different masses and accretion rates. For the sources where the peak disk temperature is higher ($>10\ev$), we obtained slopes similar to the standard disk model.  In Fig.~\ref{fig:pow_bb_n74_mr22}, we compare the best-fit \texttt{ZPOWERLAW} and the \texttt{DISKBB} models for MR~2251--178 and SWIFT1921. In the case of SWIFT1921 with $M_{BH}=3.9\times 10^7 ~\rm{M_\odot}$ and $\dot{m}=0.1$, the FUV band falls in the region of the spectrum with the power-law shape, and we indeed find a slope $\alpha\sim 0.3$ consistent with that predicted by the standard disk model. While in the case of MR~2251--178 with $M_{BH} = 3.16\times10^8 ~\rm M_\odot$ and $\dot{m} = 0.02$, the emission in the FUV band arises from the innermost disk regions responsible for the exponentially declining continuum and we find a steep power-law slope, $\alpha=-1.1$.

\begin{figure}
\epsscale{1.2}
\plotone{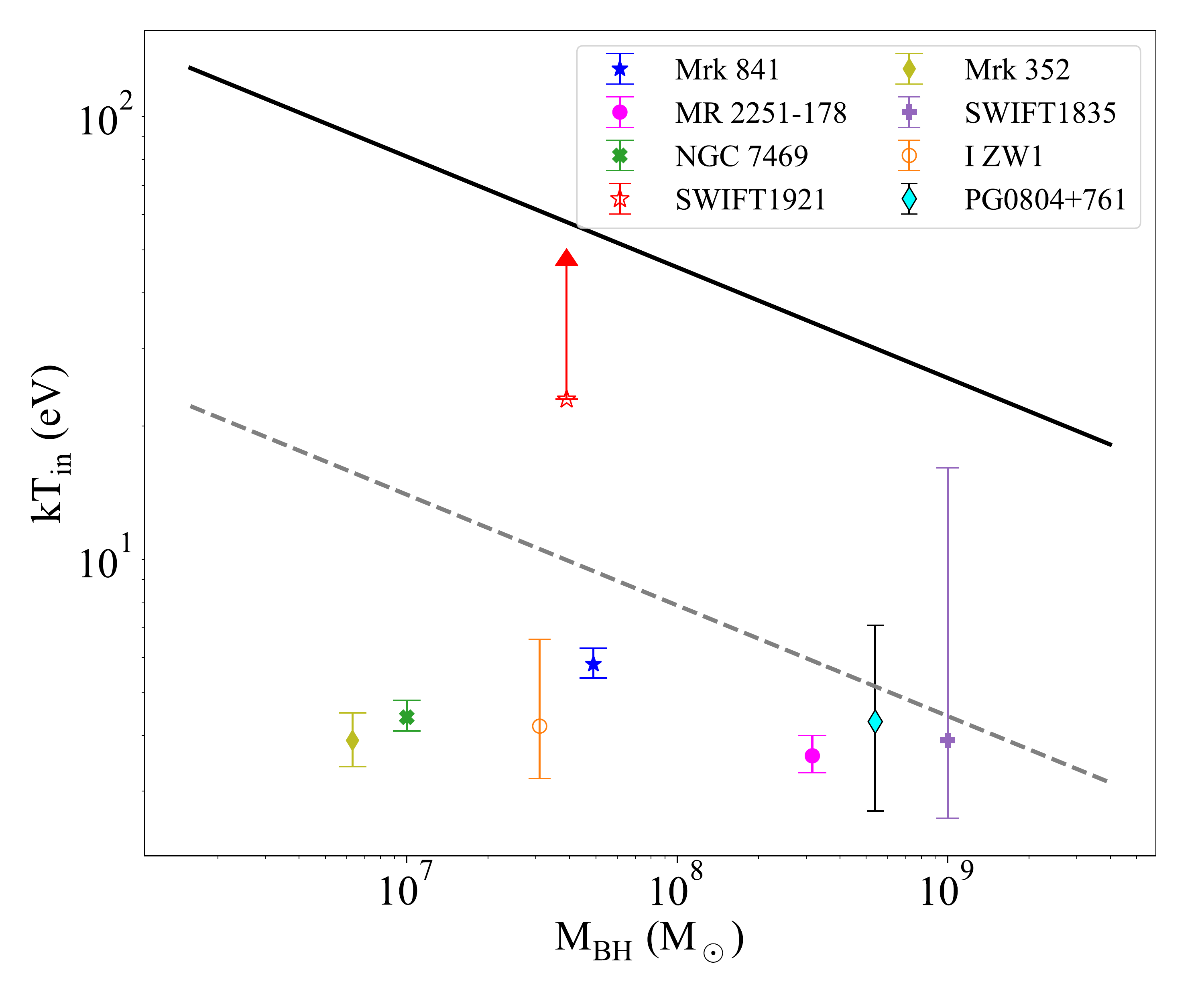}
   \caption{Peak effective disk temperatures  calculated using  Eq.~\ref{eq5:stdtemp}. The black solid line corresponds to $\dot{M}/\dot{M_E} = 1$ and $r = 1.69r_g$, while the  gray solid line corresponds to  $\dot{M}/\dot{M_E} = 0.1$ and $r = 8.17r_g$. The best-fit temperatures obtained using the \texttt{DISKBB} model are shown with different symbols. The lower limit is marked with an upward-pointed arrow (SWIFT1921). 
   %All of the observed points lie around or below the gray dashed line. The upward-pointed arrows show the upper limit to temperature is not constrained.
   }
   \label{fig:inDis_temp} 
\end{figure}

We compare the best-fit inner disk temperatures derived from our spectral analysis with the peak temperature expected from standard disks. We calculated the peak disk temperatures using the well-known relation \citep{netzer_2013}, 
 \begin{equation} \label{eq5:stdtemp}
     kT(r) \simeq 111\times~\Big(\frac{ \dot M}{\dot M_E}\Big)^{1/4} M_{8}^{-1/4} (r/r_g)^{-3/4} f(r)^{1/4}\ev
 \end{equation}
where, $f(r)$ is given by $1-\big(\frac{r_{in}}{r}\big)^{1/2}$, $r_g = \frac{GM_{BH}}{c^2}$ and $M_8$ is black hole mass ($M_{BH}$) in the unit of $10^{8}\rm M_\odot$. We used the known black hole masses, and calculated the peak disk temperatures (observed at radius $r\sim 8.17r_g$ for Schwarzschild black hole and $\sim 1.69r_g$ for Kerr black hole)  for two different accretion rates $\dot{m} = \frac{\dot M}{\dot M_E} = 0.1$  at $8.17r_g$ and $\dot{m} = 1$ at $1.69r_g$. Fig.~\ref{fig:inDis_temp} compares the predicted peak temperatures for the two values of the accretion rates and the best-fit inner disk temperatures derived from the FUV data using the \texttt{DISKBB} model. Clearly, our best-fit inner-disk temperatures are lower (possibly except for Mrk~841 and SWIFT1921) than the peak temperatures predicted for accretion disks around maximally rotating SMBHs accreting at the Eddington rates. Except for NGC~7469 and Mrk~352, the derived temperatures are  similar to that predicted for non-rotating SMBHs accreting at $10\%$ of the Eddington rate. This suggests that either the AGN have truncated accretion disks around highly spinning SMBHs  or full accretion disks  around slowly or non-rotating SMBHs with low $\dot{m} \lesssim 0.1$. The best-fit temperatures for NGC~7469 and Mrk~352 are much lower than that predicted for non-rotating SMBHs accreting at $\dot{m}=0.1$ (see Fig.~\ref{fig:inDis_temp}). The relativistic accretion disk models further support the possibility of truncated disks in these AGN. 
%It is possible that the standard disks do not extend to the innermost  stable circular orbit but the inner regions are filled with a warm corona that is thought to be responsible for the soft X-ray excess observed in many AGN \citep[see e.g.,][]{done2012intrinsic,2018MNRAS.480.1247K}.  

 %According to Eq.~\ref{eq5:stdtemp}, assuming  material is accreting at Eddington limit, ($\dot M = \dot M_{Edd}$) and radius $\rm r = 8r_g$, the peak disk temperatures ($\rm kT_{in}$) are $25$ eV (Mrk~352), $22.5$ eV (NGC~7469), $16.9$ eV (I~Zw~1), $16$ eV (SWIFT1921), $15$ eV (Mrk 841), $9.48$ eV (MR~2251-178), $7.8$ eV (PG 0804+761) and $7.1$ eV (SWIFT1835). In\texttt{DISKBB}, f(r) = 1, i.e., temperature estimated for $\rm r > r_{in}$. We observed inner disk temperatures mostly lie below that predicted by Eq.~8. If we assume $\dot M = 0.1\times\dot M_{Edd}$ and r $= 6r_g$, most of the observed values of temperature lie around this region (See the gray straight line in Fig.~\ref{fig:inDis_temp}).  It can be inferred that the disk is accreting at a low accretion rate ($\sim 0.1 \dot{M_E}$) assuming the black holes are non-rotating. The model predicted a normalized mass accretion rate in the range of 0.0001 to 0.5 for an inclination angle of 0$^\circ$. Only I~Zw~1 resulted in the accretion ratio of $\sim 5$. In real scenarios, most of the black holes are observed to be rotating. In that case, the best-fit observed temperatures reflect either lower mass accretion rates ($\frac{\dot M }{\dot M_{Edd}}<1$) or truncated disk scenario or both. 

 The emission from the inner disk regions is subjected to relativistic effects which the multi-color disk model does not account for. The \texttt{DISKBB} model also does not account for the color-correction factor which can make the inner disk temperature higher than that obtained using the \texttt{DISKBB} model. These effects are accounted for in the relativistic accretion disk models which we discuss below.
 %Another effect that is not considered here is the color correction factor on the disk emission resulting in a modified multi-color black body. The inner disk temperature can be higher than that obtained using\texttt{DISKBB}. 

 %Also, the inner disk emission will be subjected to various relativistic effect which is not considered by\texttt{DISKBB}. 
 %In the next sub-section we discuss the effect of color-correction and relativistic effects on the intrinsic disk emission.
 
\subsection{Relativistic Standard disks}
We used two different relativistic disk models - \texttt{OPTXAGNF} and \texttt{ZKERRBB}. While the \texttt{ZKERRBB} is a fully relativistic model  with radius independent color-correction factor and no provision for truncation, the \texttt{OPTXAGNF} model uses the temperature profile of a relativistic disk which allows for radius dependent color-correction and disk truncation but does not account for light bending. For I~Zw~1, SWIFT1921 and SWIFT1835,  the \texttt{ZKERRBB} model provided either better or similar quality fit as the \texttt{OPTXAGNF} model, and our data suggested untruncated accretion disks in these AGN.  We were not able to constrain the spin parameter and disk inclination, therefore we used reasonable or previously estimated values. 

For Mrk~841, MR~2251--178, PG0804, NGC~7469, and Mrk~352 we obtained a better fit with the \texttt{OPTXAGNF} model compared to the \texttt{ZKERRBB} model.
  We were able to well constrain the inner disk radii ($r_{cor}$) of $35-125r_g$ and $50-135r_g$ for  NGC~7469 and Mrk~352, respectively, irrespective of the spin parameter.  The disk in MR~2251--178 may also be truncated at a smaller radius ($\sim 16r_g$)  if the SMBH is spinning at a high rate. 
  %The data on these AGN suggested a truncated disk with an inner radius of $16.5_{-4}^{+3.5} r_{g}$, $99_{-40}^{+27}r_g$ and $101_{-40}^{+33}r_g$. 
  The large inner radii for NGC~7469 and Mrk~352 are consistent with the absence of broad, relativistic iron K$\alpha$ lines. 
 X-ray spectroscopy of NGC~7469 has shown only a  narrow iron line (FWHM $\sim 2700 - 3400 ~\rm kms^{-1}$)   \citep{mehdipour2018multi,andonie2022localizing}. A narrow Fe K$\alpha$ line has been observed from Mrk~352 \citep{winter2008x}. 
 The disk truncation we inferred in NGC~7469 and Mrk~352 may be similar to the low/hard state of Galactic black hole X-ray binaries (BHB)
 \citep[see][]{esin1997advection,done2007modelling,bollimpalli2020looking} 
 %In this
%case, the inner disc is replaced by a hot, optically thin, Advection-
%Dominated Accretion Flow (ADAF)-like flow (Narayan & Yi 1994),
%which emits in X-rays. This could also be the case with low-
%luminosity AGN, like Low Ionisation Nuclear Emission Line Region
%galaxies (LINERS). 

The low/hard state of BHB is observed at $\sim 1-2\%$ of the Eddington limit.
We found $L/L_{Edd} \sim 1.6\%$ for Mrk~352 using the \texttt{OPTXAGNF} model, this may be similar to the low/hard state of BHB. However, NGC~7469 has a much higher bolometric fraction ($L/L_{Edd} \sim 57\%$), and it appears to be similar to IC~4329A with a possible truncated disk at high bolometric fraction \citep{dewangan2021astrosat}.
Another possibility for the apparent truncation of accretion disks could be due to the presence of soft X-ray excess emitting region in the innermost regions of the disk such as that predicted by the intrinsic Comptonised disk models \citep{done2012intrinsic,2018MNRAS.480.1247K}. The simultaneously acquired X-ray data along with the UV data presented here  with \astrosat{} will help us to investigate the connection between the inner disk and the soft excess emitting region, which we plan in a future paper.

%{\color{blue}By using the fully relativistic accretion disk model \texttt{ZKERRBB}, we obtained similar quality fits as  \texttt{DISKBB} for Mrk~841 and SWIFT1921. These sources also showed high inner disk temperature, therefore requiring larger color-correction on the emitted spectrum. }
%However, the inclination angle and the mass accretion rate were not well constrained (See Table~\ref{??}).
%
%found the  best-fit free parameters in\texttt{ZKERRBB} inclination angle ($i$) varies from $0^\circ-52^\circ$ and mass accretion rate ($\dot M$) from $0.01-1.33 M_\odot~yr^{-1}$. This resulted in a similar quality fit as with\texttt{DISKBB} or power-law. 
%However, the inclination angle and $\dot{M}$ are not well constrained, we could only derive upper limits. 
%Eddington ratio ($L_{Bol}/L_{Edd}$)  ranges from $0.05-20$ for $a^\star=0$ and $0.09-61$ for $a^\star=0.998$. 
Since the relativistic effects are expected to dominate in the innermost regions of the disk that emits at shortest UV wavelengths, it is difficult to constrain all the parameters with the limited band-pass of the UVIT/FUV gratings. It is also possible that the thermal Comptonisation of UV photons into the broadband X-ray power-law component and the possible presence of optically thick and warm Comptonising component giving rise to the soft excess emission can modify the intrinsic UV continuum from the disk. Broadband UV, soft, and hard X-ray spectral modeling can help investigate the innermost disk regions. We hope to address these issues by using the far UV grating spectra, the soft and hard X-ray data simultaneously acquired with \astrosat{}.

Although the disk emits in the EUV/optical bands, the intrinsic disk emission may be modified by X-ray reprocessing. Observations of optical/UV variations lagging behind the variations in the X-ray emission by a few days in several AGN have been interpreted in terms of the  X-ray reprocessing into UV/optical \citep{haardt1991two,mchardy2014swift}. It is thought that the X-rays from the hot corona illuminate the disk, and a part of the X-ray emission is absorbed, reprocessed, and re-radiated in the UV/optical band. Such reprocessing can alter the shape of the accretion disk continuum. We plan to use physical disk/corona models that  include the X-ray reprocessing and reflection e.g., the \texttt{KYNSED} model \citep{2022A&A...661A.135D} and study the disk/corona emission using our \astrosat{} UV/X-ray observations.

\section{Conclusion}\label{sec:concl}

We performed far UV grating spectroscopy of eight AGN using \astrosat{}/UVIT observations, and derived intrinsic continuum emission after accounting for the Galactic and intrinsic extinction, contributions from the NLR/BLR including Fe~II complex, etc. We then investigate the shape of the intrinsic continuum in the context of accretion disk emission. The main results of our work are as follows. 
The far UV spectral slopes of the intrinsic continua of the eight AGN are in the range of $\sim -1.1$ to $\sim0.3$. For most AGN, these slopes are redder than that expected from the standard disk models in the optical/UV band. 
   The limited-band far UV data are consistent with the accretion disk models. The derived temperatures are generally lower than the highest possible disk temperatures expected for accretion disks around  maximally spinning SMBHs accreting at the Eddington limit. 
Observed peak effective temperatures  indicate the possibility of truncated disks if the SMBHs are spinning at the maximal rate. Irrespective of the spin parameter, we find possible evidence for truncated disks in NGC~7469 and Mrk~352. 
%The color factor can play crucial role in shaping the spectrum from the inner accretion disk. 
The bolometric fraction  of the disk emission ranges from $\sim 0.003$ (SWIFT1835) to $\sim 5.9$ (SWIFT1921).
  A number of effects such as the X-ray reprocessing, Thermal Comptonisation, soft X-ray excess may affect the shape of the UV continuum. Joint spectral modeling of the simultaneous acquired UV/X-ray data, which we plan in a future paper, may be necessary to further investigate the emission from the innermost disk/corona regions.

\begin{acknowledgments}
This publication uses the data from Indian
Space Science Data Centre (ISSDC) of the\textit{AstroSat} mission of the Indian Space Research Organisation (ISRO). This publication uses UVIT data processed by the payload operations center at IIA. The UVIT is built in collaboration between IIA, IUCAA, TIFR, ISRO, and CSA.  UVIT data were reprocessed by CCDLAB pipeline \citep{Postma_2017}. This publication used archival COS spectra from \textit{HST} data archive (\url{https://archive.stsci.edu/hst/search.php}) and FOS spectra from  \citep{kuraszkiewicz2004emission} (\url{http://hea-www.harvard.edu/~pgreen/HRCULES.html}). This research has used the Python and Julia packages. This research has used the SIMBAD/NED database. S.K. acknowledges the University Grant Commission (UGC),
Government of India, for financial support. K. P. Singh thanks the Indian National Science Academy
for support under the INSA Senior Scientist Programme.\\

\noindent{}
Facility: \astrosat{}, \hst{}.\\

\noindent{}
Software: CCDLAB \citep{Postma_2017}, {\sc Sherpa }\citep{2001SPIE.4477...76F}, SAOImageDS9 \citep{joye2003new}, Julia \citep{bezanson2017julia}, Astropy \citep{astropy2013astropy}.
\end{acknowledgments}

%% For this sample we use BibTeX plus aasjournals.bst to generate the
%% the bibliography. The sample631.bib file was populated from ADS. To
%% get the citations to show in the compiled file do the following:
%%
%% pdflatex sample631.tex
%% bibtext sample631
%% pdflatex sample631.tex
%% pdflatex sample631.tex

\bibliography{faruv_grating_agn}{}
\bibliographystyle{aasjournal}
\vspace{10mm}

\appendix

\section{Absorption lines used in the \hst{} spectra }
\label{sec:apendix}

\begin{deluxetable*}{cccccccccccc}[h]
\tablenum{A1}
\tablecaption{Absorption lines in the HST spectra. $E_{o}$ is the observed line centroid, $\sigma$ is in the unit of $10^{-6}$ keV, and Strength is in the unit of $10^{-6}$ keV.}
\label{tab:abs_lin1}
%\tablewidth{650pt}
\tablehead{
\multicolumn4l{Mrk 841}&\multicolumn4l{MR 2251-178} &\multicolumn4l{ PG0804}\\
\colhead{$\lambda_{rest}$} & \colhead{$E_{o}$} &  \colhead{$\sigma$} & \colhead{Strength} &\colhead{$\lambda_{rest}$} & \colhead{$E_{o}$} &  \colhead{$\sigma$} & \colhead{Strength} &\colhead{$\lambda_{rest}$} & \colhead{$E_{o}$} &  \colhead{$\sigma$} & \colhead{Strength}\\
(\AA)&(eV)&(keV)&(keV)&(\AA)&(eV)&(keV)&(keV)&(\AA)&(eV)&(keV)&(keV)}
\startdata
1204 & 9.935 & $0.78_{-0.06}^{+0.06}$& $1.04_{-0.06}^{+0.06}$ & 1206.2 & 9.659 & $2.62_{-0.57}^{+0.57}$ & $1.14_{-0.21}^{+0.21}$ & 1183.6 & 9.522 & $0.95_{-0.01}^{+0.01}$ & $4.8_{-0.1}^{+0.1}$ \\
1206.5& 9.914 & $0.87_{-0.05}^{+0.05}$ & $1.65_{-0.06}^{+0.06}$ &1214 &9.598 & $5.4_{-0.1}^{+0.1}$ & $1.2_{-0.3}^{+0.3}$ & 1185.6 & 9.506 & $0.84_{-0.01}^{+0.01}$ & $6.4_{-0.1}^{+0.1}$  \\
1208.8 & 9.895 & $1.07_{-0.01}^{+0.01}$ & $7.75_{-0.11}^{+0.11}$&1215 & 9.587 & $5.9_{-0.1}^{+0.1}$ & $18.2_{-0.4}^{+0.4}$ &1213 & 9.292& $1.228_{-0.007}^{+0.007}$& $12.14_{-0.09}^{+0.09}$ \\
1209.6 & 9.889 & $0.59_{-0.02}^{+0.02}$ & $1.71_{-0.05}^{+0.05}$ & 1221 & 9.539 & $4.2_{-0.5}^{+0.5}$ & $2.5_{-0.4}^{+0.4}$ & 1214 & 9.283 & $0.91_{-0.01}^{+0.01}$ & $1.89_{-0.02}^{+0.02}$ \\
1211.5 & 9.874 & $0.99_{-0.03}^{+0.03}$ & $1.38_{-0.04}^{+0.04}$ & 1224 & 9.516 & $4.9_{-0.5}^{+0.5}$ & $4.1_{-0.5}^{+0.5}$ & 1217.9 & 9.254 & $0.86_{-0.01}^{+0.01}$ & $5.13_{-0.06}^{+0.06}$  \\
1213 & 9.861 & $1.42_{-0.01}^{+0.01}$ & $9.8_{-0.1}^{+0.1}$ &1226.7 &9.499 & $2.8_{-0.3}^{+0.3}$ & $2.2_{-0.2}^{+0.2}$& 1218 & 9.252 & $0.52_{-0.01}^{+0.01}$ & $1.43_{-0.04}^{+0.04}$   \\
1215 & 9.845 & $0.67_{-0.01}^{+0.01}$ & $2.12_{-0.03}^{+0.03}$ & 1238 & 9.409 & $4.3_{-0.2}^{+0.2}$ & $8.6_{-0.4}^{+0.4}$& 1266.8 & 8.897 &  $1.08_{-0.04}^{+0.04}$ & $2.38_{-0.07}^{+0.07}$ \\
1216 &9.838 & $1.04_{-0.01}^{+0.01}$ & $8.21_{-0.08}^{+0.08}$ & 1242.6 & 9.377 & $3.3_{-0.3}^{+0.3}$ & $5.2_{-0.4}^{+0.4}$& 1275 & 8.84 & $1.19_{-0.08}^{+0.08}$ & $1.19_{-0.06}^{+0.07}$  \\
1216.7 & 9.831 & $0.48_{-0.03}^{+0.03}$ & $0.58_{-0.03}^{+0.03}$ &  1255 & 9.284 & $4.7_{-0.5}^{+0.5}$ & $6.5_{-0.7}^{+0.7}$&  1387.6 & 8.122 & $0.75_{-0.01}^{+0.01}$ & $6.1_{-0.1}^{+0.2}$ \\
1258 & 9.507 & $0.93_{-0.03}^{+0.03}$ & $4.2_{-0.1}^{+0.1}$ & 1311& 8.888 & $2.4_{-0.9}^{+0.9}$ & $1.9_{-0.5}^{+0.5}$ &1407 & 8.01 & $1.08_{-0.03}^{+0.03}$ & $3.43_{-0.08}^{+0.08}$  \\
1287 & 9.292 & $0.97_{-0.03}^{+0.03}$ & $11.1_{-0.6}^{+0.7}$ &1435.8 &8.116& $2.1_{-0.3}^{+0.3}$ & $2.4_{-0.3}^{+0.3}$ & 1409 & 7.996 & $1.16_{-0.05}^{+0.05}$ & $1.65_{-0.07}^{+0.07}$ \\
1288 & 9.284 & $1.46_{-0.07}^{+0.08}$ & $3.6_{-0.2}^{+0.2}$ & 1456 & 7.999& $11.5_{-1.5}^{+1.5}$ & $8.1_{-0.8}^{+0.8}$ & 1518.6 & 7.422 & $0.64_{-0.02}^{+0.02}$ & $6.8_{-0.4}^{+0.4}$ \\
1344 & 8.897 & $0.85_{-0.03}^{+0.03}$ & $4.7_{-0.2}^{+0.2}$ &  1545.7 & 7.538 & $4.2_{-0.2}^{+0.2}$ & $7.2_{-0.3}^{+0.3}$ &  1551.2 & 7.265 & $0.39_{-0.02}^{+0.02}$ & $1.21_{-0.05}^{+0.05}$  \\
1353 & 8.840 & $0.86_{-0.04}^{+0.04}$ & $3.1_{-0.2}^{+0.2}$ &  1548 & 7.525 & $3.4_{-0.2}^{+0.2}$ & $4.6_{-0.3}^{+0.3}$ & 1553.8&7.253 & $0.63_{-0.05}^{+0.05}$ \\
1472.8 & 8.122 & $0.81_{-0.03}^{+0.03}$ & $6.3_{-0.3}^{+0.3}$ &1569 & 7.424 & $2.7_{-0.3}^{+0.3}$ & $2.7_{-0.2}^{+0.2}$ &\\
1493 & 8.01 & $0.82_{-0.03}^{+0.03}$ & $6.1_{-0.3}^{+0.3}$ & 1650.8 & 7.058 & $38_{-3}^{+3}$ & $124_{-34}^{+34}$ & \\
1496 & 7.996 & $0.83_{-0.03}^{+0.04}$ & $3.8_{-0.2}^{+0.2}$ & \\
1551.7 & 7.709 & $0.69_{-0.02}^{+0.02}$ & $2.68_{-0.07}^{+0.07}$ & \\
1611.8 & 7.422 & $0.068_{-0.04}^{+0.04}$ & $6.4_{-0.5}^{+0.7}$ & \\
\enddata
\end{deluxetable*}

\begin{deluxetable*}{cccccccccccc}
\tablenum{A2}
\tablecaption{Same as Table \ref{tab:abs_lin1} but for NGC~7469, I~Zw~1 and SWIFT1921.}
\label{tab:abs_lin2}
%\tablewidth{650pt}
\tablehead{
\multicolumn4l{NGC 7469}&\multicolumn4l{I Zw 1} &\multicolumn4l{SWIFT1921}\\
\colhead{$\lambda_{rest}$} & \colhead{$E_{o}$} &  \colhead{$\sigma$} & \colhead{Strength} &\colhead{$\lambda_{rest}$} & \colhead{$E_{o}$} &  \colhead{$\sigma$} & \colhead{Strength} &\colhead{$\lambda_{rest}$} & \colhead{$E_{o}$} &  \colhead{$\sigma$} & \colhead{Strength}\\
(\AA)&(eV)&(keV)&(keV)&(\AA)&(eV)&(keV)&(keV)&(\AA)&(eV)&(keV)&(keV)}
\startdata
1231.6& 9.908 & $3.0_{-0.6}^{+0.7}$ & $2.3_{-0.4}^{+0.5}$ & 1207.7&9.675&$2.3_{-0.1}^{+0.1}$ & $1.8_{-0.1}^{+0.1}$& 1205.9&9.914& $0.62_{-0.01}^{+0.02}$ & $1.26_{-0.03}^{+0.03}$ \\
1235.8& 9.874& $0.87_{-0.86}^{+0.92}$ & $0.4_{-0.2}^{+0.2}$ &1227& 9.522& $0.88_{-0.04}^{+0.04}$&$5.4_{-0.3}^{+0.3}$& 1209&9.888&$0.73_{-0.01}^{+0.01}$& $1.93_{-0.03}^{+0.03}$ \\
1240.6&9.836& $3.2_{-0.4}^{+0.4}$ & $3.8_{-0.4}^{+0.5}$& 1229&9.505& $0.65_{-0.03}^{+0.03}$& $3.3_{-0.2}^{+0.2}$& 1213&9.855 & $0.851_{-0.005}^{+0.005}$& $5.82_{-0.04}^{+0.04}$ \\
1282&9.514& $10.5_{-1.3}^{+1.4}$& $6.9_{-0.9}^{+1.0}$&1257&9.291& $1.08_{-0.08}^{+0.06}$& $9.9_{-0.7}^{+0.9}$& 1215 & 9.836 & $1.203_{-0.005}^{+0.005}$& $11.9_{-0.1}^{+0.1}$ \\
1313 & 9.2911& $8.9_{-0.9}^{+1.0}$& $10.6_{-1.2}^{+1.2}$& 1258.8&9.282& $0.42_{-0.06}^{+0.08}$ & $1.2_{-0.1}^{+0.1}$& 1231.6 & 9.707 & $0.38_{-0.02}^{+0.02}$ & $0.52_{-0.02}^{+0.02}$  \\
1503 & 8.118& $1.9_{-0.3}^{+0.4}$ & $1.7_{-0.3}^{+0.3}$& 1261&9.265& $9.2_{-1.1}^{+1.3}$& $10.9_{-2.4}^{+3.4}$& 1234& 9.685 & $0.27_{-0.04}^{+0.04}$ & $0.12_{-0.02}^{+0.02}$ \\
1524& 8.007& $7.7_{-0.8}^{+0.9}$& $5.0_{-0.5}^{+0.5}$ &1303&8.967& $7.2_{-1.0}^{+1.8}$ & $13.2_{-4.5}^{+1.5}$& 1287 & 9.29 & $1.26_{-0.01}^{+0.01}$ & $13.2_{-0.2}^{+0.2}$  \\
1539.6& 7.926& $2.4_{-0.3}^{+0.3}$& $1.9_{-0.2}^{+0.2}$& 1313& 8.896 & $1.1_{-0.2}^{+0.2}$ & $1.8_{-0.4}^{+0.3}$& 1288&9.282& $0.74_{-0.02}^{+0.02}$ & $2.14_{-0.05}^{+0.05}$ \\
1542&7.913&$2.8_{0.3}^{+0.4}$&$1.8_{-0.2}^{+0.2}$& 1394.7& 8.378&$11.2_{-0.8}^{+0.8}$ & $8.1_{-1.0}^{+1.0}$& 1321 & 9.049& $1.1_{-0.1}^{+0.1}$ & $0.57_{-0.05}^{+0.05}$  \\
& & & &1438.7&8.122& $0.66_{-0.06}^{+0.07}$ & $4.1_{-0.4}^{+0.4}$ & 1344 & 8.895 & $1.21_{-0.03}^{+0.03}$ &$3.49_{-0.07}^{+0.07}$ \\
& & & & 1458.9&8.01& $14.5_{-1.6}^{+1.9}$ & $11.2_{-1.3}^{+1.1}$& 1352.6& 8.838& $1.07_{-0.04}^{+0.04}$& $1.69_{-0.06}^{+0.06}$ \\
& & & & 1515.8&7.709& $6.62_{-0.09}^{+1.0}$ & $1.9_{-0.2}^{+0.3}$& 1472&8.121& $0.82_{-0.01}^{+0.01}$ & $6.3_{-0.1}^{+0.1}$ \\
& & & &1538&7.595& $0.9_{-0.2}^{+0.2}$& $1.1_{-0.2}^{+0.2}$& 1492.8& 8.008&$1.15_{-0.02}^{+0.02}$ & $5.3_{-0.1}^{+0.1}$\\
& & & &1574.6& 7.421& $0.70_{-0.08}^{+0.09}$ & $3.3_{-0.4}^{+0.5}$&1495&7.995& $1.08_{-0.03}^{+0.03}$ & $2.98_{-0.08}^{+0.08}$ \\
& & & & & & & & 1504.6&7.946& $0.26_{-0.03}^{+0.03}$ & $0.34_{-0.03}^{+0.03}$\\
& & & & & & & & 1551& 7.707& $0.67_{-0.01}^{-0.01}$ & $3.02_{-0.05}^{+0.05}$\\
& & & & & & & & 1597.7& 7.482& $0.23_{-0.04}^{+0.04}$ & $0.36_{-0.04}^{+0.04}$\\
& & & & & & & & 1611& 7.420& $0.72_{-0.02}^{+0.02}$ & $6.3_{-0.2}^{+0.3}$\\
\enddata
\tablecomments{We did not use any absorption line for Mrk~352.}
\end{deluxetable*}

%% This command is needed to show the entire author+affiliation list when
%% the collaboration and author truncation commands are used.  It has to
%% go at the end of the manuscript.
%\allauthors

%% Include this line if you are using the \added, \replaced, \deleted
%% commands to see a summary list of all changes at the end of the article.
%\listofchanges

\end{document}